\RequirePackage{fix-cm}
\documentclass[allclo, onecollarge, square, comma, sort&compress, natbib]{svjour2}
\bibpunct{[}{]}{,}{n}{}{,} % to get "[numbered]" references from natbib

\usepackage[utf8]{inputenc}%
\usepackage{amsmath}
\usepackage{amssymb}
\usepackage{color}
\usepackage{graphicx}
\usepackage{placeins}
\usepackage{supertabular}
\usepackage{xspace}
\usepackage{longtable}

\usepackage[mathscr,scaled=1.15]{urwchancal}
\DeclareFontFamily{OT1}{pzc}{}
\DeclareFontShape{OT1}{pzc}{m}{it}%
{<-> s * [1.15] pzcmi7t}{}
\DeclareMathAlphabet{\mathpzc}{OT1}{pzc}{m}{it}

\definecolor{purple}{rgb}{0.5,0,0.5}
\definecolor{blue}{rgb}{0.0,0,0.9}
\definecolor{prdblue}{rgb}{0.133,0.118,0.498}
\usepackage[colorlinks=true, pdfstartview=FitV, linkcolor=prdblue, citecolor= prdblue, urlcolor=prdblue]{hyperref}

\usepackage{relsize}
\def\babar{\mbox{\slshape B\kern-0.1em{\smaller A}\kern-0.1em
    B\kern-0.1em{\smaller A\kern-0.2em R}}}

\journalname{Few-Body Systems}
\begin{document}

\title{$\,$\\[-13ex]\hspace*{\fill}{\normalsize{\sf\emph{Preprint no}. NJU-INP 022/20}}\\[8.3ex]
%Selected Topics in EicC Science
Selected Science Opportunities for the EicC
 %\\ from a Poincar\'e-covariant Faddeev equation%\thanks{Grants or other notes
}

\titlerunning{Selected Science Opportunities for the EicC}

\author{Xurong Chen
    \and
        Feng-Kun Guo
    \and
        Craig D.~Roberts
    \and
        Rong Wang
}


\institute{
    Xurong Chen \at
    Institute of Modern Physics, Chinese Academy of Sciences, Lanzhou 730000, China\\
    School of Physical Sciences, University of Chinese Academy of Sciences, Beijing 100049, China\\
    \email{\href{mailto:xchen@impcas.ac.cn}{xchen@impcas.ac.cn}}\\[1ex]
    $\,$Feng-Kun Guo \at
    CAS Key Laboratory of Theoretical Physics, Institute of Theoretical Physics, Chinese Academy of Sciences,
    %Zhong Guan Cun East Street 55,
    Beijing 100190, China\\
    School of Physical Sciences, University of Chinese Academy of Sciences, Beijing 100049, China\\
    \email{\href{mailto:fkguo@itp.ac.cn}{fkguo@itp.ac.cn}}\\[-1ex]
    $\,$\\Craig D.\ Roberts \at
    School of Physics, Nanjing University, Nanjing, Jiangsu 210093, China\\
    Institute for Nonperturbative Physics, Nanjing University, Nanjing, Jiangsu 210093, China\\
    \email{\href{mailto:cdroberts@nju.edu.cn}{cdroberts@nju.edu.cn}}\\[1ex]
    $\,$Rong Wang \at
    Institute of Modern Physics, Chinese Academy of Sciences, Lanzhou 730000, China\\
    School of Physical Sciences, University of Chinese Academy of Sciences, Beijing 100049, China\\
    \email{\href{mailto:rwang@impcas.ac.cn}{rwang@impcas.ac.cn}}\\[1ex]
}

\date{Received:
25 August 2020
%31 July 2020
%10 July 2020
%04 July 2020
%24 May 2020
}

\maketitle

\begin{abstract}
An electron ion collider has been proposed in China (EicC).  It is anticipated that the facility would provide polarised electrons, protons and ion beams, in collisions with large centre-of-mass energy.  This discussion highlights its potential to address issues that are central to understanding the emergence of mass within the Standard Model, using examples that range from the exploration of light-meson structure, through measurements of near-threshold heavy-quarkonia production, and on to studies of the spectrum of exotic hadrons.
\end{abstract}

\setcounter{tocdepth}{1}
\tableofcontents

\section{Introduction}
\label{intro}
%
%\input Introduction
%%The discovery of the proton is credited to Ernest Rutherford, who proved that the nucleus of the hydrogen atom (i.e. a proton) is present in the nuclei of all other atoms in the year 1917.
%
No single date can be identified as the beginning of the modern era in nuclear and particle physics; but one can, perhaps, associate the dawning of this new age with discovery of the proton \cite{RutherfordI, RutherfordII, RutherfordIII, RutherfordIV}.  Today, one century and more than fifty Nobel Prizes later, the Standard Model of Particle Physics (SM) \cite{Politzer:2005kc} ``\ldots \emph{offers a description of all known fundamental physics except for gravity, and gravity is something that has no discernible effect when particles are studied a few at a time}.''  With discovery of the Higgs boson \cite{Aad:2012tfa, Chatrchyan:2012xdj}, the SM is now complete.  Yet, some very fundamental questions remain unanswered; arguably most important amongst them: Whence mass?

When considering the origin of mass, the Higgs boson is important, as recognised by the Nobel Prize awarded to Englert and Higgs \cite{Englert:2014zpa, Higgs:2014aqa}: ``\emph{for the theoretical discovery of a mechanism that contributes to our understanding of the origin of mass of subatomic particles} \ldots''.  However, the Higgs boson alone is responsible for $\lesssim 2$\% of the visible mass in the Universe.  The remainder is contained in nuclei.  Moreover, since the atomic weight of a given nucleus is approximately just the sum of the masses of all the neutrons and protons (nucleons) it contains, then almost all that atomic weight is lodged within the nucleons.  Each nucleon has a mass $m_N \sim 1\,$GeV, \emph{i.e}.\ approximately 2000-times the electron mass, $m_e$.  The Higgs boson produces $m_e$, but what generates $m_N$?  This is the question posed above; and it is pivotal to the development of modern physics: how can science understand and explain the emergence of hadronic mass?

Emergent hadronic mass (EHM) is seemingly ubiquitous.  For consider that chromodynamics is a local non-Abelian gauge field theory.  As with all such theories formulated in four spacetime dimensions, no mass-scale exists in the absence of Lagrangian masses for the elementary degrees of freedom. There is no dynamics in a scale-invariant theory, only kinematics: the theory looks the same at all length-scales and bound states are impossible.  Accordingly, our Universe cannot exist.
As already indicated, a spontaneous breaking of symmetry, \emph{\`{a}} \emph{la} the Higgs mechanism, does not solve this problem.  Further highlighting the issue, $m_N$ is roughly 100-times larger than the Higgs-generated current-masses of the light $u$- and $d$-quarks, the valence degrees-of-freedom which define a nucleon, and the Lagrangian gluons are massless \cite{Zyla:2020}.

On the other hand, Nature supports composite Nambu-Goldstone (NG) bosons \cite{Nambu:1960tm, Goldstone:1961eq}: pions and kaons are massless in the absence of a Higgs mechanism and possess atypically small masses when the Higgs couplings to light quarks are nonzero.  This is especially true of the pion, whose mass is similar to that of the $\mu$-lepton \cite{Zyla:2020}.  In these systems, the strong interaction's $1\,$GeV mass-scale is hidden.  Here, dynamical chiral symmetry breaking (DCSB) is the active agent; and DCSB is a material consequence of EHM.

Within the SM, nucleons, pions, kaons, \emph{etc}., collectively known as hadrons, are believed to be described by \emph{quantum} chromodynamics (QCD).  In this theory, all hadrons are composites, built from quarks and/or antiquarks (matter/antimatter fields), held together by forces produced by the exchange of gluons (gauge fields).  These forces are unlike any previously encountered.  Extraordinarily, \emph{e.g}.\ they become weaker than Coulombic when two quarks are brought close together within a nucleon \cite{Politzer:2005kc, Wilczek:2005az, Gross:2005kv}.  However, all attempts to remove a single quark from within a nucleon and isolate it in a detector have failed.  Seemingly, then, the forces become huge as the separation between quarks is increased \cite{Wilson:1974sk}.  This brings the confinement problem to the foreground, some aspects of which are covered in the Yang-Mills Millennium Problem defined by the Clay Mathematics Institute \cite{Jaffe:Clay}. Confinement is critical because it ensures absolute stability of the proton.  In the absence of confinement, protons in isolation could decay; the hydrogen atom would be unstable; nucleosynthesis would be a chance event, having no lasting consequences; and without nuclei, there would be no stars and no living Universe.  Without confinement, our Universe cannot exist.  Crucially, many sound arguments indicate that confinement and EHM are inextricably linked \cite{Roberts:2016vyn}.

\begin{figure}[t]
%\vspace*{1.5ex}

\centering
\includegraphics[width=0.60\linewidth]{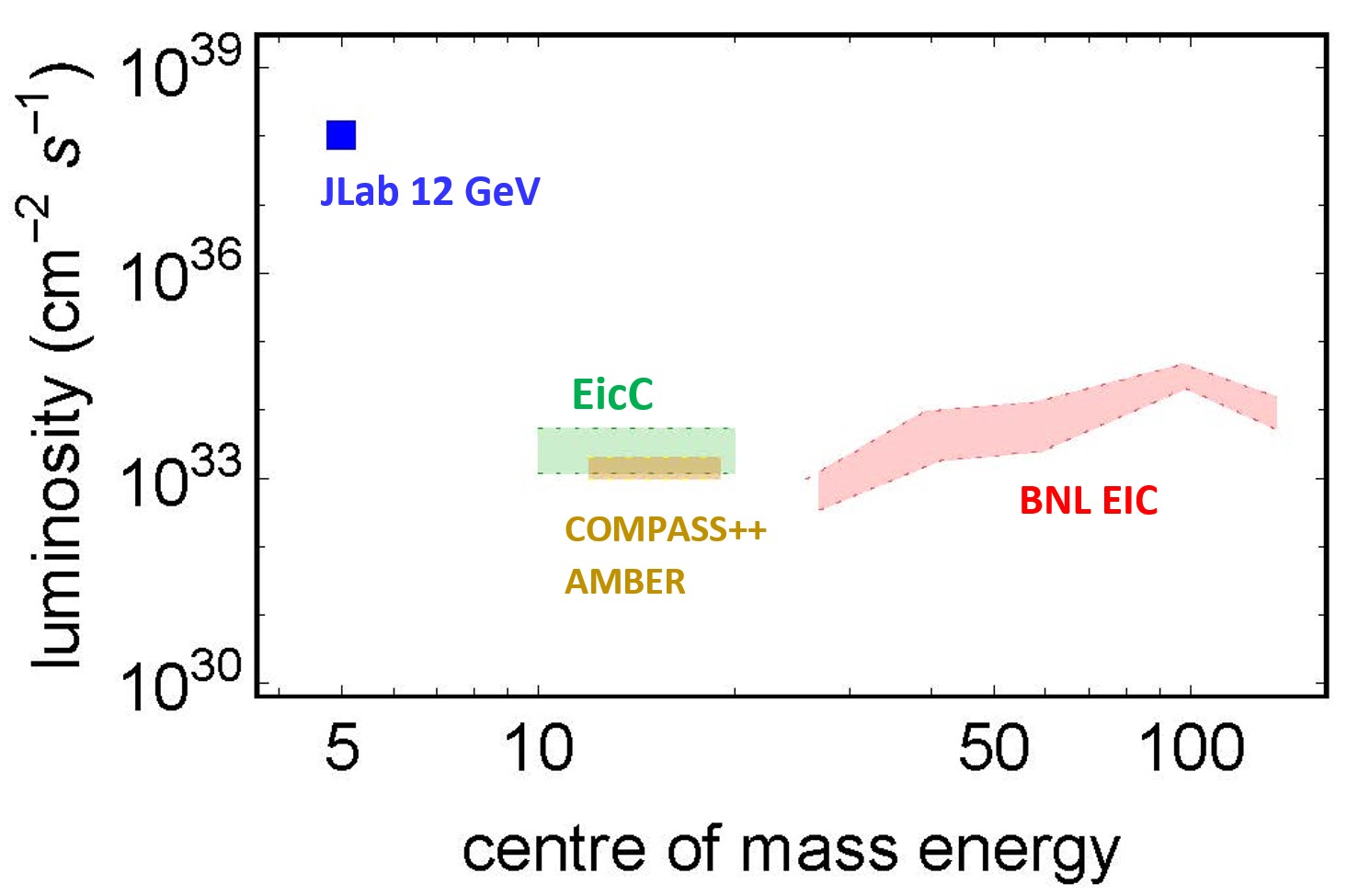}
%\end{center}
%\vspace{-.75cm}	
\caption{\label{EicCcomparison}
Comparison between the anticipated EicC luminosity and centre-of-mass (c.m.) energy coverage \cite{Chen:2018wyz, EicCWP, EicCWPEL} and the coverages of some other existing, planned, and discussed facilities.
}
\end{figure}

This is already a long list of challenges and conundrums.  There are more.  Yet, our Universe does exist.  EHM within QCD is a large part of the solution to the puzzles; and as science plans for the next thirty years, solving the problem of EHM has become a \emph{grand challenge}: it marks the SM's last frontier.  A new generation of experimental facilities are being developed in order to chart the way across.  Amongst them, the Beijing Electron Positron Collider (BEPC) \cite{doi:10.1142/11757}, the 12\,GeV Jefferson Laboratory (JLab) \cite{Dudek:2012vr, Burkert:2018nvj, Brodsky:2020vco, Carman:2020qmb} and the large hadron collider beauty experiment (LHCb) \cite{Aaij:2015tga} are already in operation and revealing surprises; and a new QCD facility (COMPASS++/AMBER) is planned at the European Laboratory for Particle Physics (CERN) \cite{Denisov:2018unjF}.  Further into the future, an electron ion collider (EIC) is expected to be built at Brookhaven National Laboratory (BNL) \cite{PhysicsToday73}, at a cost of between \$1.6-billion and \$2.6-billion, and an EIC in China (EicC) is under discussion \cite{Chen:2018wyz, EicCWP, EicCWPEL}.  The primary goal of each of these machines is to complete an array of carefully planned experiments that will provide the information necessary to understand and explain EHM.  Our discussion will describe a few of the many opportunities that might be exploited with the EicC.
As illustrated by Fig.\,\ref{EicCcomparison}, with currently anticipated design specifications, the EicC could both neatly fill a gap between JLab at 12\,GeV and the EIC at BNL and develop a powerful synergy with COMPASS++/AMBER.

%----------------------

We begin in Sect.\,\ref{massdichotomy} by discussing additional aspects of EHM and its consequences.
Section~\ref{LHstructure} then describes examples of light-hadron structure observables, accessible at EicC, which possess the capacity to reveal aspects of the mechanism(s) responsible for EHM.
Section~\ref{EStructure} complements and extends Sect.\,\ref{LHstructure}, presenting the results of simulations that demonstrate the viability of meson structure studies at EicC and highlighting additional efforts aimed at quantitative measurements of both the Higgs-boson contribution to the proton mass and proton GPDs.
The survey of opportunities is extended in Sect.\,\ref{exoticspectrum}, adding explanations of how understanding the hadron spectrum, especially the exotic states discovered in the past twenty years, and related experiments at EicC can assist in understanding EHM.
Section~\ref{epilogue} is an epilogue.

\section{Emergent Hadronic Mass}
\label{massdichotomy}
In field theory, scale invariance is expressed in conservation of the dilation current
\begin{equation}
\label{Tmumuzero}
\partial_\mu D_\mu = \partial_\mu (T_{\mu\nu} x_\nu)  = T_{\mu\mu} = 0\,,
\end{equation}
where $T_{\mu\nu}$ is the theory’s energy-momentum tensor, which satisfies $\partial_\mu T_{\mu\nu}=0$ owing to energy and momentum conservation.  The disastrous consequences of scale invariance that were explained in the Introduction, \emph{e.g}.\ the impossibility of bound states, are avoided in Nature through the instrument of quantum effects.  In quantising QCD, regularisation and renormalisation of (ultraviolet) divergences introduces a mass-scale, $\zeta$.  Consequently, mass-dimensionless quantities and other ``constants'' become dependent on $\zeta$, an outcome known as ``dimensional transmutation''.  This entails the appearance of a \emph{trace anomaly}, \emph{i.e}.\ interaction-induced violation of Eq.\,\eqref{Tmumuzero}:
\begin{equation}
\label{definetheta0}
 T_{\mu\mu} = \beta(\alpha(\zeta)) \, \tfrac{1}{4}\, G_{\alpha\beta}^a G_{\alpha\beta}^a =: \Theta_0 \,,
\end{equation}
where $\beta(\alpha(\zeta))$ is QCD's $\beta$-function, $\alpha(\zeta)$ is the associated running-coupling and $G_{\alpha\beta}^a(x)$ is the gluon field strength tensor \cite{Marciano:1977su}.  Thus, \underline{a mass-scale emerges} as an integral part of QCD's quantum definition.  This energy-scale is manifest in the gluon vacuum polarisation: a Schwinger mechanism is active in QCD owing to gauge sector dynamics \cite{Cornwall:1981zr, Binosi:2009qm, Roberts:2015lja, Aguilar:2015bud, Deur:2016tte, Fischer:2018sdj}.  It is likely that aspects of the relation between QCD's gluons (gauge-bosons) and $\Theta_0$ could be clarified by studies of hadron states in which the presence of glue determines the quantum numbers, such as hybrid hadrons.  The studies described in Sect.\,\ref{exoticspectrum} relate to this.

Having established that a trace anomaly exists, one arrives at a basic question: Can one understand the magnitude of the associated mass-scale?  For strong interactions, its size can be measured.  Consider the expectation value of the energy-momentum tensor in the proton:
\begin{equation}
\label{EPTproton}
\langle p(P) | T_{\mu\nu} | p(P) \rangle = - P_\mu P_\nu\,.
\end{equation}
The right-hand-side follows from the equations-of-motion for a one-particle proton state.  In the chiral limit, \emph{i.e}.\ absent Higgs couplings to QCD,
\begin{align}
\label{anomalyproton}
\langle p(P) | T_{\mu\mu} | p(P) \rangle  = - P^2 & = m_p^2 = \langle p(P) |  \Theta_0 | p(P) \rangle\,;
\end{align}
hence, it is possible to conclude that the entirety of the proton mass is produced by gluons.  The trace anomaly is manifestly large.  That size must logically owe to gluon self-interactions, which are also responsible for asymptotic freedom.  This is what is intended by the oft repeated statement \cite{Geesaman:2015fha}: ``\emph{The vast majority of mass comes from the energy needed to hold quarks together inside nuclei}.''   As discussed in Sect.\,\ref{ss:upsilon}, it might be possible to access current-quark-mass corrections to the $\Theta_0$ part of the proton mass via the production of $J/\psi$ and $\Upsilon$ mesons at threshold \cite{Kharzeev:1995ij, Kharzeev:1998bz, Wang:2019mza}, through which colour van der Waals forces could be accessible \cite{TarrusCastella:2018php}.

It is possible, however, to take a different perspective.  Once more, consider the chiral limit and ask the question: What is the expectation value of the trace of the energy-momentum tensor in the pion?  The answer is simple:
\begin{equation}
\label{pionmassanomaly}
  \langle \pi(k)| \Theta_0 | \pi(k)\rangle = - k_\mu k_\mu = m_\pi^2 =0 \,,
\end{equation}
\emph{i.e}.\ identically zero.  One na\"ive interpretation of Eq.\,\eqref{pionmassanomaly} is that in the chiral limit the gluons disappear; hence, contribute nothing to the pion mass.  If this were true, then one would be faced with new conundrums, \emph{e.g}.: it would mean that at large resolving scales, $\zeta \gg m_N$, the proton is full of gluons, whereas the pion contains none; and this remains true as $\zeta$ increases further.  Such a situation appears to be forbidden by QCD evolution \cite{Dokshitzer:1977sg, Gribov:1971zn, Lipatov:1974qm, Altarelli:1977zs} (DGLAP), which demands that gluons dominate within every hadron on the neighborhood $m_N/\zeta \simeq 0$ \cite{Altarelli:1981ax}.  So, are there gluons in the pion or not?  Theoretical ideas and experiments are sketched herein that may provide an answer to this fundamental question.

A more reasonable explanation of Eq.\,\eqref{pionmassanomaly} is that ``zero'' owes to destructive interference between competing effects, \emph{viz}.\ one-body dressing and two-body interactions; and the cancellation is exact in the chiral-limit pion owing to DCSB.  Phenomenological analyses of existing data \cite{Badier:1983mj, Conway:1989fs, Chekanov:2002pf, Aaron:2010ab} support this view, reporting that approximately half the pion's light-front momentum is carried by gluon partons \cite{Aicher:2010cb, Barry:2018ort}, roughly matching the proton result.

In any event, Eqs.\,\eqref{anomalyproton} and \eqref{pionmassanomaly} present Science with a remarkable dichotomy.  They entail that it is insufficient to answer only the question ``Whence the proton's mass?'' EHM is not understood unless an answer is simultaneously provided to the equally fundamental puzzle: ``Whence the \emph{absence} of a pion mass?''  The mass-scale that defines nuclear physics, $m_N$, must emerge at the same time as scale invariance is seemingly preserved in the chiral limit NG modes; and these modes must remain massless irrespective of the value taken by $m_N$.

The trace anomaly has impacts at all levels within QCD.  Perhaps the most unexpected manifestation is found with gluons.  These gauge bosons are massless in the Lagrangian and they remain massless at all orders in perturbation theory.  However, as first suggested almost forty years ago, owing to strong nonlinear dynamics in QCD's gauge sector, gluons acquire a running mass.  It is large at infrared momenta, being characterised by a renormalisation group invariant (RGI) mass-scale $m_0 \sim m_N/2$ \cite{Cui:2019dwv}.  The appearance of this mass-scale does not alter any Slavnov-Taylor identities \cite{Taylor:1971ff, Slavnov:1972fg}; hence, all aspects and consequences of QCD's BRST invariance \cite{Becchi:1975nq, Tyutin:1975qk} are preserved.

The appearance of a gluon mass is potentially critical to the existence of QCD.  This can be explained by returning to the question of confinement.  One of the earliest attempts to understand this phenomenon followed immediately upon the demonstration of asymptotic freedom \cite{Politzer:2005kc, Wilczek:2005az, Gross:2005kv} in the QCD running coupling.  The associated appearance of a Landau pole in the infrared gave rise to the notion of infrared slavery, \emph{i.e}.\ confinement owing to a divergent coupling in the far-infrared; but to go further, one needs a calculable nonperturbative running coupling \cite{Dokshitzer:1998nz, Grunberg:1982fw}.

The running coupling that characterises QED \cite{GellMann:1954fq} is the best understood \cite{Zyla:2020}.  In addition to being RGI, this ``Gell-Mann--Low'' effective charge is process-independent (PI) because it is completely determined by the photon vacuum polarisation.  Computing a QCD analogue is more difficult because ghost fields do not decouple; but joining the pinch technique \cite{Cornwall:1981zr, Cornwall:1989gv, Pilaftsis:1996fh, Binosi:2009qm} and background field method \cite{Abbott:1980hw}, one can make QCD ``look'' Abelian.   This scheme enables one to systematically rearrange classes and sums of diagrams in order to arrive at modified Schwinger functions that satisfy linear Slavnov-Taylor identities \cite{Taylor:1971ff, Slavnov:1972fg}.  Amongst them is a modified gluon dressing function from which one can compute a unique QCD running coupling, which is both RGI and PI \cite{Binosi:2016nme}.

%----------------------
\begin{figure}[t!]
\vspace*{1.5ex}

\centering
\includegraphics[width=0.60\linewidth]{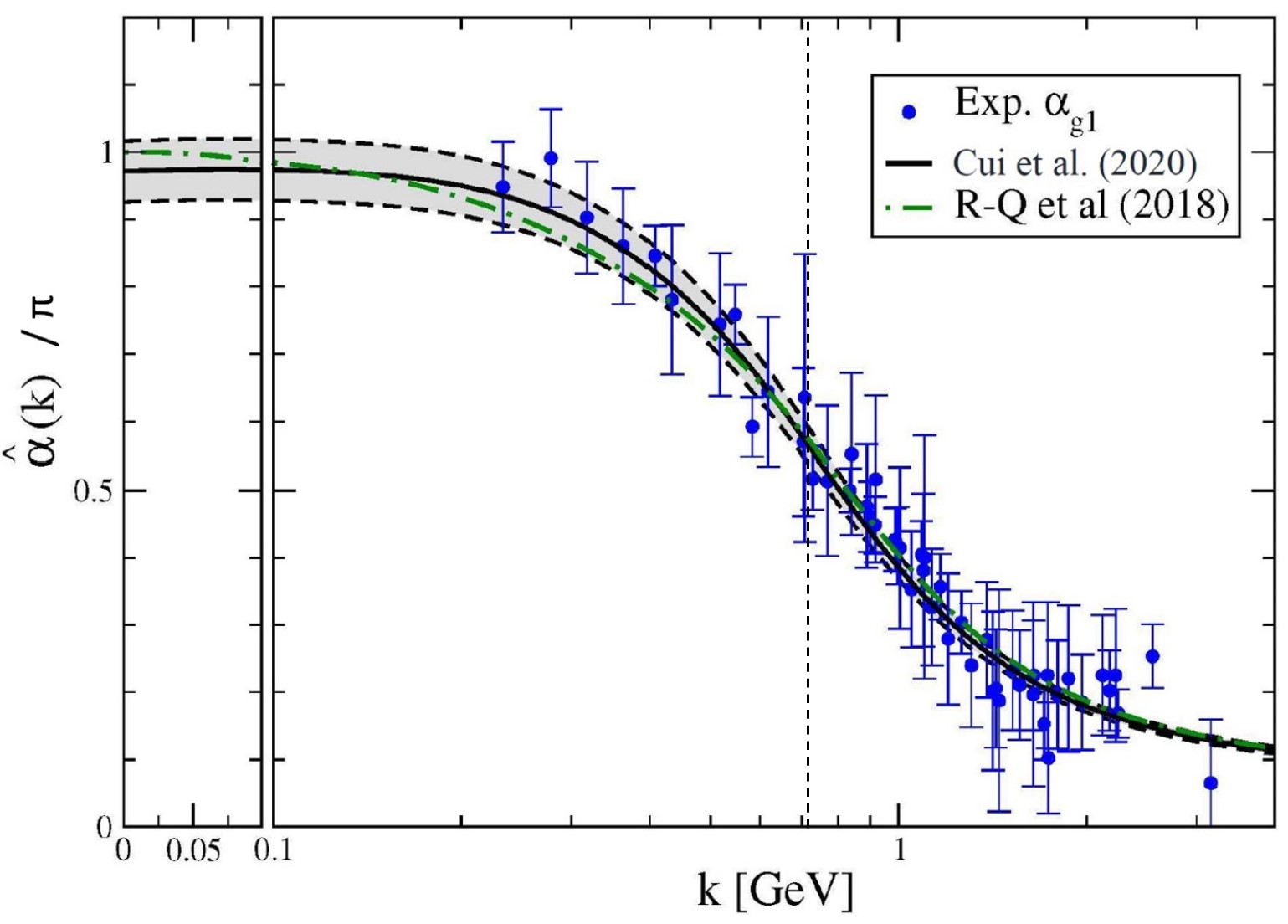}
%\end{center}
%\vspace{-.75cm}	
\caption{\label{Figwidehatalpha}
Solid black curve within grey band -- $\hat{\alpha}(k^2)/\pi$, RGI PI running-coupling computed in Ref.\,\cite{Cui:2019dwv} (Cui \emph{et al}.\ 2020); and dot-dashed green curve -- earlier result (R-Q \emph{et al}.\ 2018) \cite{Rodriguez-Quintero:2018wma}.
(The grey systematic uncertainty band bordered by dashed curves is explained in Ref.\,\cite{Cui:2019dwv}.)
For comparison, world data on the process-dependent charge, $\alpha_{g_1}$, defined via the Bjorken sum rule \cite{Bjorken:1966jh, Bjorken:1969mm, Bjorken:1969ja}, are also depicted.  (The data sources are listed elsewhere \cite{Cui:2019dwv}.
For additional details, see Refs.\,\cite{Deur:2005cf, Deur:2008rf, Deur:2016tte}.)
The k-axis scale is linear to the left of the vertical partition and logarithmic otherwise.
The vertical line,  $k=m_G$, marks the gauge sector screening mass, Eq.\,\eqref{setzetaH}.
}
\end{figure}

QCD's RGI PI effective charge is depicted in Fig.\,\ref{Figwidehatalpha}, in the form obtained using the most up-to-date results from numerical simulations of lattice-regularised QCD (lQCD) \cite{Rodriguez-Quintero:2018wma, Cui:2019dwv}.  Owing to the dynamical breakdown of scale invariance, expressed through emergence of the RGI gluon mass-scale discussed above, this running coupling saturates at infrared momenta:
\begin{equation}
\label{hatalpha0}
\hat\alpha(0)= \pi \times 0.97(4)\,.
\end{equation}

The data in Fig.\,\ref{Figwidehatalpha} are empirical information on $\alpha_{g_1}$, a process-\emph{dependent} effective-charge \cite{Grunberg:1982fw, Dokshitzer:1998nz} determined from the Bjorken sum rule \cite{Bjorken:1966jh, Bjorken:1969mm, Bjorken:1969ja}.  Sound theoretical arguments underpin the almost precise agreement between $\hat{\alpha}$ and $\alpha_{g_1}$ \cite{Binosi:2016nme, Rodriguez-Quintero:2018wma, Cui:2019dwv}.  Thus, the Bjorken sum is plausibly a near direct means by which to gain empirical insight into QCD's unique RGI PI effective charge.

Fig.\,\ref{Figwidehatalpha} reveals that $\hat\alpha(k)$ is everywhere finite; namely, there is no Landau pole.  Instead, the theory is characterised by an infrared-stable fixed point; and the emergence of a gluon mass scale has provided for a nonperturbative infrared completion of QCD.  Now, contrary to the divergence of the perturbative running coupling at $k = \Lambda_{\rm QCD}$, where $\Lambda_{\rm QCD}$ is the RGI mass scale that characterises perturbative QCD (pQCD), the finite value of the $\hat\alpha$ at this point can be used to define a screening mass \cite{Cui:2020dlm, Cui:2020piK}: $m_G \approx 1.4 \,\Lambda_{\rm QCD}$.  This line is drawn in Fig.\,\ref{Figwidehatalpha} so as to highlight that the running coupling alters character at $k \simeq m_G$: modes with $k^2 \lesssim m_G^2$ are screened from interactions and the theory enters a practically conformal domain.  Evidently, $k=m_G$ draws a natural border between soft and hard physics.  Hence, it is a natural candidate for the ``hadronic scale'', \emph{viz}.
\begin{equation}
\label{setzetaH}
\zeta_H=m_G
\end{equation}
is the infrared resolving scale at which all the properties of a hadron are expressed by the dressed quasiparticles that form bound-state kernels and emerge as the self-consistent solutions of the associated equations \cite{Cui:2020dlm, Cui:2020piK}.  The diverse features of $\hat\alpha(k)$ make it a strong contender for that object which properly represents the interaction strength in QCD at any given momentum scale \cite{Dokshitzer:1998nz}.

Having seen that the existence of a trace anomaly enables the appearance of a large gluon mass-scale, it is natural to enquire after the matter sector.  Here the answer was anticipated roughly sixty years ago \cite{Nambu:2011zz}.  It is best described in terms of the propagator for a dressed-quark with flavour $f$, which has the form
\begin{equation}
\label{genS}
  S_f(k;\zeta) = 1/[i\gamma\cdot k \, A_f(k^2;\zeta^2) + B_f(k^2;\zeta^2)] = Z_f(k^2;\zeta^2)/[i\gamma\cdot k + M_f(k^2)]\,,
\end{equation}
where $M_f(k^2)$ is the RGI running mass.  $S_f(k)$ is the solution of a QCD gap equation; and as a unique PI effective charge, $\hat{\alpha}$ sets the strength of interactions in all QCD's equations of motion, including the gap equations.  It therefore plays a crucial role in determining the fate of chiral symmetry, \emph{i.e}.\ the dynamical origin of light-quark masses in the SM even in the chiral limit.  With $\hat{\alpha}(k)$ as depicted in Fig.\,\ref{Figwidehatalpha}, a nonzero mass function, $M(k)$, is found to emerge as the solution in the absence of a Higgs coupling, with $M(0) \approx m_N/3$ \cite{Binosi:2016wcx}.  This is the basic signature of DCSB; namely, the emergence of \emph{mass from nothing}.  DCSB is expressed in hadron wave functions \cite{Brodsky:2009zd, Chang:2011mu, Brodsky:2012ku, Roberts:2015lja}; and given that $3 \times M(0) \approx m_N$, there is a firm theoretical position from which one can argue that DCSB is the immediate source for more than 98\% of the visible mass in the Universe.

One can now begin to return to Eq.\,\eqref{pionmassanomaly}.  In QCD, the pion is described by a Bethe-Salpeter amplitude with four distinct Dirac-matrix structures and associated scalar functions \cite{Maris:1997hd}.  Denote by $E(k^2,k\cdot P)$ that term which is typically described as the pion's pseudoscalar component, where $P$ is the bound-state's total momentum and $k$ is the relative momentum between the valence constituents.  Then in the absence of Higgs couplings, a massless NG pion emerges if, and only if, chiral symmetry is dynamically broken and
\begin{equation}
\label{gtrE}
f_\pi^0 E_\pi(k;0) = B(k^2)\,,
\end{equation}
where $f_\pi^0$ is the chiral-limit value of the pion's leptonic decay constant.  This identity is the most fundamental expression of the Nambu-Goldstone theorem in QCD.  It is true in any covariant gauge and independent of the renormalisation scheme.  It means the two-body problem is solved, nearly completely, once the solution to the one body problem is known.  Such a correspondence is impossible in any quantum mechanical model.  Since $B(k^2) \neq 0 $ is a direct expression of EHM in QCD's matter sector, then Eq.\,\eqref{gtrE} entails, enigmatically, that the qualities of the nearly-massless pion are the clearest expression of the mechanism responsible for virtually all visible mass in the Universe.

As detailed in Ref.\,\cite{Roberts:2016vyn}, Eq.\,\eqref{gtrE} is the basis for an algebraic proof of Eq.\,\eqref{pionmassanomaly}, \emph{viz}.\ the pion is massless in the chiral limit, irrespective of the emergence of a large gluon mass scale which drives $m_N$ to 1\,GeV.  In fact, Eq.\,\eqref{gtrE} is necessary and sufficient to ensure that the sum of the dynamically generated masses of the quark and antiquark is precisely cancelled by the attractive interaction energy between these dressed constituents in the pseudoscalar channel:
\begin{equation}
M^{\rm dressed}_{\rm quark} + M^{\rm dressed}_{\rm antiquark}
 + U^{\rm dressed}_{\rm quark-antiquark\;interaction} \stackrel{\rm chiral\;limit}{\equiv} 0\,.
\label{EasyOne}
\end{equation}
This guarantees the \emph{disappearance} of the scale anomaly in the chiral-limit pion.  Eq.\,\eqref{EasyOne} is not merely ``hand-waving''.  Rather, it sketches the cancellations that take place in the pseudoscalar projection of the fully-dressed quark+antiquark scattering matrix, which can be displayed rigorously \cite{Munczek:1994zz, Bender:1996bb, Qin:2014vya}.

Switching on the Higgs couplings, so that the light-quarks possess their small current-masses ($m \sim 7\,m_e$), then DCSB is the agent behind, amongst other things: the physical size of the pion mass ($m_\pi \approx 0.15 \,m_N$); the large mass-splitting between the pion and its valence-quark spin-flip partner, the $\rho$-meson ($m_\rho > 5 \,m_\pi$); and the natural scale of nuclear physics, $m_N \approx 1\,$GeV.  Interesting things also happen to the kaon.  Like a pion, but with one of its light quarks replaced by a strange-quark, the kaon possesses a mass $m_K \approx 0.5\,$GeV.  In this case, a competition is taking place between dynamical and Higgs-driven mass generation \cite{Ding:2015rkn, Gao:2017mmp, Ding:2018xwy, Cui:2020dlm, Cui:2020piK}: all differences between the pion and kaon owe to Higgs-induced modulations of EHM.

In closing this section it is worth recapitulating some general rules for hadron masses \cite{Roberts:2020udq}:
(\emph{i}) estimates based on notions familiar from relativistic quantum mechanics typically arrive at only $\sim 1$\% of a hadron's mass -- hence, the Higgs mechanism alone is responsible for just $\sim 1$\% of visible mass;
(\emph{ii}) the contribution of the current-mass term in QCD's Lagrangian is strongly enhanced as a consequence of EHM, in particular for the pion;
and (\emph{iii}) in all systems for which no symmetry ensures Eq.\,\eqref{EasyOne}, EHM is key to more than 98\% of a hadron's mass.
It is worth highlighting that these features lay the foundation which guarantees, \emph{inter alia}, accuracy of equal spacing rules in the hadron spectrum \cite{Okubo:1961jc, GellMann:1962xb, Qin:2018dqp, Qin:2019hgk}.
\label{Nonsense}
Finally, any attempt in quantum field theory to isolate distinct contributions to a hadron's mass is arbitrary and ambiguous.  The interpretation will depend on the frame chosen and on the scale, $\zeta$, because these two quantities determine the active degrees of freedom \cite{Roberts:2016vyn}.  For instance, each of the terms in Eq.\,\eqref{EasyOne} has a precise mathematical definition; but there is no manageable decomposition of $M^{\rm dressed}_{\rm quark}$ into contributions from gluon- and quark-parton degrees of freedom at any scale, even and especially when one chooses a light-front quantisation.  The only unequivocal quantities in Nature are Poincar\'e invariant; so to avoid confusion, it is best to focus on truly measurable quantities and eschew practitioner-dependent projections \cite{Lorce:2017xzd, Aguilar:2019teb, Brodsky:2020vco, Roberts:2020udq}.

%QCD's interactions are the same in all hadrons, so cancellations similar to those indicated by Eq.\,\eqref{EasyOne} take place within the proton.  However, as just noted, in the proton, no symmetry requires the cancellations to be complete.  Thus, the value of the proton's mass is typical of the magnitude of scale breaking in the one-body sectors, \emph{viz}.\ the gluon and quark mass scales, $m_0$ and $M(0)$, respectively.  In fact, no significant hadronic mass scale is possible unless one of similar size is expressed in the dressed-propagators of gluons and quarks.  It follows that the mechanism(s) responsible for EHM can be exposed by measurements sensitive to such dressing.  This potential is offered by a large array of observables, \emph{e.g}.: spectra and static properties; form factors -- elastic and transition; and all types of parton distributions.

\section{Light-Hadron Structure}
\label{LHstructure}
QCD's interactions are the same in all hadrons; hence, cancellations similar to those indicated by Eq.\,\eqref{EasyOne} take place within the proton.  In the proton, however, no symmetry requires the cancellations to be complete.  Thus, $m_N$ is typical of the size of scale breaking in the one-body sectors; to wit, its order-of-magnitude is set by the gluon and quark mass scales, $m_0$ and $M(0)$, respectively.  Indeed, a significant hadronic mass scale is impossible unless one of similar magnitude is manifest in the dressed-propagators of gluons and quarks.  Consequently, the mechanism(s) responsible for EHM can be revealed by measurements sensitive to such dressing.  Many observables possess this capacity, \emph{e.g}.\ \cite{Carman:2020qmb, Brodsky:2020vco}: spectra and static properties; form factors -- elastic and transition; and all types of parton distributions.  Hereafter, we highlight some apposite observables that could be the focus of high-impact experiments at the EicC.

\subsection{Pion form factor}
\label{FFspiK}
Precise measurement of the pion elastic electromagnetic form factor, $F_\pi(Q^2)$, has been a basic motivation for high-luminosity, high-energy electron-scattering experiments for forty years because QCD makes a very clean prediction for the large $Q^2$ behaviour \cite{Lepage:1979zb, Efremov:1979qk,Lepage:1980fj}:
\begin{equation}
\label{pionUV}
\exists Q_0>\Lambda_{\rm QCD} \; |\;  Q^2 F_\pi(Q^2) \stackrel{Q^2 > Q_0^2}{\approx} 16 \pi \alpha(Q^2)  f_\pi^2 \mathpzc{w}_\varphi^2,
\end{equation}
where $f_\pi=92.2\,$MeV is the pion decay constant \cite{Zyla:2020}, $\alpha(k^2)$ is QCD's running coupling, which is indistinguishable from $\hat\alpha(k^2)$ in Fig.\,\ref{Figwidehatalpha} on any domain within which perturbation theory is valid, and
\begin{equation}
\label{wphi}
\mathpzc{w}_\varphi = \frac{1}{3} \int_0^1 dx\, \frac{1}{x} \varphi_\pi(x)\,,
\end{equation}
where $\varphi_\pi(x)$ is the pion's valence-quark parton distribution amplitude (PDA).  QCD also predicts that there is a $Q_{\rm as}$ such that $\varphi_\pi(x)\approx \varphi_{\rm as}(x)=6 x(1-x)$ on $\Lambda_{\rm QCD}^2/Q_{\rm as}^2 \simeq 0$.  However, neither the value of $Q_0$ nor that of $Q_{\rm as}$ are predicted by pQCD.  On the other hand, they are computable in any nonperturbative framework that veraciously expresses EHM.

The first empirical data on $F_\pi(Q^2)$ in the modern era were obtained at JLab and published in Ref.\,\cite{Volmer:2000ek}.  They were followed by the reports in Refs.\,\cite{Horn:2006tm, Tadevosyan:2007yd, Horn:2007ug, Huber:2008id, Blok:2008jy}.  The data indicate that with momentum transfers reaching to $Q^2=2.45\,$GeV$^2$ one is still far from the resolution region wherein Eq.\,\eqref{pionUV} is valid.  This conclusion is based on the assumption that inserting $\varphi_{\rm as}(x)$ into Eq.\,\eqref{pionUV} delivers a valid approximation at $\zeta^2=Q^2=2.45\,$GeV$^2$, so that
\begin{equation}
\label{pionUV4}
Q^2 F_\pi(Q^2) \stackrel{Q^2=4\,{\rm GeV}^2}{\approx} 0.15\,.
\end{equation}
The result in Eq.\,\eqref{pionUV4} is a factor of $2.7$ smaller than the empirical value quoted at $Q^2 =2.45\,$GeV$^2$ \cite{Huber:2008id}: $0.41^{+0.04}_{-0.03}$; and a factor of three smaller than that computed at $Q^2 =4\,$GeV$^2$ in Ref.\,\cite{Maris:2000sk}.  When published, Ref.\,\cite{Maris:2000sk} provided the only prediction for the pointwise behaviour of $F_\pi(Q^2)$ that was both applicable on the entire spacelike domain then mapped reliably by experiment and confirmed thereby.  However, the algorithms used therein were inadequate for computing $F_\pi(Q^2)$ on $Q^2 > 4\,$GeV$^2$.

With the challenge posed by Eq.\,\eqref{pionUV} thus remaining, experiments were proposed for the 12\,GeV-upgraded JLab facility with the goal of reaching $Q^2=6\,$GeV$^2$ \cite{Dudek:2012vr}.  The upgrade is complete, the experiments \cite{E1206101, E12-07-105} have run, and the data are being analysed.

\begin{figure}[t]
\begin{tabular}{ccc}
\includegraphics[clip,width=0.45\textwidth]{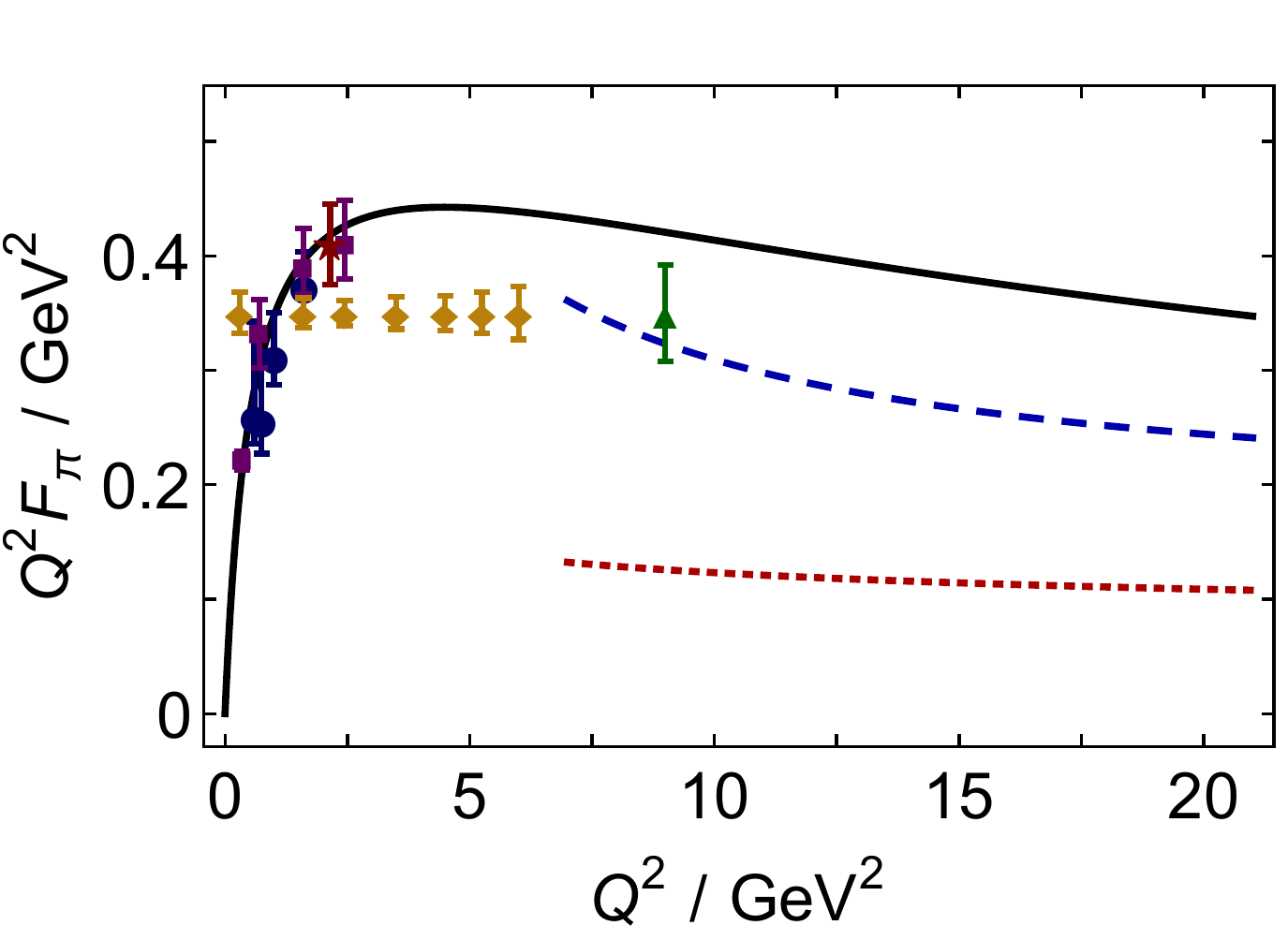} & \hspace*{1em} &
\includegraphics[clip,width=0.45\textwidth]{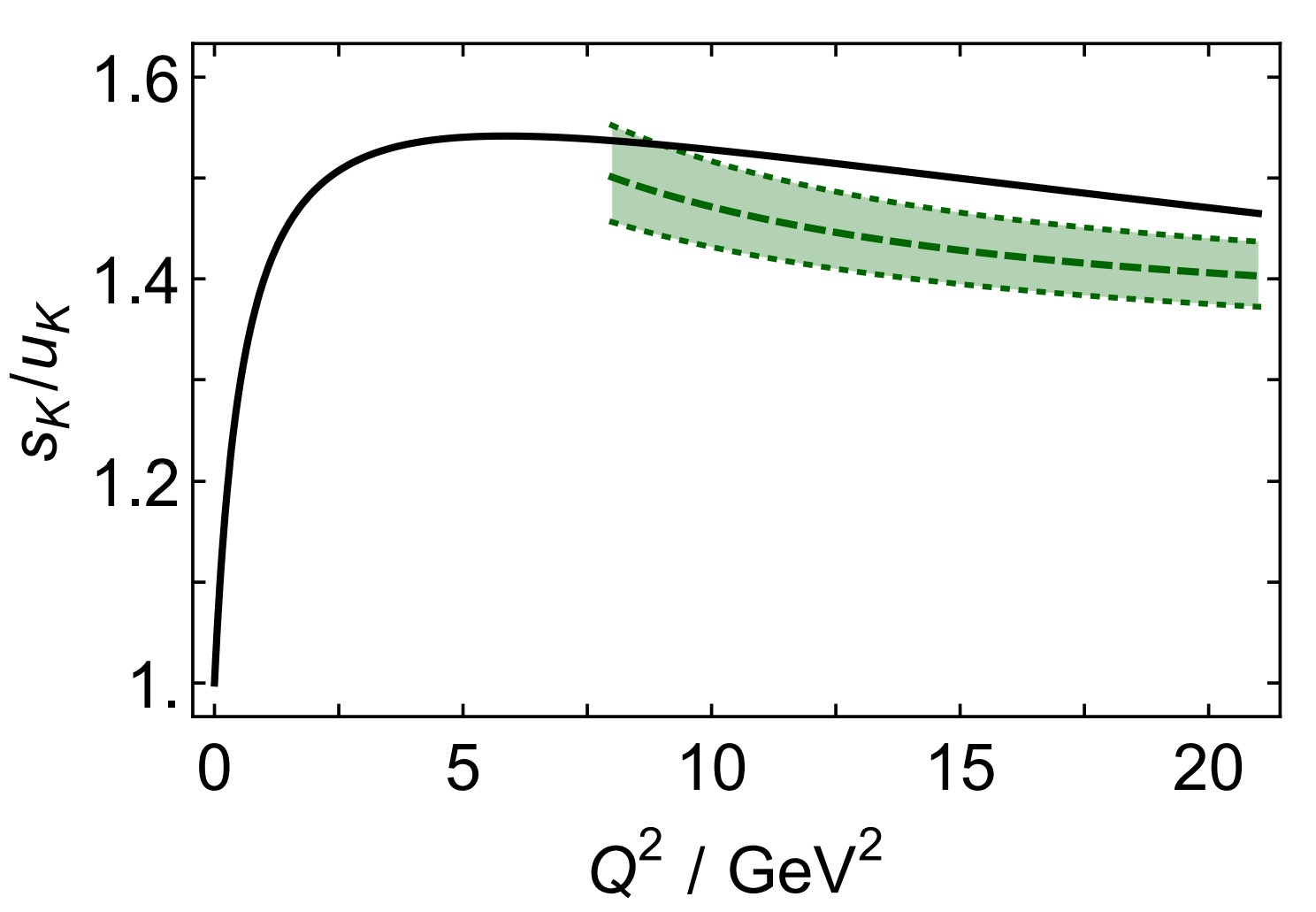} \\[-38ex]
\hspace*{-21em}{\large{\textsf{A}}} & &  \hspace*{-21em}{\large{\textsf{B}}} \\[36ex]
\end{tabular}
\caption{\label{figWPFpi}
\emph{Left panel\,--\,A}.  $Q^2 F_\pi(Q^2)$.  Solid curve (black) -- theoretical prediction \cite{Chang:2013nia, Gao:2017mmp, Chen:2018rwz}; dashed curve (blue) -- pQCD prediction computed with the modern, EHM-dilated pion PDA described in Ref.\,\cite{Chang:2013pq}; and dotted (red) curve -- pQCD prediction computed with the asymptotic profile, $\varphi_{\rm as}(x)$, which had previously been used to guide expectations for the asymptotic behaviour of $Q^2 F_\pi(Q^2)$.  The filled-circles and \mbox{-squares} represent existing JLab data \cite{Huber:2008id}; and the filled diamonds and triangle, whose normalisation is arbitrary, indicate the projected $Q^2$-reach and accuracy of forthcoming experiments \cite{E1206101, E12-07-105}.
\emph{Right panel\,--\,B}.  Predicted ratio of $\bar s$- and $u$-quark contributions to the $K^+$ elastic form factor -- solid black curve \cite{Gao:2017mmp}.
Result for this ratio produced by the QCD hard scattering formula when used with a modern EHM-dilated kaon PDA -- dashed green curve.  The associated shading reflects uncertainty in current knowledge of the kaon PDA \cite{Shi:2014uwa, Cui:2020dlm, Cui:2020piK}.
}
\end{figure}

Meanwhile, the algorithms used in Ref.\,\cite{Maris:2000sk} have been comprehensively improved, so that the continuum methods for the strong interaction bound-state problem which delivered $\hat\alpha(k^2)$ in Fig.\,\ref{Figwidehatalpha}, \emph{i.e}.\ QCD's Dyson-Schwinger equations (DSEs) \cite{Roberts:2015lja, Horn:2016rip, Eichmann:2016yit, Burkert:2019bhp, Fischer:2018sdj}, could supply a prediction for $F_\pi(Q^2)$ to arbitrarily large-$Q^2$ \cite{Chang:2013nia, Gao:2017mmp, Chen:2018rwz}.  The result is drawn in Fig.\,\ref{figWPFpi}A.
Also depicted (dashed blue curve) is the result obtained using Eq.\,\eqref{pionUV} and the PDA calculated in the DSE framework at a scale relevant to the experiment.  This PDA is very different from $\varphi_{\rm as}$, being markedly broader owing to DCSB.

On the domain depicted in Fig.\,\ref{figWPFpi}, the leading-order, leading-twist QCD prediction, computed with a pion valence-quark PDA evaluated at an experiment-appropriate scale, underestimates the full DSE computation by merely an approximately uniform 15\%.  The small mismatch is explained by a combination of higher-order, higher-twist corrections to Eq.\,\eqref{pionUV} in pQCD on the one hand and, on the other hand, shortcomings in the leading-order DSE truncation used in Refs.\,\cite{Chang:2013nia, Gao:2017mmp, Chen:2018rwz}, which predicts the correct power-law behaviour for the form factor but not precisely the right anomalous dimension (exponent on the logarithm) in the strong-coupling calculation.

The modern theory prediction has completely changed perceptions, highlighting that QCD is \emph{not} found in scaling laws; rather, since deviations from clean scaling are an essential feature of quantum field theory in four spacetime dimensions, then QCD is revealed in the presence and nature of scaling violations.  It is now anticipated that the approved JLab experiments will reveal a maximum in $Q^2 F_\pi(Q^2)$ at $Q^2 \approx 6\,$GeV$^2$.   Moreover, efforts are being made to complete a measurement at $Q^2\approx 9\,$GeV$^2$ in the hope of seeing the onset of QCD scaling violations for the first time in a hadron elastic form factor.  As shown by theory, the magnitude of any given form factor on a sizeable domain above $Q^2 \approx 10\,$GeV$^2$ is determined by the scale of EHM.  JLab's grip on this domain is tenuous; but as discussed in Sect.\,\ref{piKFFmeasurements}, EicC could reach with precision out to $Q^2 \approx 30\,$GeV$^2$.  Potentially, therefore, EicC could be the first facility to discover scaling violations and measure the size of EHM in a hard exclusive process.

\subsection{Kaon form factor}
\label{FFsK}
As described above, pions are Nature's closest approximation to a NG mode; and in the absence of Higgs couplings, there can be no distinction between pions and kaons within QCD.  Hence, knowledge of kaon structure  is crucial because it provides a window onto the interference between EHM and the Higgs mechanism for mass generation \cite{Aguilar:2019teb, Roberts:2020udq}.  Yet, today, seventy years after its discovery \cite{Rochester:1947mi}, little is known about kaon structure.  (The situation is only marginally better for the pion \cite{Horn:2016rip, Aguilar:2019teb, Roberts:2020udq} as we discuss further below.)  Much can be learnt by measuring the elastic form factors of charged and neutral kaons because, since isospin is a good symmetry of QCD and the charge of $u$- and $s$-quarks is different, this information translates into a fairly direct measure of the relative distribution of normal- and strange-matter within the kaon.  All differences result from Higgs modulation of EHM.

%%\begin{figure}[t]
%%\centerline{\includegraphics[clip,width=0.6\textwidth]{F3.jpg}}
%%\caption{\label{figWPsKuK}
%
%%Predicted ratio of $\bar s$- and $u$-quark contributions to the $K^+$ elastic form factor -- solid black curve \cite{Gao:2017mmp}.
%
%%Result for this ratio produced by the QCD hard scattering formula when used with a modern EHM-dilated kaon PDA -- dashed green curve.  The associated shading reflects uncertainty in current knowledge of the kaon PDA \cite{Shi:2014uwa, Cui:2020dlm, Cui:2020piK}.}
%%\end{figure}

An example is presented in Fig.\,\ref{figWPFpi}B, which describes the charge distribution ratio of strange-to-normal matter in the $K^+\!$.  Current conservation requires that both these distributions are unity at $Q^2=0$; and QCD predicts that the ratio is unity on $\Lambda_{\rm QCD}^2/Q_{\rm as}^2 \simeq 0$; so the interesting phenomena are displayed on the domain between these limits.

The prediction depicted in Fig.\,\ref{figWPFpi}B is a companion to that for the pion in Fig.\,\ref{figWPFpi}A, obtained using the same techniques.  Once again, when used with a meson PDA appropriate to the scale of the experiment, broadened by EHM and skewed slightly by the Higgs mechanism so that its peak is shifted away from $x=1/2$ \cite{Shi:2014uwa, Cui:2020dlm, Cui:2020piK}, the QCD hard scattering formula delivers a result in fair agreement with the direct calculation on $Q^2 \gtrsim 8\,$GeV$^2$.  Naturally, it becomes more reliable as $\ln Q^2$ is increased.  The strange-to-normal charge distribution ratio rises to a peak value of roughly 1.5 at $Q^2\approx 6\,$GeV$^2$, and thereafter returns logarithmically to unity, in accordance with pQCD.   These outcomes are typical of EHM dominance in flavour-symmetry breaking.  EHM tames the large Higgs-produced current-quark mass difference: $m_s/m_u \sim 30$ but $M_s(0)/M_u(0) \sim 1.25$; and elastic form factors are sensitive to $[M_s(0)/M_u(0)]^2 \sim 1.6$.  All mass-scales are only subdued logarithmically by parton splitting effects, so the deviation of $\bar s_K/u_K$ from unity persists to large $Q^2$.

Such are the theory predictions; and kaon form factor measurements are also planned at JLab \cite{E12-09-011} to test them.  The experiments will reach $Q^2 \approx 5\,$GeV$^2$; but this range might not be sufficient to discover the peak in the ratio of charge distributions and confirm the size of Higgs-boson effects in the elastic form factor.  On the other hand, as discussed in Sect.\,\ref{piKFFmeasurements}, EicC could also reach with precision out to $Q^2 \approx 30\,$GeV$^2$ for the kaon form factor.  Thus, EicC could be the first facility to measure the size and range of nonperturbative EHM--Higgs-boson interference effects in hard exclusive processes.

% In any event, one pion experiment and one kaon experiment on Q2 > 10 GeV2 cannot alone solve an issue that began in 1979-1980 with the papers by Lepage+Brodsky & Efremov+Radyushkin.
% Conclusion: in bridging the gap between JLab 12 and BNL-EIC, EicC is uniquely placed for discovery of scaling violations in hard exclusive processes.  This will validate a large and diverse array of rigorous analyses in QCD and, very importantly, provide the information necessary to validate Nambu's formula for the relationship between mass and structure of Goldstone bosons.

\subsection{Nucleon form factors}
\label{FFsN}
Neutron and proton elastic electromagnetic form factors are also the focus of extensive programmes in both experiment and theory; again, because they can provide insights into key features of nucleon structure, such as the role played by EHM in determining the proton's size and fixing both the location and rate of the transition between the strong and perturbative domains of QCD.  Experiments completed during the past twenty years have had a huge impact, revealing that despite its simple valence-quark content, the proton's internal structure is very complex.  Marked differences between the distributions of total charge and magnetisation have been exposed \cite{Jones:1999rz, Gayou:2001qd, Punjabi:2005wqS, Puckett:2010ac, Puckett:2011xgS, Puckett:2017flj} and also between the charge distributions generated by the different quark flavours \cite{Cates:2011pz, Wojtsekhowski:2020tlo}.  New experiments are approved at JLab that will acquire data at unprecedented photon virtualities, \emph{e.g}.\ \cite{Gilfoyle:2018xsa, Wojtsekhowski:2020tlo}:
proton electric form factor to $Q^2= 12\,$GeV$^2$ \cite{E12-07-109};
proton magnetic form factor to $Q^2= 15.5\,$GeV$^2$ \cite{E12-07-108};
neutron electric form factor to $10.2\,$GeV$^2$ \cite{E12-09-016};
and neutron magnetic form factor to $13.5\,$GeV$^2$ \cite{E12-09-019}.

An issue here is that because there are three valence quarks in a baryon, which typically share the momentum delivered by an incoming probe, then compared with mesons, one needs to reach higher values of $Q^2$ before all signals of EHM can be uncovered.  Contemporary predictions for the large-$Q^2$ behaviour of nucleon form factors are presented in Ref.\,\cite{Cui:2020rmu}.  Amongst other things, they suggest that the neutron's electric form factor will exhibit a zero at $Q^2 \approx 20\,$GeV$^2$ and that the $d$-quark contribution to the proton's Paul form factor will likewise vanish at $Q^2 \approx 14\,m_N^2$.  Such behaviour is largely a consequence of strong EHM-induced quark+quark (diquark) correlations within the nucleon \cite{Barabanov:2020jvnS}.  Predictions such as these can only be tested using a high-luminosity accelerator capable of producing $e N$ collisions at energies beyond the range of existing facilities.  Here, too, EicC could be the answer.
%%%Roberts:2013mja, Segovia:2015ufa, 

\begin{figure}[t]
\centering
\includegraphics[clip, width=0.60\textwidth]{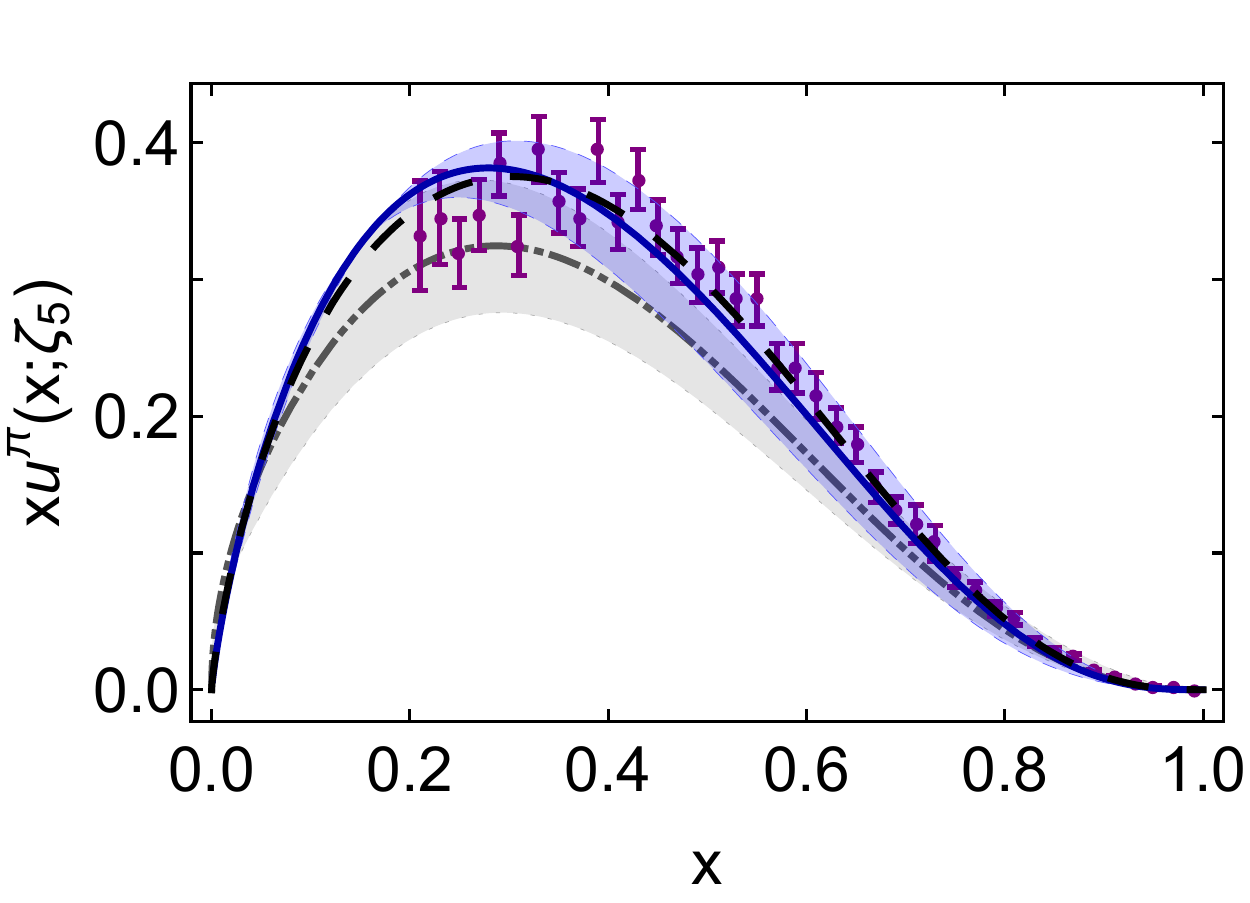}
\caption{\label{figF12}
Pion valence-quark momentum distribution function, $x {\mathpzc q}^\pi(x;\zeta_5=5.2\,{\rm GeV})$:
solid blue curve  -- modern continuum calculation \cite{Cui:2020dlm, Cui:2020piK};
long-dashed black curve -- early continuum analysis \cite{Hecht:2000xa};
and dot-dot-dashed grey curve -- lQCD result \cite{Sufian:2019bol}.
Data (purple) from Ref.\,\cite{Conway:1989fs}, rescaled according to the analysis in Ref.\,\cite{Aicher:2010cb}.
Comparing the central modern continuum prediction with the plotted data, one obtains $\chi^2/{\rm d.o.f.} = 1.66$.
}
\end{figure}

\subsection{Pion distribution functions}
\label{pionDFs}
Basic to any discussion of hadron structure is the pion valence quark distribution function (DF), ${\mathpzc q}^\pi(x;\zeta)$.  It is a density, which maps the probability that a valence ${\mathpzc q}$-quark in the pion carries a light-front fraction $x$ of the system's total momentum \cite{Ellis:1991qj}; and one of the earliest predictions of QCD is \cite{Ezawa:1974wm, Farrar:1975yb, Berger:1979du}:
\begin{equation}
\label{PDFQCD}
{\mathpzc q}^{\pi}(x;\zeta =\zeta_H) \sim (1-x)^{\beta}\,,\quad \beta = 2\,.
\end{equation}
The hadronic scale, $\zeta_H$, is not accessible in experiment because certain kinematic conditions must be met in order for the data to be interpreted in terms of ${\mathpzc q}^{\pi}(x;\zeta)$ \cite{Ellis:1991qj}.  These conditions require experiments with $Q^2 \sim \zeta_E^2 > m_N^2$.  Hence, any result for a DF at $\zeta_H$ must be evolved to $\zeta_E$ for comparison with experiment \cite{Dokshitzer:1977sg, Gribov:1971zn, Lipatov:1974qm, Altarelli:1977zs}.  Under that evolution, the exponent grows, becoming $2+\gamma$, where $\gamma\gtrsim 0$ is an anomalous dimension that increases logarithmically with $\zeta$.

The above remarks merely repeat that the parton model gives scaling laws; QCD provides scaling violations; and such scaling violations serve to increase the integer number specified by the scaling law \cite{Lepage:1980fj}.  They entail that any analysis of a Drell-Yan (DY) or deep inelastic scattering (DIS) experiment (or similar) which returns a value of $\beta < 2$ is in conflict with QCD.  Here sits a longstanding controversy \cite{Holt:2010vj}.  Experiments interpretable in terms of ${\mathpzc q}^{\pi}(x;\zeta)$ were completed more than thirty years ago \cite{Badier:1983mj, Conway:1989fs}.   All existing phenomenological analyses that fail to incorporate soft-gluon (threshold) resummation effects return $\beta \sim 1$ \cite{Barry:2018ort}, whereas the sole, consistent analysis, which includes soft gluon resummation, yields $\beta > 2$ \cite{Aicher:2010cb}.

Consider, therefore, QCD theory.  Algorithms have developed to the point that lQCD is beginning to yield results for the pointwise behaviour of the pion's valence-quark distribution \cite{Xu:2018eii, Chen:2018fwa, Karthik:2018wmj, Sufian:2019bol}.  Furthermore, the DSE approach has delivered parameter-free predictions of the valence, glue and sea distributions within the pion \cite{Ding:2019qlr, Ding:2019lwe, Cui:2020dlm, Cui:2020piK}, revealing that, like the pion's leading-twist PDA, the valence-quark DF is hardened by DCSB, \emph{i.e}.\ as an immediate consequence of EHM.

The continuum predictions from Refs.\,\cite{Cui:2020dlm, Cui:2020piK} are depicted in Fig.\,\ref{figF12}.  Evidently, the result for ${\mathpzc u}^\pi(x;\zeta_5)$, \emph{i.e}.\ the solid blue curve in Fig.\,\ref{figF12}, matches that obtained using lQCD \cite{Sufian:2019bol}.  (Included here is the continuum theory uncertainty band described in Refs.\,\cite{Cui:2020dlm, Cui:2020piK}, which reflects the precision in $\hat\alpha(k^2)$, Fig.\,\ref{Figwidehatalpha}.  Note, too, that both modern calculations are consistent with the prediction in Ref.\,\cite{Hecht:2000xa}, made twenty years ago.)  This shows that two disparate treatments of the pion bound-state problem have arrived at the \emph{same} prediction for the pion's valence-quark distribution function.  Plainly, theory has made real strides toward understanding pion structure; the SM prediction, Eq.\,\eqref{PDFQCD}, is stronger than ever before; and this places great pressure on phenomenology to address the issue of threshold resummation.

The existing controversy and modern theory developments have made new measurements of ${\mathpzc q}^{\pi}(x;\zeta)$ a high priority.  Experiments are planned at JLab and CERN \cite{Keppel:2015, C12-15-006A, Denisov:2018unjF} and being developed for the EIC \cite{Aguilar:2019teb}.   There are good reasons to expect that EicC measurements could also make important contributions, so simulations are in train, as disclosed in Sect.\ref{ss:measurepiKSF}.

\begin{figure}[t]
\centering
\begin{tabular}{cc}
\hspace*{14em}{\large{\textsf{A}}} &
\hspace*{14em}{\large{\textsf{B}}} \\[-8ex]
\includegraphics[clip, width=0.46\textwidth]{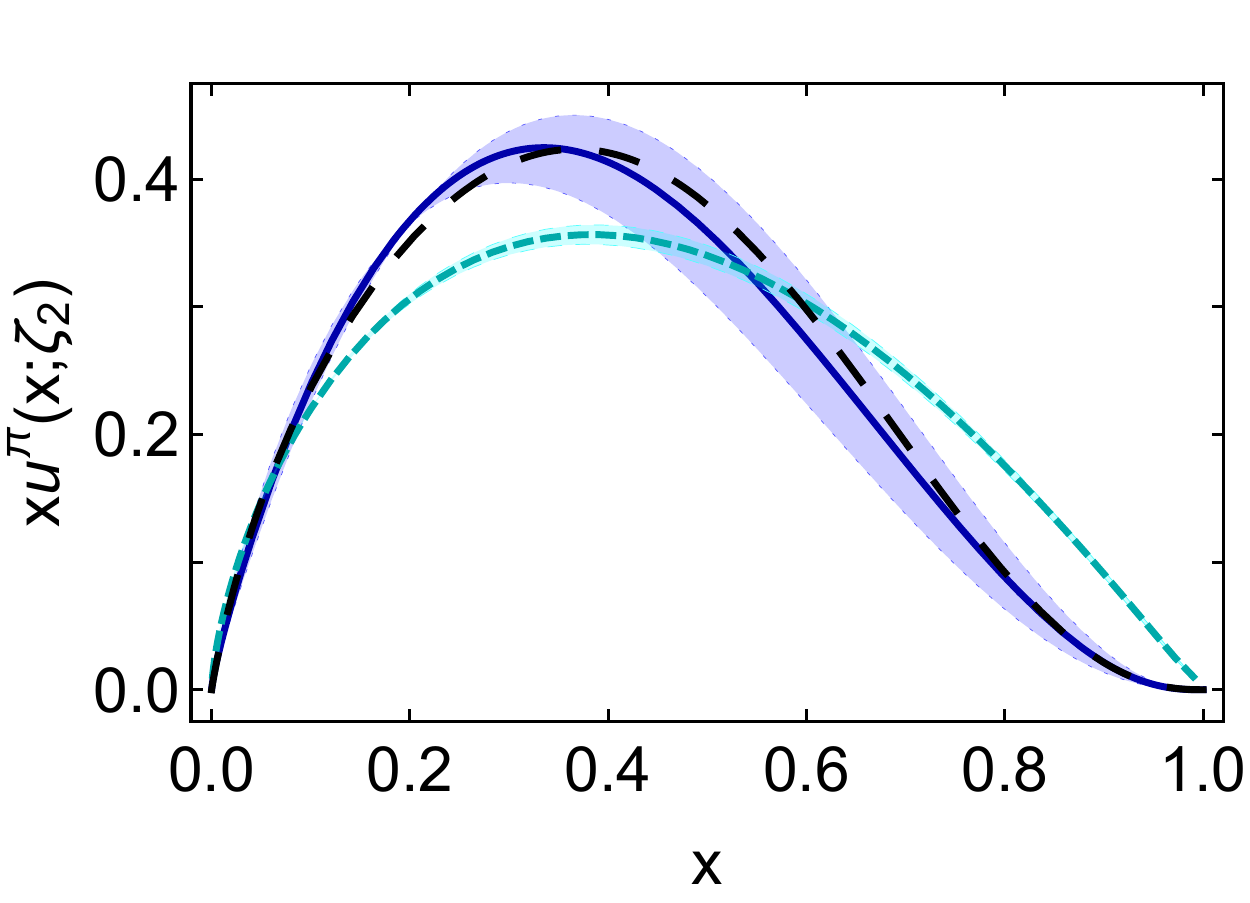} &
\includegraphics[clip, width=0.455\textwidth]{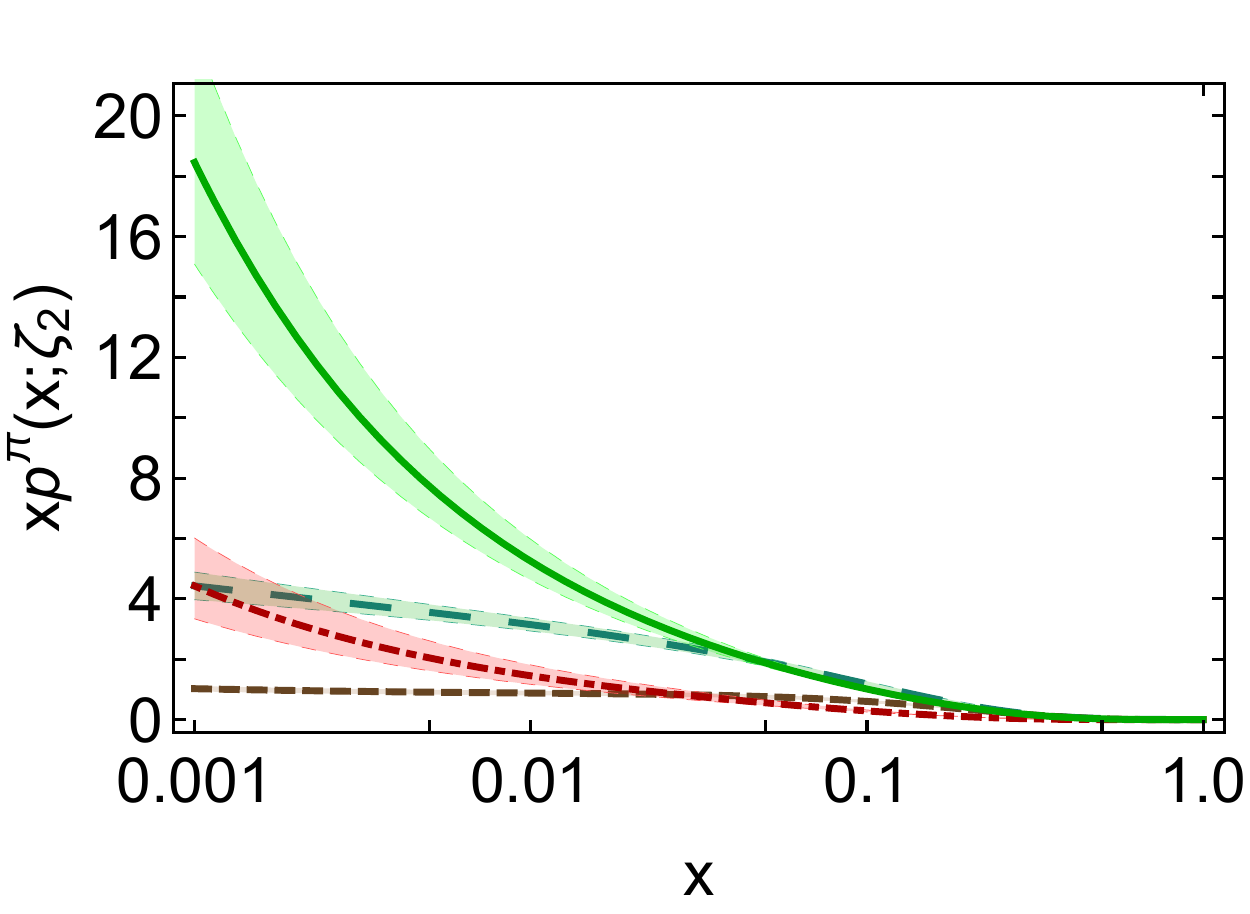}
\end{tabular}
\caption{\label{qpizeta2}
\emph{Left panel--A}.
Solid blue curve -- pion's valence-quark distribution at $\zeta=\zeta_2$, \cite{Cui:2020dlm, Cui:2020piK};
long-dashed black curve -- analogous result from Ref.\,\cite{Hecht:2000xa};
and short-dashed cyan -- phenomenological result from Ref.\,\cite{Barry:2018ort}, also at $\zeta=\zeta_2$.
\emph{Right panel--B}. Solid green curve, $p=g$ -- prediction for the pion's glue distribution; and dot-dashed red curve, $p=S$ -- predicted sea-quark distribution.   Both drawn from Refs.\,\cite{Cui:2020dlm, Cui:2020piK}.
Phenomenological results from Ref.\,\cite{Barry:2018ort} are plotted for comparison:
$p=\,$glue -- long-dashed dark-green; and
$p=\,$sea -- dashed brown.
Normalisation convention: $\langle x[2 {\mathpzc u}^\pi(x;\zeta_2)+g^\pi(x;\zeta_2)+S^\pi(x;\zeta_2)]\rangle=1$.
Notably, $2 {\mathpzc u}^\pi(x;\zeta_2) > [{g}^\pi(x;\zeta_2)+S^\pi(x;\zeta_2)]$ on $x>0.2$, marking this as the valence domain within the pion.
(The uncertainty bands, explained in Refs.\,\cite{Cui:2020dlm, Cui:2020piK}, express the DF uncertainty owing to that in $\hat\alpha(k^2=0)$, Fig.\,\ref{Figwidehatalpha}.)
}
\end{figure}

%As already noted, EicC will provide good access to 0.1 < x < 0.4.  It is on this domain that the sea and glue distributions become large, being power-law suppressed with respect to the valence distributions on x>0.5.
%Conclusion: EicC is potentially uniquely capable of providing information on the partonic structure of Goldstone bosons, adding significantly to data on the valence structure and providing essentially new information on the sea and glue distributions.

A unique feature of Refs.\,\cite{Ding:2019qlr, Ding:2019lwe, Cui:2020dlm, Cui:2020piK} is that they supply parameter-free predictions for all pion DFs, including glue and sea.  They are displayed in Fig.\,\ref{qpizeta2}, as computed at the resolving scale $\zeta=\zeta_2=2\,$GeV, whereat these DFs yield the following momentum fractions:
\begin{equation}
\label{pionMFs}
\mbox{valence:}\,
\langle 2 x {\mathpzc q}^\pi(x;\zeta_2) \rangle = 0.48(4)\,, \quad
\mbox{glue:}\,
\langle x {\mathpzc g}^\pi(x;\zeta_2) \rangle = 0.41(2)\,, \quad
\mbox{sea:}\,
\langle x {\mathpzc S}^\pi(x;\zeta_2) \rangle = 0.11(2)\,.
\end{equation}
Also displayed in Fig.\,\ref{qpizeta2} are the phenomenological extractions from Ref.\,\cite{Barry:2018ort}.  Even though the valence-quark distribution function fitted in Ref.\,\cite{Barry:2018ort} yields a momentum fraction compatible with that in Eq.\,\eqref{pionMFs}, its $x$-profile is very different.  In fact, as already noted, the phenomenological analysis in Ref.\,\cite{Barry:2018ort} neglected threshold resummation effects, which are important at large $x$ \cite{Aicher:2010cb, Westmark:2017uig}, and produced a valence-quark DF that does not satisfy the QCD large-$x$ constraint, Eq.\,\eqref{PDFQCD}.  (Similar remarks apply to the analysis in Ref.\,\cite{Novikov:2020snp}.)

Regarding the glue and sea, the size-ordering of the predictions in Refs.\,\cite{Ding:2019qlr, Ding:2019lwe, Cui:2020dlm, Cui:2020piK} agrees with that in Ref.\,\cite{Barry:2018ort}, but the gluon fraction is $\sim 20\%$ larger and the sea fraction is $\sim 30$\% smaller.

Referring to Fig.\,\ref{qpizeta2}, the predicted glue distribution and the phenomenological result agree semiquantitatively on $x\gtrsim 0.05$; but they are very different on the complementary domain.  Additionally, both glue DFs in Fig.\,\ref{qpizeta2} conflict with those estimated previously \cite{Gluck:1999xe, Sutton:1991ay}.  These discrepancies stress that the pion's gluon content is poorly known; but new prompt photon and $J/\psi$ production measurements could address this problem  \cite{Denisov:2018unjF, Chang:2020rdy}.

The sea DFs in Fig.\,\ref{qpizeta2}B disagree on the complete $x$-domain; so if knowledge of the pion's gluon DF is poor, then one can say that the sea quark distribution is experimentally uncharted.  This could be corrected by obtaining DY data with $\pi^\pm$ beams on isoscalar targets \cite{Londergan:1995wp, Denisov:2018unjF} or at EicC via corresponding tagged DIS measurements.

\subsection{Kaon distribution functions}
Given that knowledge of kaon structure provides a unique window onto the interference between Higgs boson effects and EHM, Refs.\,\cite{Cui:2020dlm, Cui:2020piK} developed parameter-free predictions for the pointwise behaviour of all $K$ distribution functions (DFs), including glue and sea, and comparisons with the analogous $\pi$ distributions.  The latter are important because, concerning kaon structure functions, the only available empirical information is the ratio ${\mathpzc u}^K(x)/{\mathpzc u}^\pi(x)$, which was measured at $\zeta \approx \zeta_5$ using the DY process forty years ago \cite{Badier:1980jq}.

\begin{figure}[t]
\centering
\begin{tabular}{cc}
\hspace*{14em}{\large{\textsf{A}}} &
\hspace*{14em}{\large{\textsf{B}}} \\[-8ex]
\includegraphics[clip, width=0.46\textwidth]{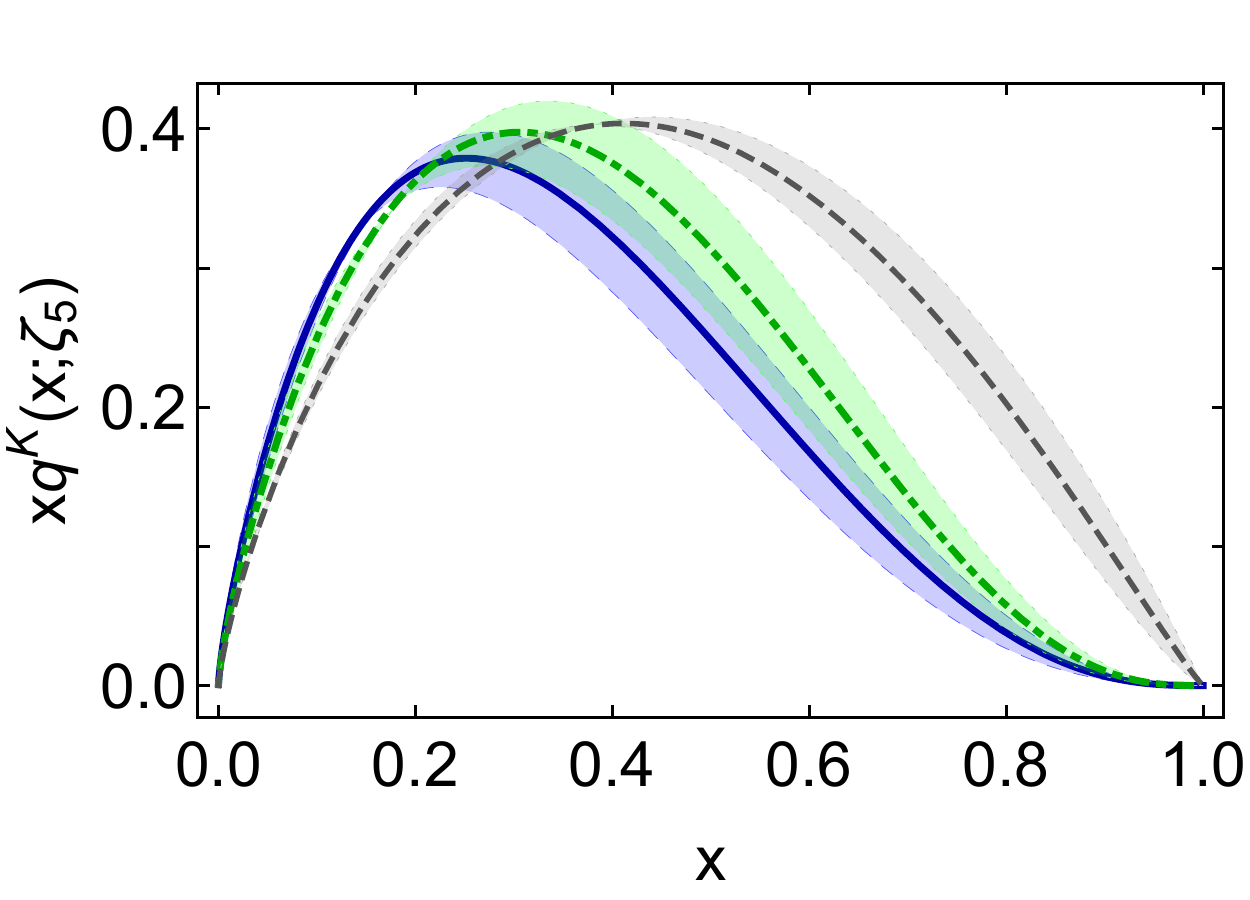} &
\includegraphics[clip, width=0.455\textwidth]{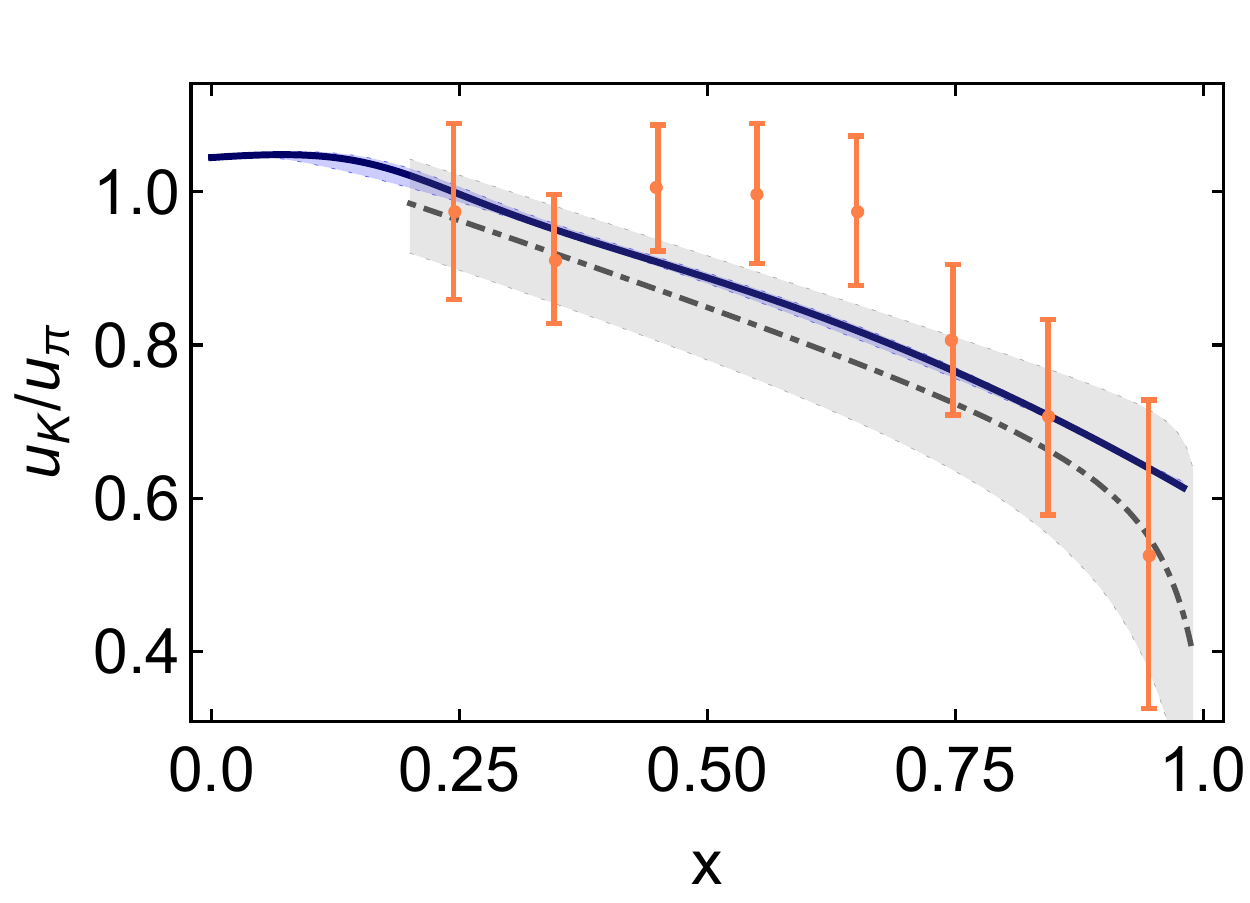}
\end{tabular}
\caption{\label{qKzeta5}
\emph{Left panel --A}.
Solid blue curve -- kaon's valence $u$-quark distribution; and dot-dashed green curve -- analogous result for the kaon's valence $\bar s$ distribution.  Both drawn from Refs.\,\cite{Cui:2020dlm, Cui:2020piK}.
Dashed grey curve within grey bands -- kaon $\bar s$ valence-quark distribution obtained in a recent lQCD study \cite{Lin:2020ssv}.
\emph{Right panel--B}.
$u^K(x;\zeta_5)/u^\pi(x;\zeta_5)$.
Solid blue curve -- prediction from Refs.\,\cite{Cui:2020dlm, Cui:2020piK}.
Dot-dashed grey curve within grey band -- lQCD result \cite{Lin:2020ssv}.
Data (orange) from Ref.\,\cite{Badier:1980jq}.
(In both panels, the bands bracketing the central DF curves from Refs.\,\cite{Cui:2020dlm, Cui:2020piK} reveal the effect of uncertainty in $\hat\alpha(0)$.  It is negligible for the ratio.)
}
\end{figure}

The valence-quark DFs obtained in Refs.\,\cite{Cui:2020dlm, Cui:2020piK} using mass-independent evolution $\zeta_H\to \zeta_5$ are depicted in Fig.\,\ref{qKzeta5}A.  Both are consistent with Eq.\,\eqref{PDFQCD}; and they produce the following low-order moments:
\begin{equation}
\begin{array}{l|c|c|c}
{\mathpzc q}\backslash\zeta_5 & \langle x{\mathpzc q}^K \rangle & \langle x^2{\mathpzc q}^K \rangle & \langle x^3{\mathpzc q}^K \rangle\\\hline
u & 0.19(2) & 0.067(09) & 0.030(5)\\
{\bar s}_{\not m} & 0.22(2) & 0.081(11) & 0.038(7)
\end{array}\,;
\label{uKsKmoments}
\end{equation}
hence, $\langle x[{\mathpzc u}^K(x;\zeta_5) + \bar{\mathpzc s}_{\not m}^K(x;\zeta_5)] \rangle = 0.41(4)$, reproducing the pion result at this scale.

First lQCD results for the kaon's valence-quark DFs are also now available \cite{Lin:2020ssv}.  The study finds the following moments, listed here in the order of appearance in Eq.\,\eqref{uKsKmoments}:
$u$ -- $0.193(8)$, $0.080(7)$, $0.042(6)$; and
$\bar s$ -- $0.267(8)$, $0.123(7)$, $0.070(6)$.
These values are systematically larger than the continuum predictions, especially for the $\bar s$.  The excesses are: $u$ -- $0.6(4.8)$\%, $21(6)$\%, $40(4)$\%; and $\bar s$ -- $24(7)$\%, $53(13)$\%, $84(16)$\%; and appear because, when compared with the continuum DFs, the lQCD DFs are much harder.  This feature is highlighted in Fig.\,\ref{qKzeta5}A.  In fact, the lQCD results are inconsistent with the QCD prediction, Eq.\,\eqref{PDFQCD}: on $x\simeq 1$, the lQCD DF behaves as $(1-x)^{\beta}$, $\beta = 1.13(16)$.  One may reasonably expect that future refinements of lQCD setups, algorithms and analyses will move the lattice results closer to the continuum predictions.

Figure~\ref{qKzeta5}B depicts the ratio ${\mathpzc u}^K(x;\zeta_5)/{\mathpzc u}^\pi(x;\zeta_5)$ calculated in Refs.\,\cite{Cui:2020dlm, Cui:2020piK}.  Referred to the published data \cite{Badier:1980jq}, $\chi^2/{\rm d.o.f}=0.86$.
The first lQCD results for this ratio are also drawn in Fig.\,\ref{qKzeta5}B.  Compared with experiment, $\chi^2/{\rm d.o.f}=1.81(1.38)$.  Relative to the continuum prediction, the central lQCD result deviates by only $\approx 5$\% despite the fact that the individual lQCD DFs are qualitatively and quantitatively different from the continuum DFs, \emph{e.g}.\ Fig.\,\ref{qKzeta5}A.  This feature highlights that ${\mathpzc u}^K(x;\zeta_5)/{\mathpzc u}^\pi(x;\zeta_5)$ is forgiving of even large variations between the individual DFs used to produce the ratio; see, \emph{e.g}.\ Refs.\,\cite{Martin:1980mi, Gluck:1997ww, Davidson:2001cc, Alberg:2011yr, Chen:2016sno, Lan:2019rba}.  More precise data is crucial if this ratio is to be used effectively to test the modern understanding of SM NG modes; and results for ${\mathpzc u}^\pi(x;\zeta_5)$, ${\mathpzc u}^K(x;\zeta_5)$ separately have greater discriminating power \cite{Keppel:2015, C12-15-006A, Denisov:2018unjF}.

In any symmetry-preserving study, which begins at $\zeta_H$ with a bound-state constituted solely from dressed quasiparticles and implements physical constraints on $\pi$ and $K$ wave functions, the exercise of DGLAP evolution using massless splitting functions will yield glue and sea distributions in the kaon that are practically identical to those in the pion \cite{Cui:2020dlm, Cui:2020piK}.  Of course, the $\bar s$ quark is more massive than the $u$ quark.  Hence,  \cite{Landau:1953um, Migdal:1956tc}: valence $\bar s$ quarks must in reality produce less gluons than valence $u$ quarks; and gluon splitting must produce less $\bar s s$ pairs than light-quark pairs.  Such effects are expressed in mass-dependent splitting functions.

\begin{figure}[t]
\centering
\begin{tabular}{cc}
\hspace*{14em}{\large{\textsf{A}}} &
\hspace*{14em}{\large{\textsf{B}}} \\[-8ex]
\includegraphics[clip, width=0.46\textwidth]{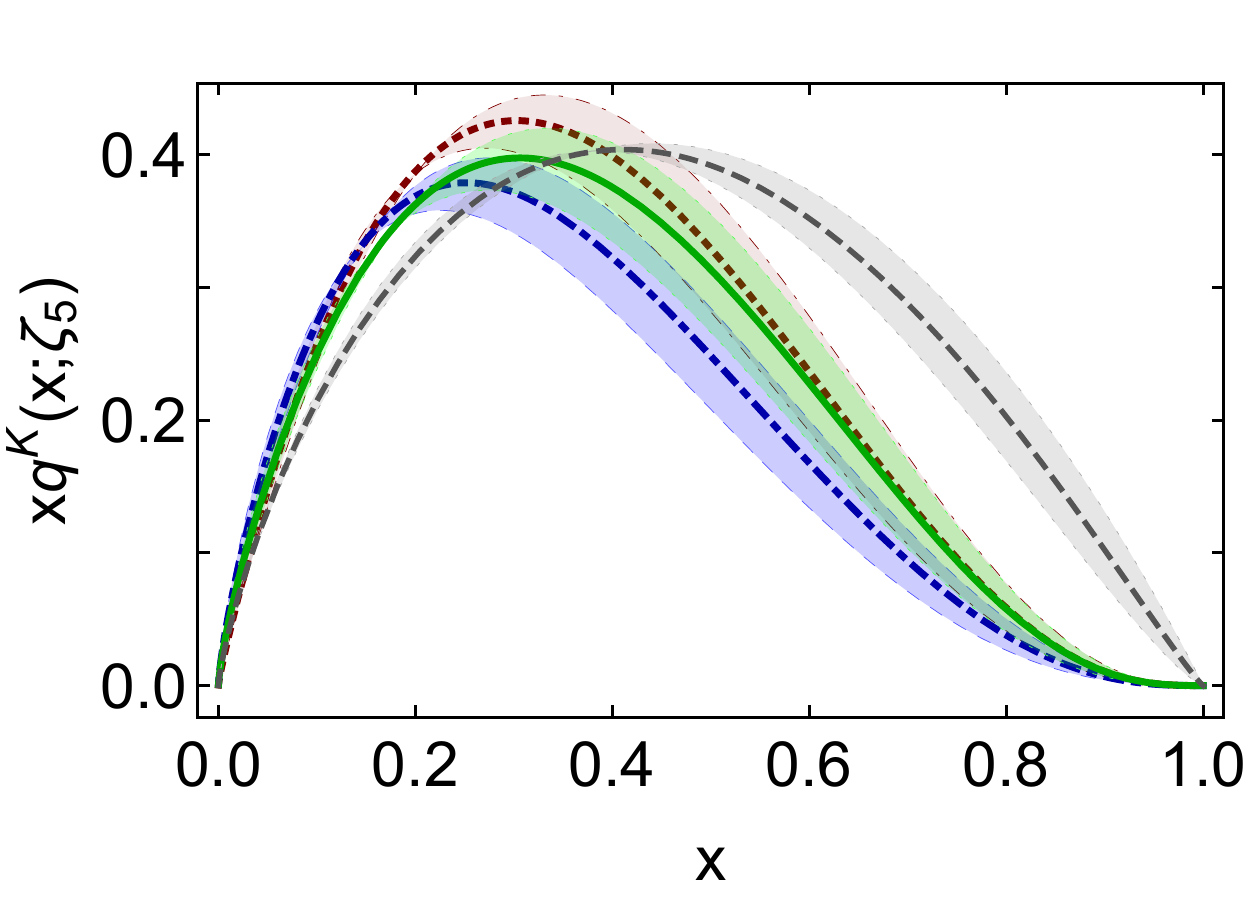} &
\includegraphics[clip, width=0.455\textwidth]{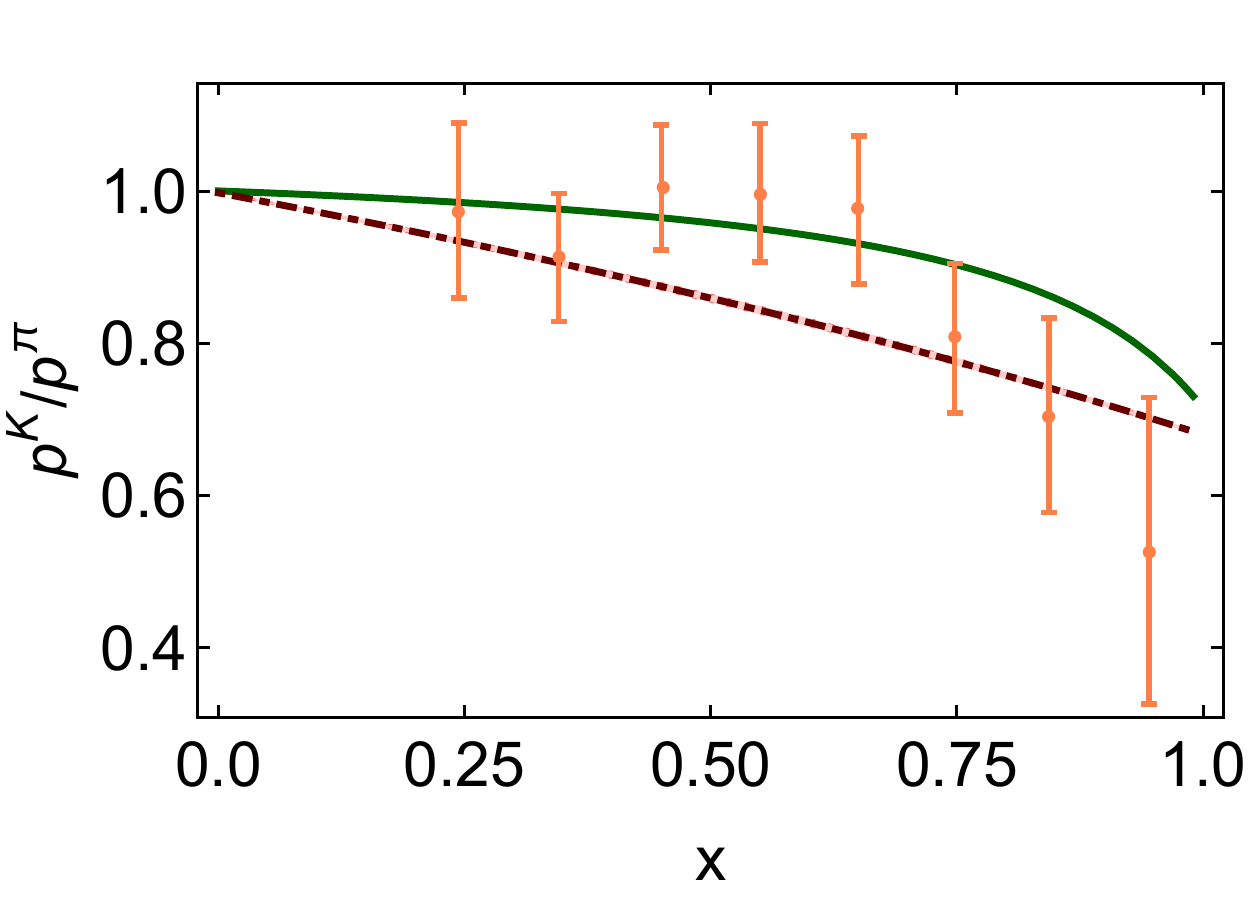}
\end{tabular}
\caption{\label{KDFmassive}
\emph{Left panel --A}.  From Refs.\,\cite{Cui:2020dlm, Cui:2020piK}:
$u^K(x;\zeta_5)$ -- dot-dashed blue curve;
$\bar s_{\not m}^K(x;\zeta_5)$ [mass-independent splitting] -- solid green;
and $\bar s_m^K(x;\zeta_5)$ [mass-dependent splitting] -- dotted maroon.
Dashed grey curve within grey bands -- lQCD result for $\bar s^K(x;\zeta_5)$ \cite{Lin:2020ssv}.
(In this panel, the bands bracketing the central continuum DF curves reflect the uncertainty in $\hat \alpha(0)$, Fig.\,\ref{Figwidehatalpha}.)
%%%
\emph{Right panel--B}.
Predictions: ${\mathpzc g}^K(x;\zeta_5)/{\mathpzc g}^\pi(x;\zeta_5)$ -- solid green curve;
and ${\mathpzc S}^K(x;\zeta_5)/{\mathpzc S}^\pi(x;\zeta_5)$ -- dot-dashed red curve.
Data on ${\mathpzc u}^K(x;\zeta_5)/{\mathpzc u}^\pi(x;\zeta_5)$ (orange) from Ref.\,\cite{Badier:1980jq} are included to guide comparisons.
(In this panel, the $\hat\alpha(0)$-induced uncertainty is negligible.)
}
\end{figure}

The impact of mass-dependent splitting was estimated in Refs.\,\cite{Cui:2020dlm, Cui:2020piK}, with the results drawn in Fig.\,\ref{KDFmassive}.  By construction, $u^K(x;\zeta_5)$ is unaffected.  However, the $\bar s$-quark DF is enlarged -- Fig.\,\ref{KDFmassive}A, with the lowest three nontrivial moments growing by approximately 5\%, as one can see by comparing Eq.\,\eqref{uKsKmoments} with the following:
\begin{equation}
\begin{array}{l|c|c|c}
{\mathpzc q}\backslash\zeta_5 & \langle x{\mathpzc q}^K \rangle & \langle x^2{\mathpzc q}^K \rangle & \langle x^3{\mathpzc q}^K \rangle\\\hline
u & 0.19(2) & 0.067(09) & 0.030(05)\\
{\bar s}_m & 0.23(2) & 0.085(11) & 0.040(07)\\\hline
%\bar s_m^K(x;\zeta_5)
%
u + {\bar s}_m & 0.42(3) & 0.152(20) &0.070(12) \\
\end{array}\,.
\label{uKsKmoments2}
\end{equation}

Figure~\ref{KDFmassive}B depicts the kaon-to-pion ratio of glue and sea distributions as obtained with the mass-dependent splitting functions described in Refs.\,\cite{Cui:2020dlm, Cui:2020piK}.
Evidently, the kaon's glue (${\mathpzc g}$) and sea (${\mathpzc S}$) distributions differ from those of the pion only on the valence region $x\gtrsim0.2$.  This is not surprising because:
mass-dependent splitting functions act primarily to modify the valence DF of the heavier quark;
valence DFs are negligible at low-$x$, where glue and sea distributions are large, and vice versa;
hence the biggest impact of a change in the valence DFs must lie at large-$x$.
Notably, each of the predicted ratios in Fig.\,\ref{KDFmassive}B is pointwise similar to the measured value of ${\mathpzc u}^K(x;\zeta_5)/{\mathpzc u}^\pi(x;\zeta_5)$.
On the complementary domain, $x\lesssim 0.2$, the glue and sea DFs in the kaon and pion are practically identical.
Using the computed DFs, one finds ($\zeta=\zeta_5$):
%\begin{equation}
$\langle x\rangle^K_g = 0.44(2)$,
$\langle x\rangle^K_{\rm sea} = 0.14(2)$,
%\end{equation}
with $\langle x\rangle^K_{{\rm sea}_{{\mathpzc l}}} = 0.091(11)$, $\langle x\rangle^K_{{\rm sea}_{s}} = 0.045(06)$, where ${\mathpzc l}$ denotes the light-quarks.
Comparing these results with those for the pion, then accounting for mass-dependent splitting functions, the gluon light-front momentum fraction in the kaon is $\sim 1$\% less than that in the pion and the sea fraction is $\sim 2$\% less.

Regarding kaon valence distributions, it is worth recording here that there is a single, recent lQCD study \cite{Lin:2020ssv} and model estimates exist, \emph{e.g}.\, Refs.\,\cite{Davidson:2001cc, Alberg:2011yr, Chen:2016sno}; but there are no results for the pointwise behaviour of the kaon's glue and sea distributions.  Thus, the predictions in Refs.\,\cite{Cui:2020dlm, Cui:2020piK} for the entire array of $\pi$ and $K$ DFs stand alone.

The Standard Model's (pseudo-) Nambu-Goldstone modes ($\pi$ and $K$) are basic to the formation of everything, from nucleons to nuclei, and on to neutron stars.  Hence, new-era experiments capable of testing the modern theory predictions reviewed above should have high priority.  Such measurements present a great opportunity to the EicC.
%% EicC is potentially uniquely capable of providing information on the partonic structure of Goldstone bosons, adding significantly to data on the valence structure and providing essentially new information on the sea and glue distributions.

\subsection{Nucleon distribution functions}
As noted in Sect.\,\ref{FFsN}, modern $ep$ elastic scattering experiments can be understood to indicate that the proton's internal structure is very complex, with EHM producing a Poincar\'e-covariant proton wave function which is distinguished by the presence of strong, nonpointlike, fully-interacting diquark correlations and significant quark-diquark orbital angular momentum \cite{Barabanov:2020jvnS}.  
%  Roberts:2013mja, Segovia:2015ufa
These features are also expressed in the proton's DFs, \emph{e.g}.\ in the large-$x$ behaviour of the ratio $F_2^n/F_2^p$, which is a surrogate for the ratio $d_{\rm v}/u_{\rm v}$, where $d_{\rm v}$, $u_{\rm v}$ are the proton's valence-quark DFs \cite[Sect.\,II.G]{Holt:2010vj}.  In this context, the MARATHON experiment was proposed at JLab \cite{MARATHON}.  It aimed to use DIS off the mirror nuclei $^3$H and $^3$He to determine $F_2^n/F_2^p$ on the valence-quark domain.

\begin{figure}[t]
\centering
\begin{tabular}{cc}
\hspace*{14em}{\large{\textsf{A}}} &
\hspace*{14em}{\large{\textsf{B}}} \\[-8ex]
\includegraphics[clip, width=0.46\textwidth]{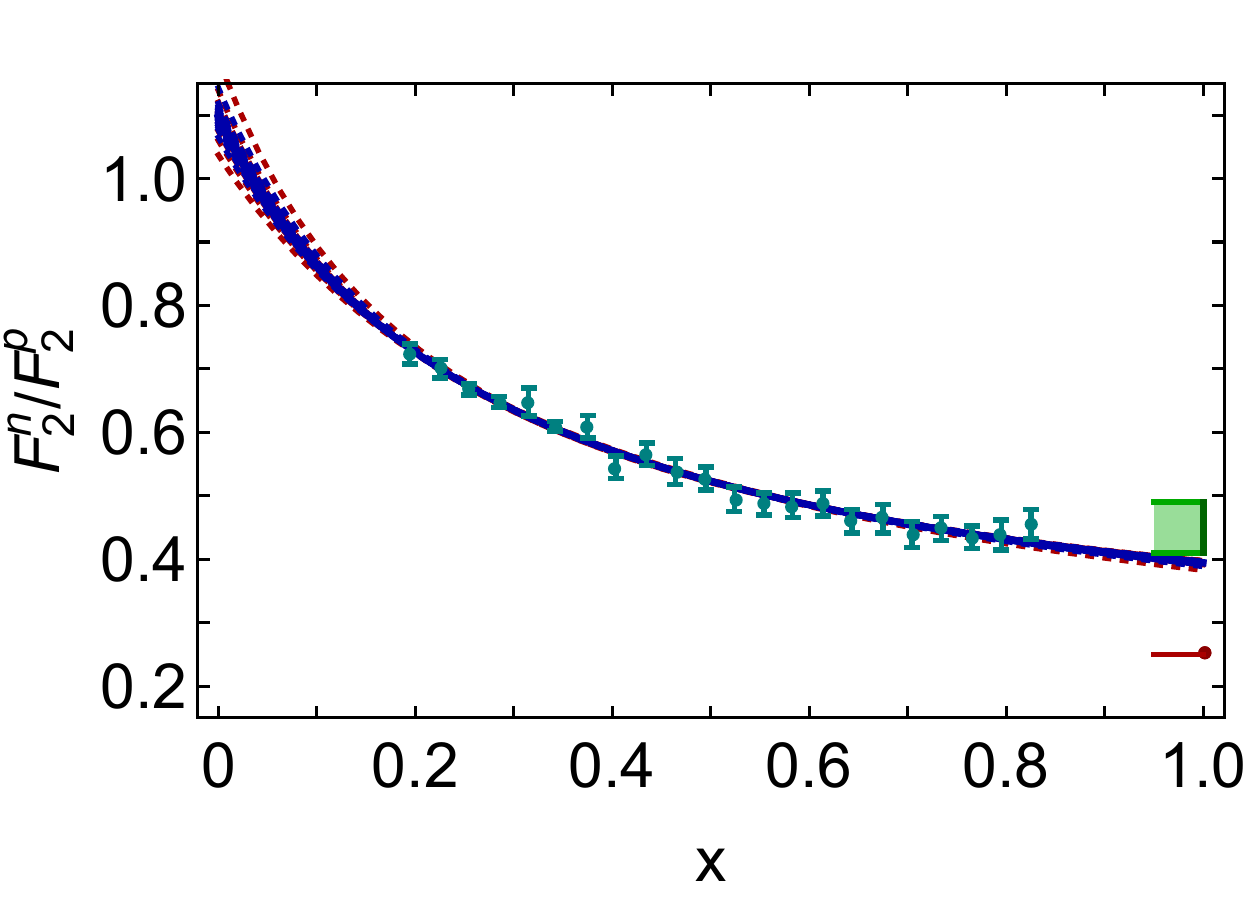} &
\includegraphics[clip, width=0.455\textwidth]{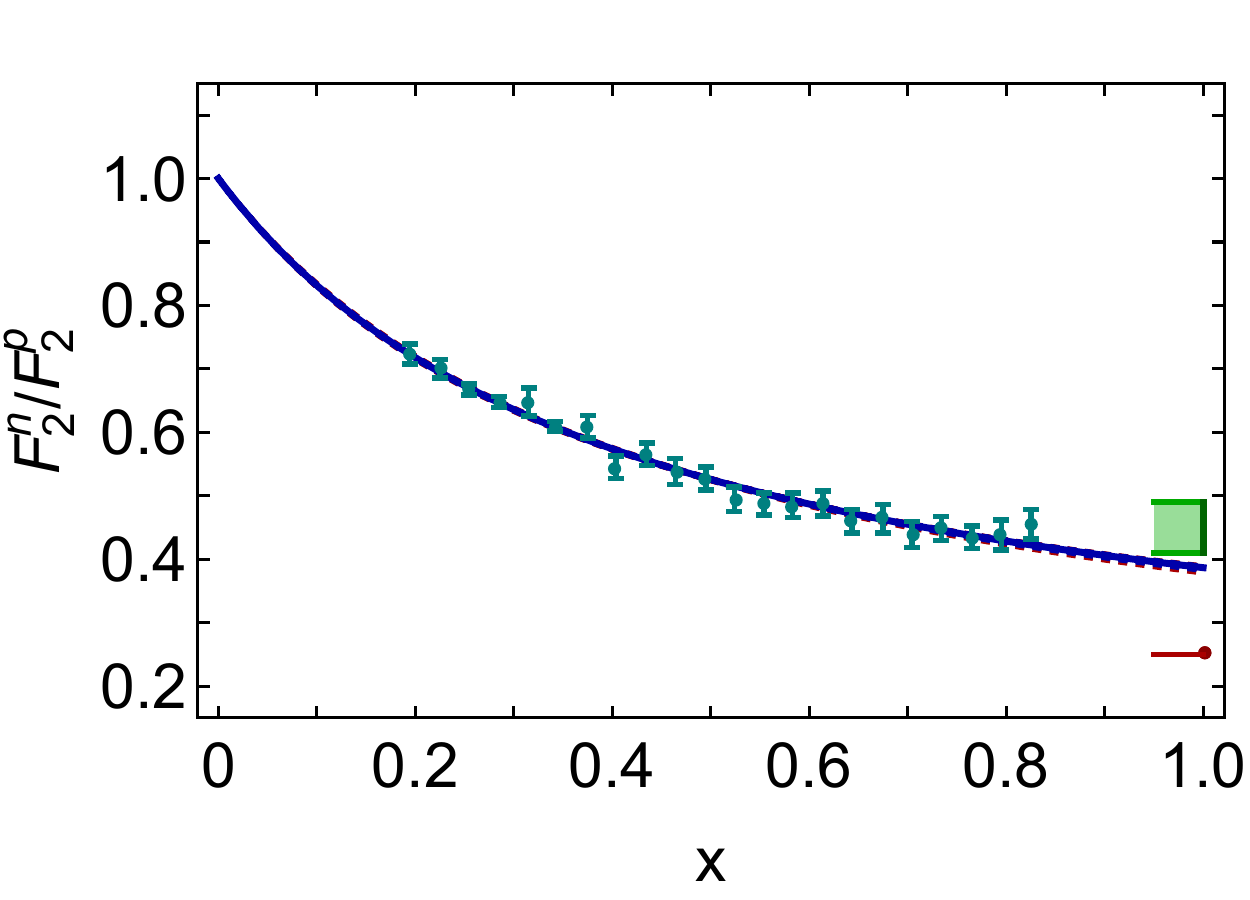}
\end{tabular}
\caption{\label{FMarathon}
Preliminary data from the MARATHON experiment (teal).
\emph{Left panel -- A}.
Dotted red and blue curves: array of $[1,1]$ Pad\'e fits obtained from a jackknife analysis of the data.  Extrapolated, these curves yield $\left. F_2^n/F_2^p\right|_{x=0}=1.10(3)$, consistent with dominance of sea- over valence-quarks on $x\simeq 0$; and $\left. F_2^n/F_2^p\right|_{x=1} = 0.395(3)$.
\emph{Right panel -- B}.  Array of $[1,1]$ Pad\'e fits obtained from a jackknife analysis of the data constrained by the assumption of sea-quark dominance, \emph{i.e}.\ enforcing $\left. F_2^n/F_2^p\right|_{x=0}=1$.  Extrapolated, these curves yield $\left. F_2^n/F_2^p\right|_{x=1} = 0.387(2)$.
Theory predictions in both panels: (\emph{i}) red line and circle at $x=1$, the value $\tfrac{1}{4}$, which is obtained if the proton's valence structure is simply $u$-quark + isoscalar-scalar $[ud]$-diquark; and (\emph{ii}) green band, range of values obtained when axial-vector diquark correlations contribute $25$-$35$\% of the proton’s normalization.
Evidently, current indications are that the MARATHON data and the reanalysis in Ref.\,\cite{Segarra:2019gbp} lean heavily in favour of scenario (\emph{ii}).
}
\end{figure}

The MARATHON experiment is complete.  Preliminary results were released in 2019 (see, \emph{e.g}.\ Ref.\,\cite[page\,5]{MARATHONMcK}) and are displayed in Fig.\,\ref{FMarathon}.  They appear quantitatively consistent with new reanalyses of existing data on a wide variety of nuclei \cite{Segarra:2019gbp}.  This agreement increases confidence in the preliminary MARATHON analysis.

A modern theoretical interpretation of $F_2^n/F_2^p$ is provided in Ref.\,\cite{Roberts:2013mja}, from which the theoretical predictions in Fig.\,\ref{FMarathon} are taken.  Evidently, the new experimental results are consistent with longstanding continuum predictions that, as a corollary of EHM, the proton contains both scalar and axial-vector diquark correlations, with the latter being significant.  The MARATHON data are a crucial step forward in understanding hadron structure.  They could have a far-reaching impact on developing a solution to the puzzle of EHM.  This is strong motivation for the development of experiments at EicC aimed at extracting $F_2^n/F_2^p$ on the valence-quark domain.

\section{Testing Structure Predictions}
\label{EStructure}
% proton mass and anomaly ... pi FF ... pi SF ... pi DVMP (GPD)
%
%\input ExperimentStructure
%
Owing to the high resolution and the well-known character of electromagnetic interactions, a clean and precise way to test hadron structure predictions is to use photon probes with high virtuality in high-energy $eN$ scattering.  The EicC, which can be regarded as a ``super electron microscope'', should provide an excellent opportunity for such studies, owing to its anticipated high c.m.\ energy, high luminosity and diverse polarisation settings, and the expected nearly full-acceptance of the associated spectrometer.

Given that the proton serves as a meson source and mesons are the force carriers within nuclei, one expects to find a cloud of virtual mesons (pions and kaons) in the neighbourhood of any proton.  These virtual states can be exploited as ``meson targets'' via Sullivan-like processes \cite{Sullivan:1971kd} (detailed in connection with Fig.\,\ref{fig:SullvianDiagrams} below).  As opposed to $ep$ DIS experiments, in Sullivan-like processes one focuses on those events in which the final-state nucleon has small transverse momentum, $p_T$, and a large fraction of the incoming nucleon's longitudinal momentum.  This leading nucleon in the final state is tagged in order to isolate the events of interest.

%%%In experimental particle physics, pseudorapidity, {\displaystyle \eta }\eta , is a commonly used spatial coordinate describing the angle of a particle relative to the beam axis. It is defined as
%{\displaystyle \eta \equiv -\ln \left[\tan \left({\frac {\theta }{2}}\right)\right],}\eta \equiv -\ln \left[\tan \left({\frac  {\theta }{2}}\right)\right],
%where {\displaystyle \theta }\theta  is the angle between the particle three-momentum {\displaystyle \mathbf {p} }\mathbf {p}  and the positive direction of the beam axis.
%%% larger eta => more forward the scattering event

With the current conceptual design for the EicC spectrometer, it is envisioned that the barrel and end-cap detectors will cover a pseudorapidity range $|\eta|<3$, with the limiting coverage reaching $|\eta|\sim3.5$ owing to the limited space near the interaction point.  Nevertheless, zero-degree detectors are being considered for both the electron and the ion forward regions.  Therefore, the largest pseudorapidity could reach around $5$ with the far-forward detectors.  Moreover, with the time-of-flight and Cherenkov detectors, reliable identification of $\pi /K /N$ can be expected in a wide momentum range.  Finally and notably, a forward-neutron calorimeter will likely be integrated into the detector system.  Hence, the proposed EicC detector should meet the requirements of a very diverse experimental programme.

%%% subsection - 1
\subsection{Proton mass and $\rm\Upsilon$ electroproduction}
\label{ss:upsilon}
As discussed in Sect.\,\ref{massdichotomy}, uncovering the origin and distribution of the proton's mass has become one of the highest priorities in modern nuclear and particle physics.  An early analysis of the QCD energy momentum tensor \cite{Ji:1994av, Ji:1995sv} separates the proton mass into four distinct contributions:
quark mass, $M_m$;
quark energy, $M_q$;
gluon energy, $M_g$;
and trace anomaly, $M_a$.
Using a parton model basis, the trace anomaly part is a pure-glue quantum effect; and thus isolated, it seems to contribute roughly one quarter of the proton mass, $M_a=(1-\varsigma_h)M_N/4$.  The precise size of this contribution is evidently characterised by the parameter $\varsigma_h$, which, from the same perspective, measures the contribution to the proton mass that originates with the current-quark mass-term in the standard QCD Lagrangian.  $\varsigma_h$ is only loosely constrained by existing experiments.  (Caveats on the interpretation of such decompositions are listed on page\,\pageref{Nonsense} and elsewhere \cite{Aguilar:2019teb}.)

Similar to the way a hydrogen atom interacts with an external electromagnetic field, an $\Upsilon$ meson, which is significantly smaller than the proton, interacts with the proton's gluon field, providing an opportunity to study aspects of the proton's mass distribution.  Indeed, for low energy $\Upsilon N$ interactions, using the operator product expansion and low-energy theorems, Refs.\,\cite{Kharzeev:1995ij, Kharzeev:1998bz} relate the scattering amplitude at threshold to the energy-momentum tensor evaluated in the proton state:
\begin{align}
\label{UpsilonNAmplitude}
{\cal M}_{\rm{\Upsilon N}} &\simeq 2m_{\rm\Upsilon} r_0^3 d_2 \frac{2\pi^2}{27}\left(2m_{N}^2 -\left<N\bigg|\sum_{i=u,d,s}m_i\bar{q}_i q_i\bigg|N\right>\right)\simeq 2m_{\rm\Upsilon}  r_0^3 d_2 \frac{2\pi^2}{27}2m_{N}^2 (1 - \varsigma_h),
\end{align}
%%% \varsigma_h * m_N = normal sigma term
where $r_0$ is the Bohr radius of the heavy quarkonium $\rm{\Upsilon(1S)}$ \cite{Kharzeev:1995ij} and the Wilson coefficient, $d_2$, is given in Ref. \cite{Peskin:1979va}.  Tracing the derivation, one learns from Eq.\,\eqref{UpsilonNAmplitude} that, interpreted in a parton model basis, ${\cal M}_{\rm{\Upsilon N}}$ is completely determined by $\Theta_0$ evaluated in the proton state, Eqs.\,\eqref{definetheta0}, \eqref{anomalyproton}, with a (small) correction from the QCD Lagrangian's current-quark mass term, $\varsigma_h$.
Note that $\varsigma_h = (\sigma_{\pi N} + \sigma_s)/m_N$, where $\sigma_{\pi N}$ and $\sigma_s$ are the pion-nucleon and strangeness $\sigma$ terms, respectively.  While a value of $\sigma_{\pi N}$ has been  extracted from a careful analysis of pion-nucleon scattering data \cite{Hoferichter:2015dsa, Hoferichter:2016ocj, RuizdeElvira:2017stg}, $\sigma_s$ remains poorly constrained.

Assuming validity of Eq.\,\eqref{UpsilonNAmplitude} for charmonium production, Ref.\,\cite{Wang:2019mza} analysed GlueX data \cite{Ali:2019lzf}, with the result $\varsigma_h= 0.07(17)$, similar to a lQCD estimate \cite{Yang:2018nqn}.  Ref.\,\cite{Ali:2019lzf} also included an analysis motivated by Eq.\,\eqref{UpsilonNAmplitude}, concluding that the $\varsigma_h$-correction to the $\Theta_0$ contribution is small.

It is worth remarking that Eq.\,\eqref{UpsilonNAmplitude} features $r_0$, any estimate of which depends on the heavy-quark mass and the strong running coupling evaluated at a relevant scale.  The heavy-quark mass and ``relevant'' scale are model dependent.
Furthermore, given the closeness of the $\Lambda_c\bar D$ and $J/\psi N$ thresholds, there could be important coupled-channel effects which are not captured by a simple gluon-exchange picture of the $J/\psi$-nucleon interaction \cite{Baru:2020}.
Hence, improved theory guidance would greatly assist with future such analyses.

\begin{figure}[t]
\begin{tabular}{lcl}
\includegraphics[width=0.42\textwidth]{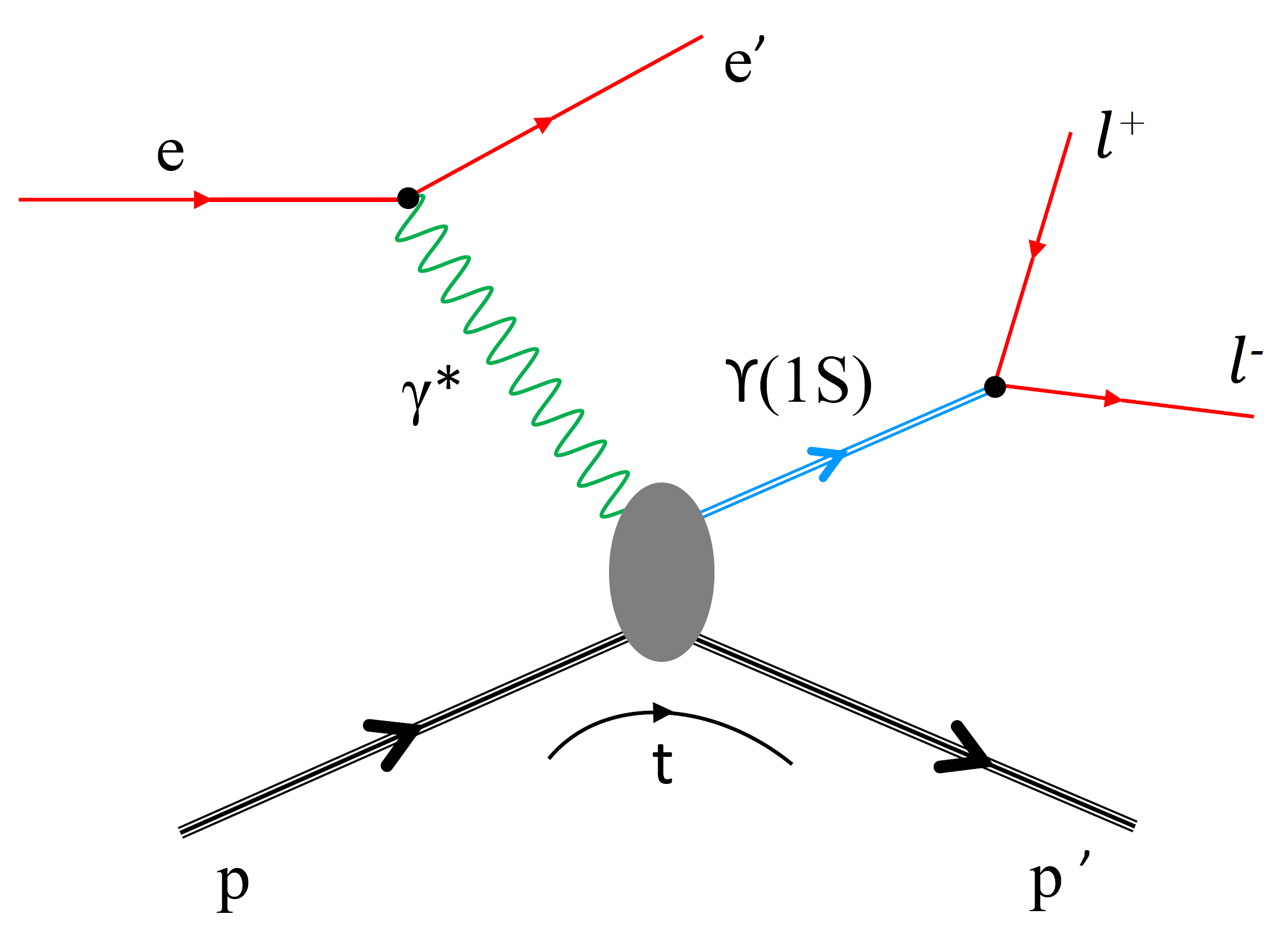} & \hspace*{1em} &
\includegraphics[width=0.42\textwidth]{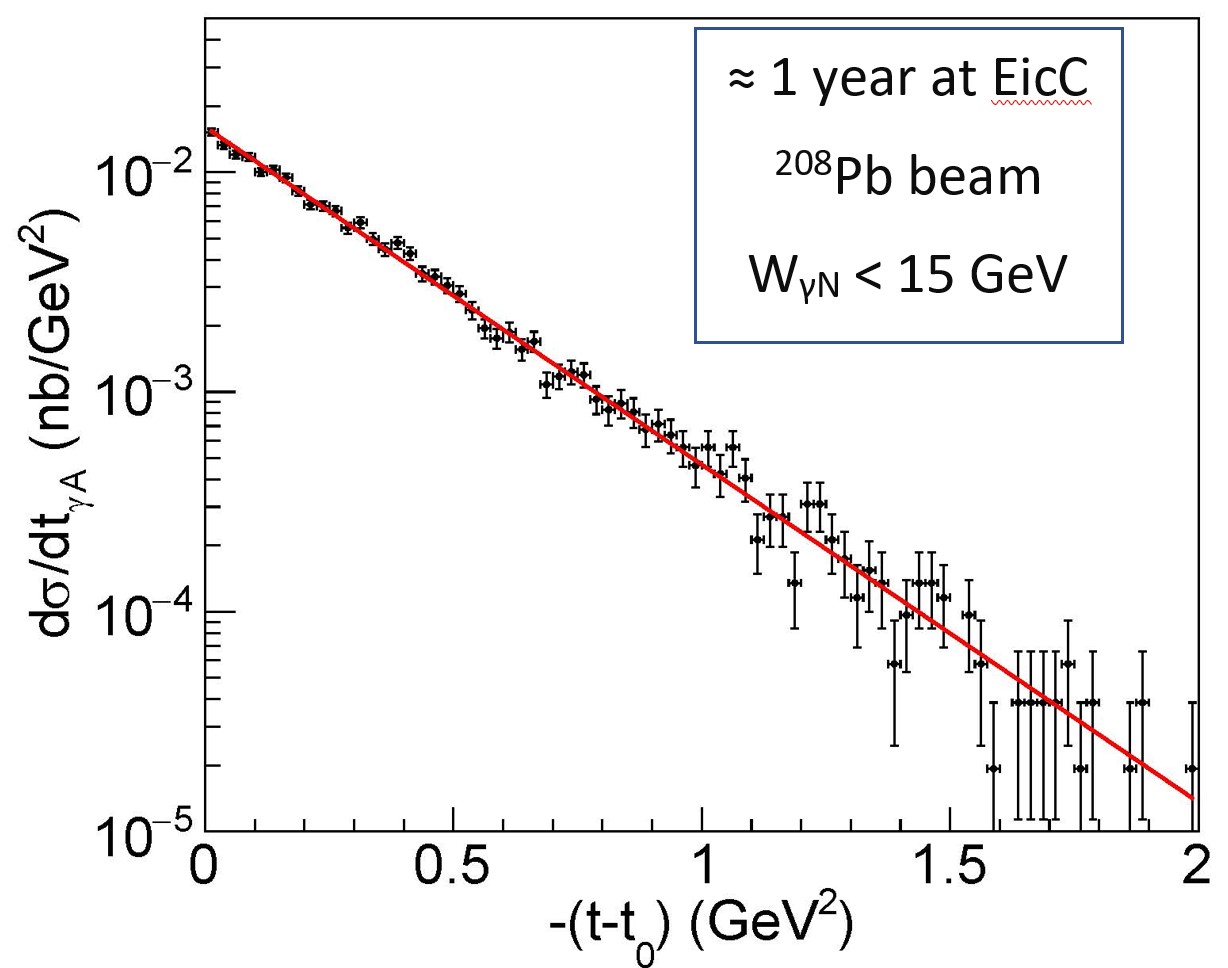}\\[-38ex]
\hspace*{1em}{\large{\textsf{A}}} & \hspace*{1em} & \hspace*{1em}{\large{\textsf{B}}}
\end{tabular}
\vspace*{38ex}
\caption{
\emph{Left panel--A}.
Schematic diagram of ${\Upsilon(1S)}$ electroproduction: $t$ is the square of the four momentum exchange between the virtual photon and the proton.
\label{fig:UpsilonProductionDiagram}
\emph{Right panel--B}.
A simulation of the extracted ${\Upsilon(1S)}$ per nucleon photoproduction cross-section in
$e$-Pb collisions.  The $W$-range is from the threshold to $\sim 15\,$GeV.
%\label{fig:UpsilonSimulation}
}
\end{figure}

%%\begin{figure}[t]
%%\centering
%%\includegraphics[width=0.46\textwidth]{Upsilon-production-crosssection-at-EicC.pdf}
%%\caption{
%%A simulation of the extracted ${\Upsilon(1S)}$ per nucleon photoproduction cross-section in$e$-Pb collsions.  The $W$-range is from the threshold to $\sim 15\,$GeV.\label{fig:UpsilonSimulation}}
%%\end{figure}

With the expected strengths of the EicC, one could constrain $\varsigma_h$ via precise measurement of $\Upsilon$ electroproduction near the threshold.  Consider Fig. \ref{fig:UpsilonProductionDiagram}A, which depicts a high-energy photon exchanged between the electron and the proton.  By isolating the virtual photon flux $\Gamma(Q^2<1, E_{\gamma}>E_0)$ \cite{Budnev:1974de}, where $E_0$ is the minimum photon energy required to produce the $\Upsilon$, the differential cross-section for $\Upsilon$ photoproduction can be separated from electroproduction data $ e p \to e^{\prime} p^{\prime} \Upsilon \to e^{\prime} p^{\prime} l^+ l^-$.  Assuming vector meson dominance (VMD), the differential cross-section $\gamma N \to \Upsilon N$ is directly connected to the differential cross-section for $\Upsilon N \to \Upsilon N$:
\begin{align}
\label{GammaNCrossSection}
\frac{d\sigma_{\rm{\gamma N\to \Upsilon N}}}{dt}\bigg|_{t=0}
=\frac{3\Gamma(\rm{\Upsilon\to e^+e^-})}{\alpha m_{\rm\Upsilon}}
\left(\frac{k_{\rm\Upsilon N}}{k_{\rm\gamma N}}\right)^2
\frac{d\sigma_{\rm{\Upsilon N\to \Upsilon N}}}{dt}\bigg|_{t=0}.
\end{align}
Therefore, exploiting the photoproduction cross-section of $\Upsilon$ near threshold, one has access to the low-energy dynamics of the $\Upsilon$-$N$ interaction; enabling an extraction of the $\varsigma_h$ correction to the QCD trace anomaly in the proton.

It should be remarked that VMD is questionable for $J/\psi$ and $\Upsilon$ systems \cite{Wu:2019adv, Xu:2019ilh}.  In particular, the $\Upsilon$'s in the near-threshold $\Upsilon$-nucleon scattering are on-shell, while that connected to the photon in the VMD model is very far off-shell.  So here, too, improved theory guidance is necessary to better inform the connection between $\Upsilon$ photoproduction and the proton mass decomposition.

To estimate the statistical uncertainty of the extracted value for $\varsigma_h$, it is necessary to know how many events to expect.
The EicC is anticipated to have an optimal c.m.\ energy of 16.75\,GeV \cite{Chen:2018wyz, EicCWP, EicCWPEL}.  At this energy, the cross-section for $\Upsilon$ electroproduction off the proton is predicted to lie in the range 48-85\,fb \cite{Wang:2016sfq, Xu:2020uaa, Gryniuk:2020mlh}.  According to current thinking, one full year of EicC running would deliver an integrated luminosity of approximately 50\,fb$^{-1}$.
Under these circumstances, it is advisable to use a Pb-beam in order to accumulate enough events because the cross-section for a nuclear beam is much larger: $\sigma_A = A \sigma_N$.
With one year of running, a statistical uncertainty of around $0.011$ is achievable in the measurement of a trace-anomaly correction $\varsigma_h \sim 0.10$, as can be seen from the least-squares fit to a Monte-Carlo simulation drawn in Fig.\,\ref{fig:UpsilonProductionDiagram}B.  This would improve to $ \approx 0.004$ over ten years,
%% ten years of running, a statistical uncertainty of around $0.004$ is achievable in the measurement of a trace-anomaly correction $\varsigma_h \sim 0.10$.  This estimate is based on a least-squares fit to a Monte-Carlo simulation, depicted in Fig.\,\ref{fig:UpsilonProductionDiagram}B.

Since the $\Upsilon$ is much heavier than the $J/\psi$, the need to study the quarkonium-nucleon interaction at sufficiently low energy is easily satisfied.  Moreover, the theory uncertainties from the $b$-quark mass and the strong running coupling at the $\rm\Upsilon$ scale are smaller.  Nonetheless, based on current estimates of the $b$-quark mass, the strong running coupling, and the uncertainty in each, the combined uncertainty from these QCD parameters is approximately $\delta_{\varsigma_h}^{\rm th}=0.05$.   Plainly, theoretical efforts aimed at reducing the size of $\delta_{\varsigma_h}^{\rm th}$ should have a high priority.  This is not just an issue for EicC measurements, but also for those underway or planned at other facilities.
Hence, with an EicC that achieved the c.m.\ energy under discussion, which lies just above the $\Upsilon$(1S) production threshold \cite{Chen:2018wyz, EicCWP, EicCWPEL}, the facility should prove very competitive in the race to determine $\varsigma_h$, the quark-mass correction to the QCD trace anomaly in the proton.

\begin{figure}[t]
\centering
\includegraphics[width=0.5\textwidth]{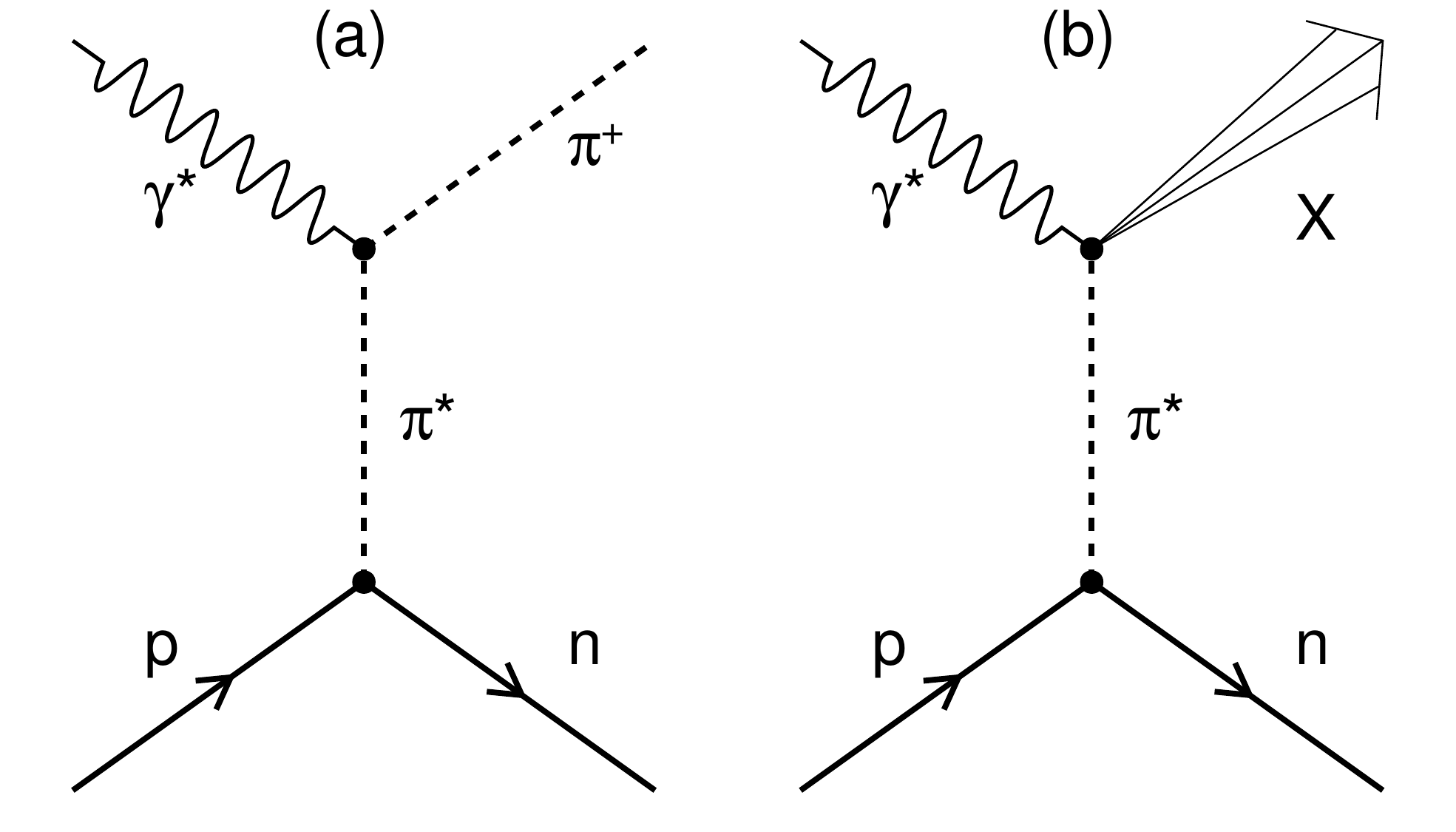}
\caption{\label{fig:SullvianDiagrams}
Sullivan processes.  In these examples, a nucleon's pion cloud is used to provide access to the pion's (\emph{a}) elastic form factor and (\emph{b}) parton distribution functions.  The intermediate pion, $\pi^\ast(P)$, is off-shell, with $P^2= –t$.  It has been estimated \cite{Qin:2017lcd} that such processes provide reliable access to a pion target on $-t\lesssim 0.6\,$GeV$^2$; and for the kaon, on $-t\lesssim 0.9\,$GeV$^2$.
}
\end{figure}

%%%  subsection - 2
\subsection{Pion and kaon from factor measurements}
\label{piKFFmeasurements}
To delve deep into the structure of QCD's NG modes (pions and kaons), the modern approach capitalises on the prominent $\pi$/$K$ pole contribution to the exclusive electroproduction of $\pi$/$K$-mesons when $t$ is kept small.  Assuming dominance of single meson exchange with such kinematics, the elastic scattering process can be inferred from careful analysis of $ep\to e\pi^+ n$ or $ep\to eK^+ \Lambda$ reactions.  The theoretical basis for such elastic form factor extractions is provided by the Sullivan process \cite{Sullivan:1971kd} illustrated in Fig.\,\ref{fig:SullvianDiagrams}(a).  Here, the pion remains intact in the final state and the pion elastic form factor enters through the $\gamma\pi^\ast\pi$ coupling.  For $\pi^+$ production, neutron exchange in the $u$ channel is suppressed because the $\gamma n n$  coupling disappears at tree level.

The method of extracting the pion elastic form factor, $F_\pi(Q^2)$, from pion electroproduction reactions has reached a mature level \cite{Guidal:1997by, Vanderhaeghen:1997ts, Choi:2015yia, Perry:2020hli}.  It capitalises on the fact that a good description of the $t$- and $W$-dependence in these reactions is provided by a gauge invariant model based on exchanges along the $\pi$ and $\rho$ Regge trajectories, with $\rho$-exchange making no contribution to the longitudinal cross-section at $t_{\rm min}$.  The framework allows the longitudinal electroproduction cross-section to be formulated such that $F_\pi(Q^2)$ is the only unknown, which can be determined by comparison between data and the model \cite{Guidal:1997by, Vanderhaeghen:1997ts, Choi:2015yia, Perry:2020hli}.  The certainty with which the longitudinal cross-section has been isolated can be checked using the $\pi^+/\pi^-$ production ratio extracted from electron-deuteron collisions in the same kinematics as charged pion data from $ep$ collisions \cite{Horn:2016rip}.

The statistical error for pion/kaon from factor extractions at EicC may be estimated using a one-pion exchange model.  At Born level, the differential cross-section is \cite{Sullivan:1971kd, Actor:1974ym}:
\begin{align}
\label{BornP}
{\mathpzc P}\frac{d\sigma_L}{dt}=4\hbar c(eg_{\pi NN})^2\frac{-t}{(t-m_{\pi}^2)^2}Q^2F_{\pi}^2(Q^2),
\end{align}
where $g_{\pi NN}$ is the pion-nucleon coupling and ${\mathpzc P}$ is a phase-space factor.  In this context, it is reasonable to suppose that the pion/kaon form factor is well described by a monopole form:
\begin{align}
\label{BornF}
F_{\pi,K}(Q^2)=\frac{1}{1+Q^2/\Lambda_{\pi,K}^2},
\end{align}
where $\Lambda_{\pi,K}$ is around $0.7\sim 0.8\,$GeV.  The cross-section for exclusive meson electroproduction is then obtained by adding the photon flux factor for $eN$ scattering.

\begin{figure}[t]
\centering
\includegraphics[width=0.60\textwidth]{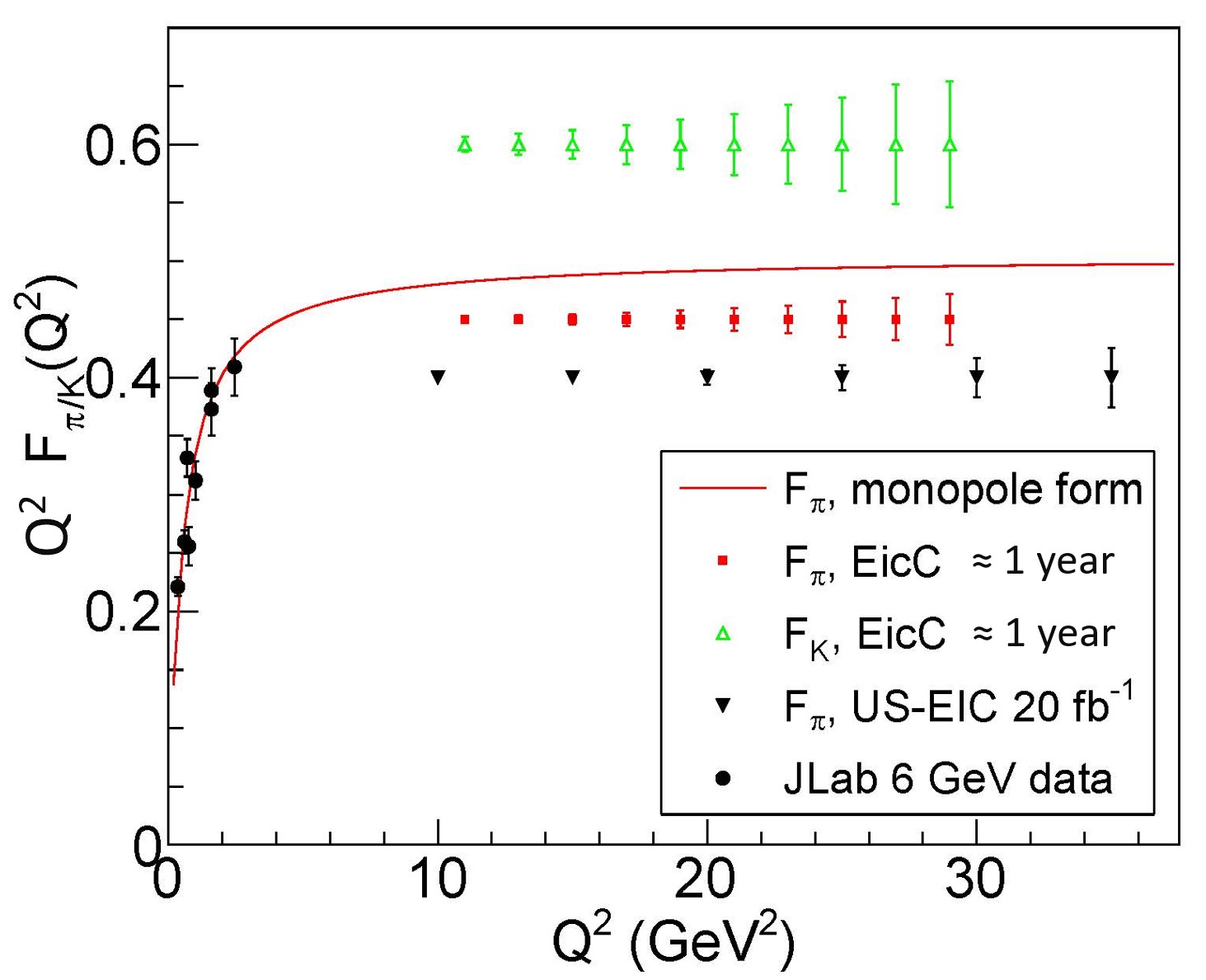}
\caption{
Statistical error estimations for pion and kaon form factor measurements at EicC and the US EIC \cite{Aguilar:2019teb}.
The JLab data plotted in this figure were produced by the F$_{\pi}$-Collaboration \cite{Horn:2006tm, Tadevosyan:2007yd, Horn:2007ug, Huber:2008id, Blok:2008jy}.
}
\label{fig:FormFactorSimulation}
\end{figure}

The number of events can now be estimated using the anticipated EicC parameters \cite{Chen:2018wyz, EicCWP, EicCWPEL}.
%one-year integrated luminosity of approximately 50\,fb$^{-1}$.
This leads to the statistical error projections shown in Fig.\,\ref{fig:FormFactorSimulation}.  In drawing this figure, the following cuts were used for event selection: $|t|<0.6\,$GeV and $W/{\rm GeV}\in (2,10)$.  Separating the data into ten bins on the domain $10 \leq Q^2/{\rm GeV}^2 \leq 30$, one finds statistical errors on the pion form factor data that are comparable with all other planned measurements.  Similar analysis reveals that EicC could also deliver a competitive measurement of the kaon form factor.
%, the statistical error increases with $Q^2$, but is always competitive with other existing proposals.

%%%  subsection - 3
\subsection{Measuring pion and kaon structure functions}
\label{ss:measurepiKSF}
As illustrated in Fig.\,\ref{fig:SullvianDiagrams}(b), one can also use a Sullivan process to measure NG mode structure functions.  In contrast to elastic form factor measurements, however, the meson target is disintegrated by the interaction and the cross-section is much larger.  In order to ensure that the DIS interaction is $e\pi$, the transverse momentum of the tagged final-state neutron needs to be small and its longitudinal momentum must exceed 50\% of that of the incoming proton.  The longitudinal momentum fraction $x_L := P_n\cdot q / P_p\cdot q$ \cite{Holtmann:1994rs, Chekanov:2002pf, Aaron:2010ab}, where $q$ is the photon momentum and $P_{p,n}$ are the momenta of the incoming proton and tagged outgoing neutron.
Kaon structure functions can be measured by tagging a $\Lambda$-baryon in the final state.

The differential cross-section for the leading-neutron tagged-DIS process is given by \cite{Sullivan:1971kd, Bishari:1972tx, Holtmann:1994rs, Chekanov:2002pf, Aaron:2010ab}:
\begin{align}
\label{dcsTDIS}
\frac{d^4\sigma}{dx_BdQ^2dx_Ldt} = \frac{4\pi\alpha^2}{x_BQ^4}\left(1-y+\frac{y^2}{2}\right)F_2^{LN(4)}(Q^2,x_B,x_L,t).
\end{align}
Here, $F_2^{LN(4)}(Q^2,x_B,x_L,t)$ is the tagged leading-neutron structure function, which can be factorised:
\begin{align}
\label{F2LN}
F_2^{LN(4)}(Q^2,x_B,x_L,t)=F_2^{\pi}(x_{\pi},Q^2, t)f_{\pi/p}(x_L,t) \,,
\end{align}
where $F_2^{\pi}(x_{\pi},Q^2,t)$ is the pion structure function and $f_{\pi/p}(x_L,t)$ is the pion flux in the proton.  The momentum fraction in the pion structure function is $x_\pi = Q^2/(2p_{\pi}\cdot q)=x_B/(1-x_L)$.

%
%The $(x_\pi,Q^2)$ domain accessible to EicC is shown in Fig.\,\ref{fig:NeutTagged-DIS-Domain}.  In order to focus on those events relevant for leading-neutron tagged DIS, one should apply the following cuts to the Monte Carlo simulation: $x_L>0.5$, $M_{X}^2=(P_p+P_e-P_n)^2>4\,$GeV$^2$.  As evident in the lower panel, EicC can map the domain $x_\pi \in (0.01,0.9)$ with precision; and this would certainly be sufficient to address the longstanding pion structure function controversy described in Sec.\,2.6.3.2.

\begin{figure}[t]
\includegraphics[width=0.95\textwidth]{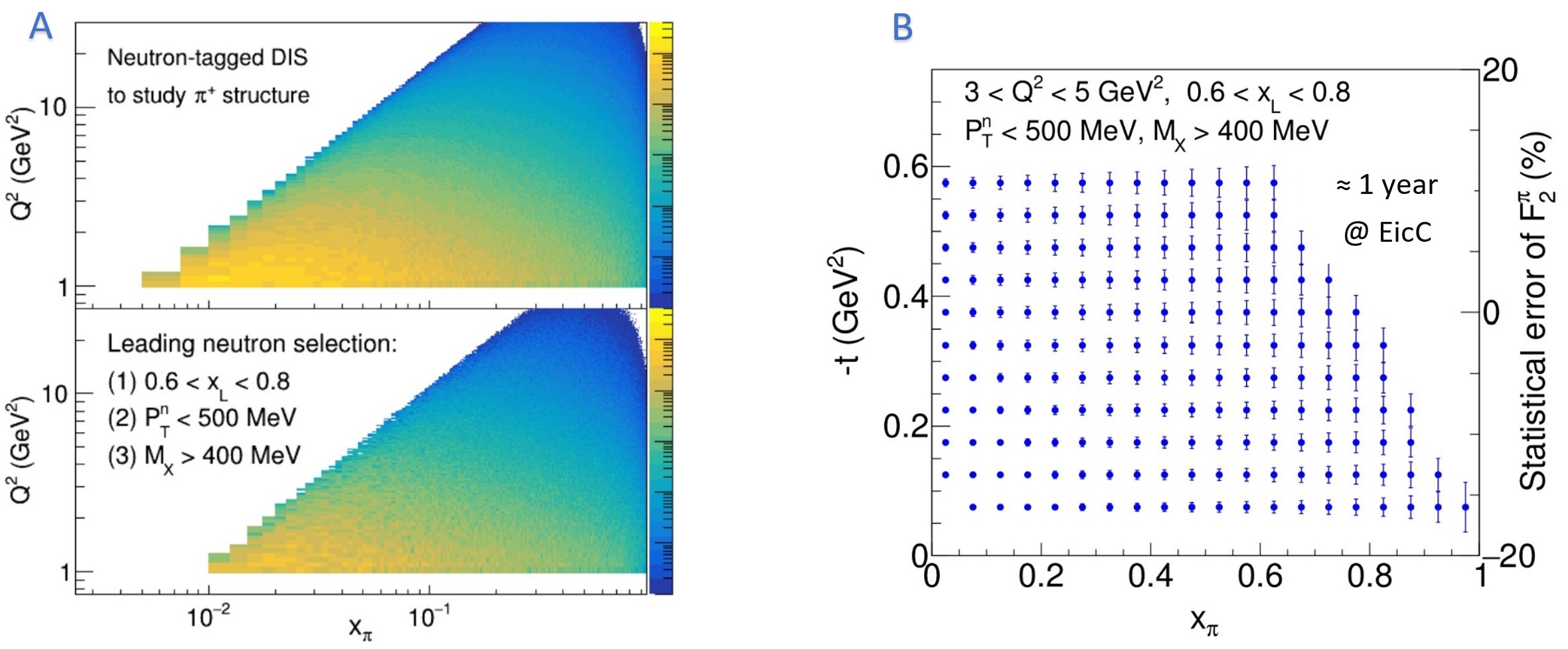}
\caption{\label{fig:NeutTagged-DIS-Domain}
\emph{Left panels}\,--\,A.  Kinematic coverage in leading-neutron tagged DIS at EicC as measured by the cross-section weighted event distribution in the $(x_{\pi},Q^2)$ plane: \emph{upper left} -- complete coverage; and \emph{lower left} -- domain available for pion structure function extraction.
\emph{Right panel}\,--\,B.  Statistical precision estimates for a pion structure function measurement at $Q^2\sim 4$ GeV$^2$ from roughly one year of running at EicC.  The centre of the bin is indicated by the left and bottom axis.  The right axis provides a reference for the size of the associated statistical error, which is consistently $\lesssim 5$\%.
}
\end{figure}

The $(x_\pi,Q^2)$ domain that is accessible to EicC is shown in Fig.\,\ref{fig:NeutTagged-DIS-Domain}A.  The estimate is based on a Monte-Carlo simulation of leading-neutron tagged DIS with the pion valence-quark distribution function taken from Ref.\,\cite{Gluck:1999xe}.  In order to focus on those events relevant for leading-neutron tagged DIS, the following cuts were applied: $0.6< x_L<0.8$,
%%  this cut "0.6 < x_L < 0.8" is taken as the cut for HERA experiment, to make sure the model above is solid & accurate.
%%
$M_{X}^2=(P_p+P_e-P_n - P_{e^\prime})^2>(0.4\,{\rm GeV})^2$, with the reduced event sample displayed in the lower panel of Fig.\,\ref{fig:NeutTagged-DIS-Domain}A.  Evidently, an EicC with the parameters under discussion could deliver precise results on the domain $x_\pi \in (0.01,0.95)$.  This is highlighted by Fig.\,\ref{fig:NeutTagged-DIS-Domain}B, which depicts the projected statistical precision of an EicC pion structure function measurement assuming one year of running.
Such data quality would certainly be sufficient to address the longstanding pion structure function controversy described in Sect.\,\ref{pionDFs}.

%%%  subsection - 4
\subsection{Deeply virtual meson production}
\begin{figure}[t]
\begin{tabular}{lcr}
\includegraphics[width=0.45\textwidth]{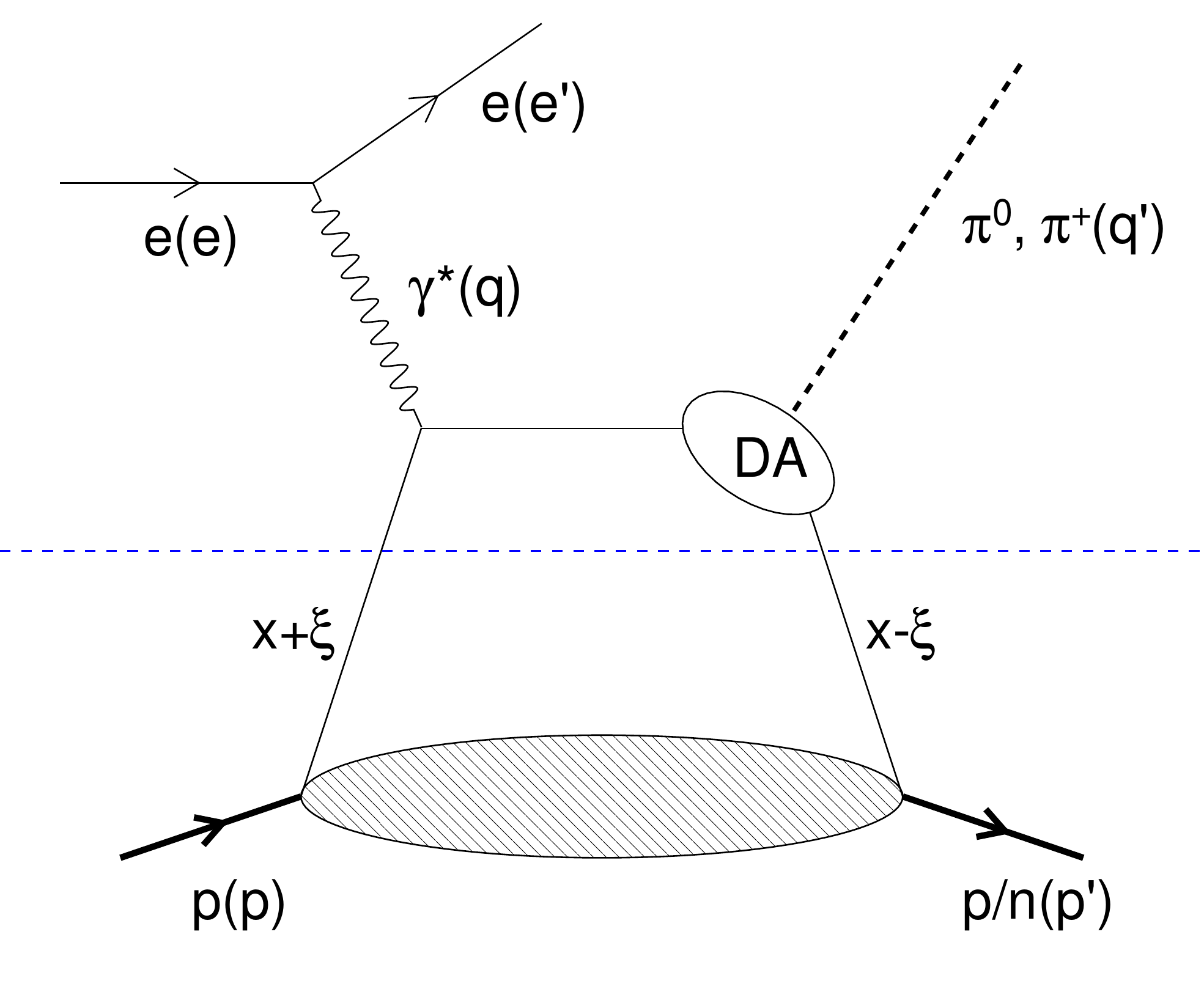} & \hspace*{1em} &
\includegraphics[width=0.45\textwidth]{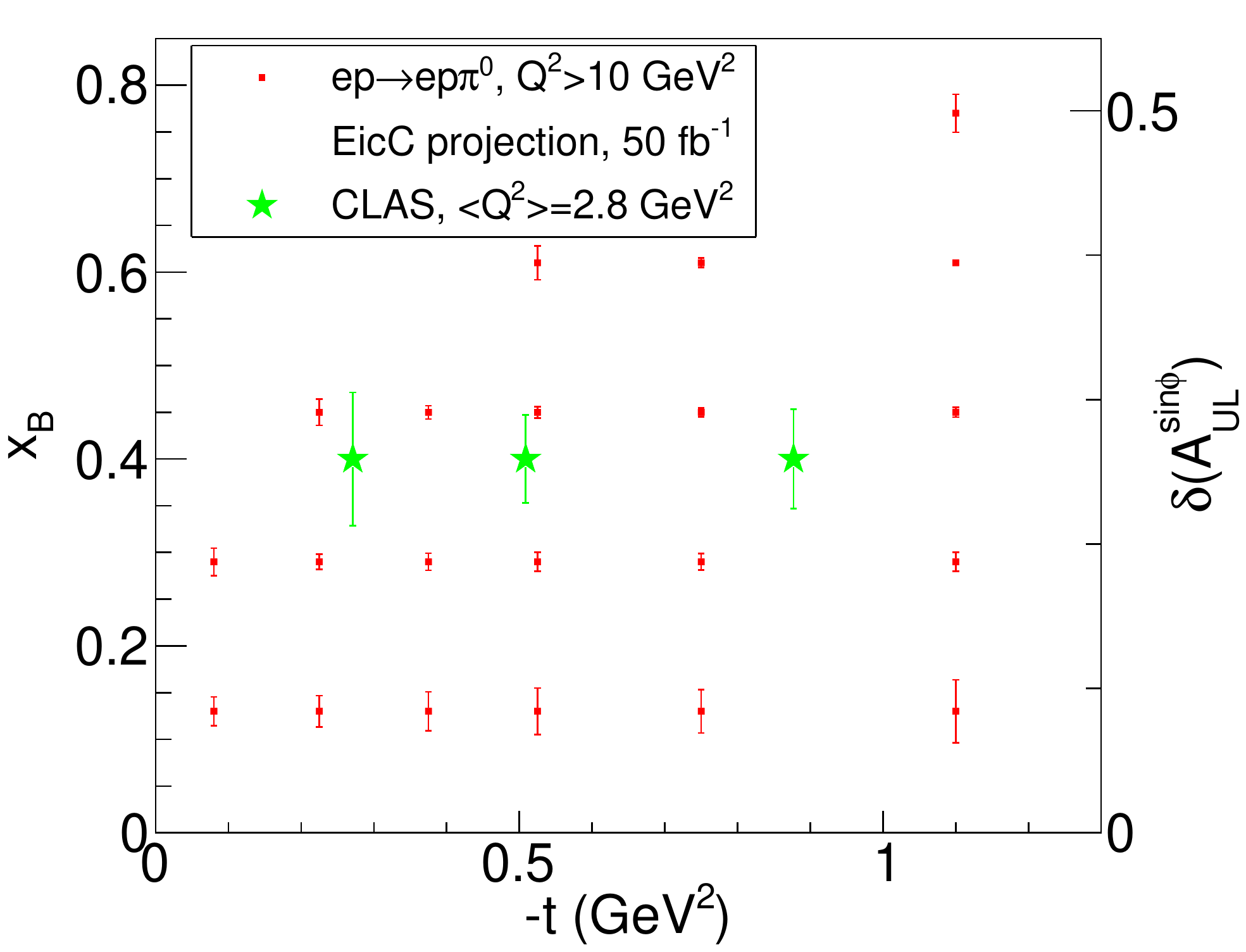}\\[-42ex]
\hspace*{1em}{\large{\textsf{A}}} & \hspace*{1em}{\large{\textsf{B}}} \hspace*{-1em} &
\vspace*{42ex}
\end{tabular}
\caption{\label{fig:DVMPDiagram}
\emph{Left panel--A}.  Handbag diagram that is typically imagined to be dominant in the deeply virtual production of $\pi^0$ and $\pi^+$.
\emph{Right panel--B}. Statistical error projections for the proton spin asymmetries in the $\pi^0$-DVMP process at EicC, with $Q^2>10\,$GeV$^2$.
}
\end{figure}

A three-dimensional (3D) imaging of the nucleon can be realised by measuring its generalised parton distribution functions (GPDs).  In fact, GPDs provide access not only to a nucleon tomography, but also, \emph{e.g}.\ to decompositions of its spin and a confinement pressure distribution \cite{Brodsky:2020vco}.   Experimentally, deeply virtual meson production (DVMP) $ep\to e p M$ is an important complement to deeply virtual Compton scattering (DVCS) measurements  \cite{Vanderhaeghen:1998uc, Goloskokov:2007nt, Goloskokov:2009ia, Goloskokov:2011rd}.  Moreover, deeply virtual pseudoscalar meson production provides a more sensitive way to extract polarised GPDs and is indispensable in constraining transverse GPDs.  Fig.\,\ref{fig:DVMPDiagram}A shows the handbag diagram that is typically assumed to describe DVMP in analyses of data for the extraction of nucleon GPDs.

For DVMP processes, EicC \cite{Chen:2018wyz, EicCWP, EicCWPEL} would provide access to both the region of transition between strong and perturbative QCD and the purely perturbative domain, making it ideal for testing a diverse array of nonperturbative and perturbative features in the theory of hard exclusive meson production \cite{Favart:2015umi}.  Importantly, light-meson DVMP channels have high event rates at EicC.  Such processes involving pseudoscalar mesons are of great interest because they select the polarised GPDs $\tilde{H}$, $\tilde{E}$, and transversity GPDs, $H_T$,  $\bar{E}_{T}$ \cite{Goloskokov:2009ia, Goloskokov:2011rd},
which cannot be measured well in other processes, such as DVCS.  As an example, consider that measurement of proton spin asymmetry $A_{UL}^{sin\phi}$ in pion DVMP is connected to $\langle\tilde{H}\rangle$ and $\langle\bar{E}_T\rangle$ \cite{Goloskokov:2011rd, Kim:2015pkf}.  Clean separation of different GPDs can be achieved by simultaneously measuring a diverse array of observables.

Using Monte-Carlo methods, one can study the energy and angular distributions of final-state particles and also estimate statistical errors in $\pi^0$ DVMP experiments at EicC.
The electrons and photons are concentrated in the region $|\eta|<3$ \cite{Chen:2018wyz, EicCWP, EicCWPEL}, which are covered with the barrel and end-cap detectors implemented in the current EicC design.  The largest number of final-state protons will be collected using the far-forward proton detector.
Regarding statistics, consider that one may decompose the unpolarised differential cross-section into a sum of four terms \cite{Goldstein:2013gra}:
\begin{align}
\frac{d^4\sigma}{dQ^2dx_Bdtd\phi_{\pi}} & =\Gamma(Q^2,x_B,s)  \nonumber \\
& \times \frac{1}{2\pi}\left[ \frac{d\sigma_T}{dt}+\epsilon\frac{d\sigma_L}{dt}
+\sqrt{2\epsilon(1+\epsilon)}\cos(\phi_{\pi})\frac{d\sigma_{LT}}{dt}
+\epsilon \, \cos(2\phi_{\pi})\frac{d\sigma_{TT}}{dt} \right].
\label{eq:XSection}
\end{align}
Here, the photon flux is
\begin{equation}
\Gamma(Q^2,x_B,s) = \frac{\alpha y^2 (1-x_B)}{2\pi x_B (1-\epsilon)Q^2},
\label{eq:PhotonFlux}
\end{equation}
where $y$ is the inelasticity, and $Q^2$ and $x_B$ are, respectively, the familiar square of the virtual-photon momentum and the Bjorken variable.  Then parametrisations of the response structure functions ($d\sigma_{T}/dt$, $d\sigma_{L}/dt$, $d\sigma_{LT}/dt$, $d\sigma_{TT}/dt$) can be developed from existing JLab data \cite{Bedlinskiy:2012be, Defurne:2016eiy}.  Finally, extrapolations to EicC energies are accomplished by assuming the following $W$- and $Q^2$-dependence:  $1/(W^2-M_p^2)^2$, $(Q^2)^{-3}$.  Combining these things with the anticipated EicC one-year integrated luminosity, one obtains the statistical uncertainty projections on the $A_{UL}^{sin\phi}$ asymmetry depicted in Fig.\,\ref{fig:DVMPDiagram}B.  Comparison with CLAS uncertainties very much favours EicC.  Furthermore and at least equally important, the much higher EicC interaction scales ($>10\,$GeV$^2$) mean that data analyses using the leading-twist handbag formalism are likely to be much better justified, in which case the resulting GPDs will be more realistic.

%%\begin{figure}[htp]
%%\centering
%%\includegraphics[width=0.48\textwidth]{DVMP-Asymmetry-projections.pdf}
%%\caption{
%%The statistical error projections of the proton spin asymmetries of the $\pi^0$-DVMP process on EicC, with $Q^2$ above 10 GeV$^2$.
%%}
%%\label{fig:DVMPSimulation}
%%\end{figure}

\section{Exotic Hadron Spectroscopy}
\label{exoticspectrum}
In addition to the lightest mesons and baryons, \textit{i.e}.\ pions, nucleons, \emph{etc}.,  discussed in the previous sections, there exists a rich zoo of hadrons with different flavour valence quarks.  The experimental discoveries of hadrons with $u$, $d$ and $s$ quarks spurred invention of the quark model \cite{GellMann:1964nj,Zweig:1964jf}, and finally led to the establishment of QCD as the fundamental theory of the strong interaction, which was followed by the discoveries of the heavier charm ($c$) and bottom ($b$) quarks.  The emergence of massive-hadron spectroscopy as a consequence of the interactions of these quarks and gluons is a prominent feature of nonperturbative QCD. Understanding how the hadron spectrum is organised presents another facet of the EHM problem discussed herein.

Historically, the constituent quark model has been quite successful in describing the hadron spectrum \cite{Godfrey:1985xj,Capstick:1986bm}, with some exceptions, such as the lightest scalar mesons, the nucleon's first positive- and negative-parity excitations, and the lightest negative-parity strange baryon $\Lambda(1405)$.
Qualitatively, this may be understood as a consequence of constituent light quarks acquiring a mass from DCSB \cite{Lane:1974he, Politzer:1976tv, Manohar:1983md} and Eq.\,\eqref{genS}, of the order $\Lambda_{\rm QCD}$, which sets a scale for hadron masses and their excitation energies.
However, this is not the whole story, as many resonant structures have been reported since 2003, when the \babar\ and Belle experiments at $B$ factories reported their unexpected observations of two very narrow resonances, the $D_{s0}^*(2317)$ \cite{Aubert:2003fg} and $X(3872)$ \cite{Choi:2003ue}, respectively.
Their particularity is marked by the fact that their masses are about 100\,MeV lower than the predictions of quark models. In particular, the mass of the $X(3872)$, which will hereafter be called $\chi_{c1}(3872)$ in accordance with the new nomenclature of the Particle Data Group (PDG) \cite{Zyla:2020}, precisely coincides with the $D^0\bar D^{*0}$ threshold.
These ground-breaking discoveries were followed by observations of many other hadronic resonances or resonance-like structures in various high-energy experiments. In addition to the $B$ factories already mentioned, the experiments include BESIII and CLEO at electron-positron colliders as well as ATLAS, CDF, CMS, D0 and LHCb at hadron colliders.

\begin{figure}[htp]
\centering
\includegraphics[width=0.87\textwidth]{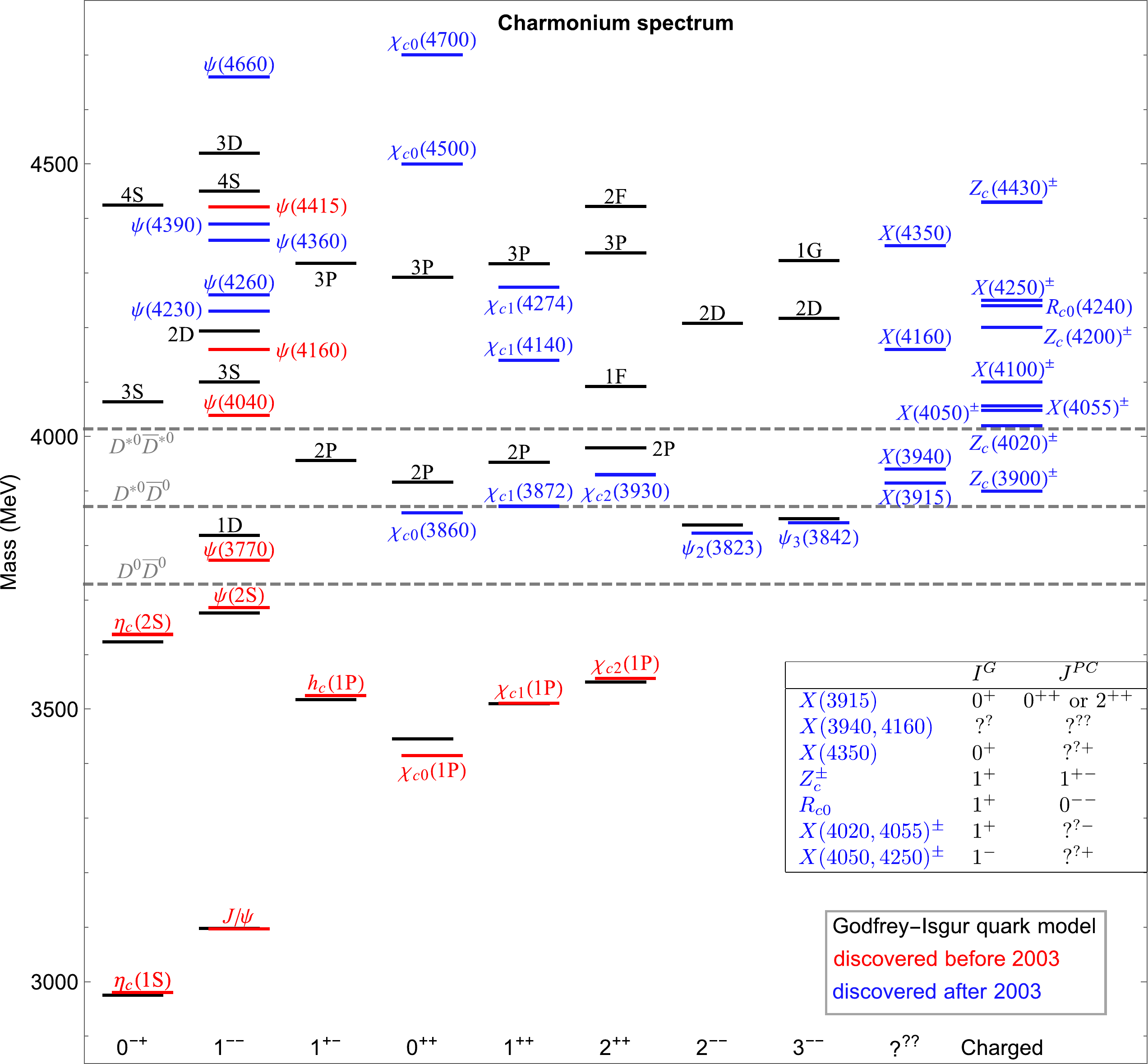}\\
\includegraphics[width=0.87\textwidth]{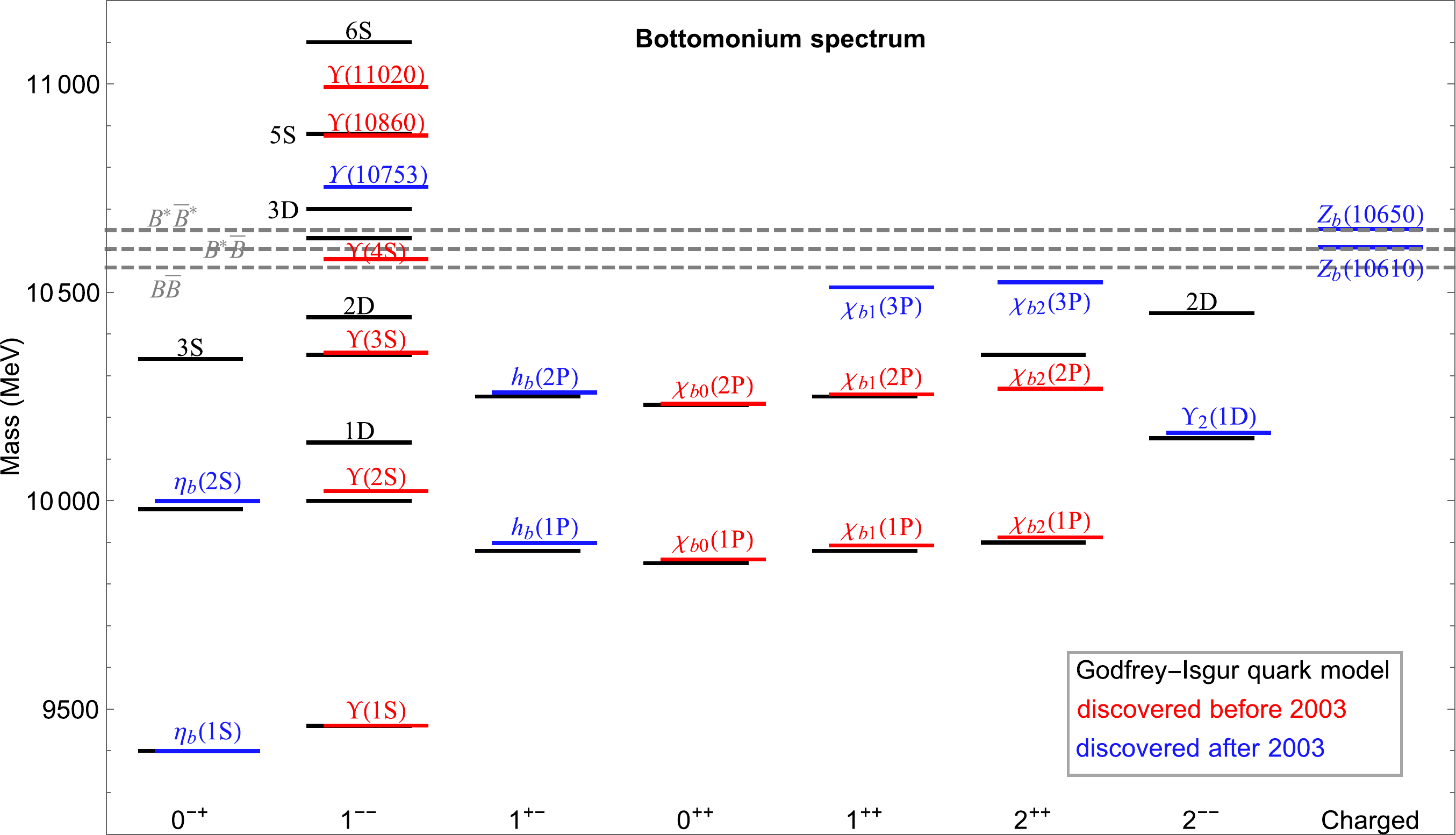}
\caption{The spectrum of the charmonia/bottomonia and charmonium-like/bottomonium-like structures collected in the full version of RPP \cite{Zyla:2020}. The states are named according to their quantum numbers $I^G(J^{PC})$, and $J^{PC}$ are shown for each column of neutral mesons. The red/blue lines correspond to the structures observed before/since the year 2003. For comparison, quark model predictions \cite{Godfrey:1985xj,Barnes:2005pb} are drawn as black lines, which are labelled by the orbital and radial excitations as $nL$.
The lowest three open-charm and open-bottom thresholds are shown as dashed lines.
\label{fig:spectrum}
}
\end{figure}

In particular, the hadron mass spectrum in the charmonium mass region has been remarkably enriched. This can be seen plainly from Fig.\,\ref{fig:spectrum}.
The spectrum is over-populated (see the vector sector which can be directly produced in $e^+e^-$ collisions and thus has been the best measured) in comparison with the expectations from any quark model of $c\bar c$ systems, which, since the $c$-quark mass is much larger than $\Lambda_{\rm QCD}$, were expected to be relatively cleanly described by quantum mechanical models like that developed at Cornell University \cite{Eichten:1978tg, Eichten:1979ms}.

In Fig.\,\ref{fig:spectrum}, quark model predictions \cite{Godfrey:1985xj,Barnes:2005pb} are presented for comparison.
Most of these new structures need to find explanations other than being regular $c\bar c$ mesons.
Thus, they are regarded as pronounced candidates for exotic hadrons, \emph{viz}.\ hadrons beyond the na\"{\i}ve quark-antiquark and three-quark pictures for mesons and baryons, respectively.
The puzzling charmonium-like systems are collectively called $XYZ$ states; yet some of them have well-measured quantum numbers, which have been used to name them in the latest version of the Review of Particle Physics (RPP) \cite{Zyla:2020}.

Although the classification of exotic and nonexotic hadrons is a quark-model notation, and thus not directly grounded in QCD, understanding how these excited charmonium-like states get their masses is a big challenge, and is a manifestation of the EHM problem in the excited heavy-flavour sector.
The current $c\bar c$(-like) spectrum is a mess and little consensus has been reached on how each individual amongst these structures should be interpreted. Explanations that have been considered include regular charmonia; hybrid states; compact tetraquarks; hadronic molecules; mixtures of different components; and non-particle explanations such as kinematical effects and resonance interference.
The experimental measurements and theoretical models have been summarised in a large number of review articles \cite{Jaffe:2004ph, Swanson:2006st, Klempt:2007cp, Klempt:2009pi, Brambilla:2010cs, Chen:2016qju, Hosaka:2016pey, Richard:2016eis, Chen:2016spr, Lebed:2016hpi, Esposito:2016noz, Dong:2017gaw, Guo:2017jvc, Ali:2017jda, Olsen:2017bmm, Kalashnikova:2018vkv, Liu:2019zoy, Brambilla:2019esw, Yamaguchi:2019vea, Guo:2019twa}.
Readers are referred to these comprehensive reports for more details. Herein, the focus is on how EicC can contribute to the understanding of heavy-flavour exotic hadron spectroscopy.

\subsection{Complementarity between EicC and other experiments}

The study of exotic hadrons, in particular in the heavy-flavour sector, has been and will remain an important element in ongoing and planned experiments, \emph{e.g}.\ BESIII \cite{Ablikim:2019hff, Yuan:2019zfo}, Belle-II \cite{Kou:2018nap}, upgrades of LHC \cite{Cerri:2018ypt} and \={P}ANDA \cite{Lutz:2009ff}.
The main physical processes for these experiments and those which have primarily contributed to the discoveries of the new hadron spectrum are as follows:
\begin{itemize}
  \item \emph{Weak decay processes}.  The main hadronic processes corresponding to $b\to c\bar c s$ are the decays of $b$-flavoured hadrons, such as the decays of $B$ mesons into a kaon plus a pair of charm and anti-charm mesons or a kaon plus a charmonium and other light mesons, and the decays of the $\Lambda_b$ into $Kp$ and a charmonium.
  The hidden-charm mesons and pentaquark(-like) structures can then be sought in the corresponding final states. Thus, the maximum mass for the charmonium(-like) structures in such processes is constrained to be $m_B-m_K\approx4.8\,$GeV, and that for hidden-charm pentaquarks is $m_{\Lambda_b}-m_K\approx5.1\,$GeV.\\[0.5ex]
  The most interesting charmonium-like state $\chi_{c1}(3872)$ was discovered in $B^\pm\to J/\psi\, \pi^+\pi^-K^\pm$ \cite{Choi:2003ue}, and the highest state in the charmonium spectrum shown in Fig.\,\ref{fig:spectrum} is the $\chi_{c0}(4700)$, observed by the LHCb Collaboration in $B^+\to J/\psi\, \phi \, K^+$\cite{Aaij:2016iza}. \\[0.5ex]
  Thus far, only three hidden-charm pentaquark candidates $P_c(4312)$, $P_c(4440)$ and $P_c(4457)$ have been reported by LHCb in $\Lambda_b\to J/\psi \, p \, K^-$ \cite{Aaij:2015tga, Aaij:2019vzc}. These weak decay processes have fixed masses for the initial particles, hence lack a way to study the initial-state energy dependence and thereby distinguish a genuine resonance from the signal produced by the so-called triangle-singularity kinematical effect, which owes to the on-shell propagation of all intermediate particles in a triangle diagram and is energy-sensitive.  (For a review, see Ref.\,\cite{Guo:2019twa}.)\vspace*{0.5ex}

  \item \emph{$e^+e^-$ collisions}.  Charmonium and bottomonium states with vector quantum numbers can directly be produced through a virtual photon. They can also be produced via initial-state radiation (ISR), the cross-section for which, however, is smaller by two orders of magnitude than direct production because of the suppression factor $\alpha_{e}\approx 1/137$.
  The $\psi(4260)$ was discovered by the \babar\ Collaboration in the ISR process $e^+e^-\to \gamma_\text{ISR}\,J/\psi\, \pi^+\pi^-$ \cite{Aubert:2005rm}, and a precise measurement by BESIII in direct production reduces the extracted mass to about 4.23\,GeV \cite{Ablikim:2016qzw}. (The state is now labelled as $\psi(4230)$ in the latest RPP version \cite{Zyla:2020}, although the $\psi(4260)$ remains.)  \\[0.5ex]
  States with other quantum numbers are much less observed as they need to be produced through the decays of higher states or two-photon collisions. The production rates from two-photon collisions and in radiative transitions from higher vector charmonia are suppressed by powers of $\alpha_e$ compared to that of the direct production of the vector states.  BEPC, at which BESIII is located, does not have enough energy to produce the $X$ and $\chi_{cJ}$ states with mass $\gtrsim3.9\,$GeV in association with light mesons, such as $\rho$ and $\phi$.\vspace*{0.5ex}
  % , and producing bottomonium-like structures through the decays of the $\Upsilon(10890,11020)$.

  \item \emph{Hadron-hadron/nuclei collisions}.  The LHC has enough energy to produce any hadrons, and the cross-sections are larger than those obtained with virtual photons. However, the large energy also leads to an overabundance of final-state particles. Consequently, in prompt production, the exotic hadron candidates need to be hunted in semi-inclusive processes and are often obscured by huge backgrounds. Furthermore, the final states that can efficiently be detected are also limited.
\end{itemize}

As one can see, there are pros and cons for each of the different types of experiment. An understanding of the spectroscopy in the heavy-flavour sector needs complementary inputs from them all. The high-luminosity EicC, whose energy range covers both the charm and bottom sectors, is expected to play an indispensable role here, especially in the following areas.
\begin{itemize}
  \item \emph{Charmonium(-like) states}. The spectrum of charmonium(-like) states above the lowest open-charm threshold shown in Fig.\,\ref{fig:spectrum} lacks a clear pattern and poses a serious challenge to the understanding of excited hadron spectroscopy, even at a qualitative level. Thus, the study of such states will be a major focus of EicC in the hadron spectroscopy sector.\\[0.5ex]
  Many such states were seen in decays into a charmonium (in most cases $J/\psi$) and light mesons. Yet, their masses are large enough to allow decays into open-charm channels. In particular, the most interesting states, $\chi_{c1}(3872)$ \cite{Gokhroo:2006bt, Aubert:2007rva, Adachi:2008sua} and $Z_c(3900)$ \cite{Ablikim:2013xfr, Ablikim:2015swa}, couple strongly to open-charm channels, and meson-meson channels should play an important role in forming these structures.\\[0.5ex]
  However, it is more difficult for the LHC experiments to search for structures in final states with open-charm mesons than in those with $J/\psi$ plus light mesons using inclusive prompt production processes.
  Moreover, some structures seen in $e^+e^-$ experiments, such as the $Z_c(3900)$ \cite{Ablikim:2013xfr, Liu:2013dau} and $Z_c(4020)$ \cite{Ablikim:2013wzq}, elude detection in $B$-meson decays for reasons unknown. It is thus important for these states to be sought in other experiments, such as EicC and \={P}ANDA, especially since their signals in $e^+e^-$ collisions may have contamination from kinematical singularity effects, as discussed in Refs.\,\cite{Wang:2013cya, Wang:2013hga, Liu:2014spa, Albaladejo:2015lob, Szczepaniak:2015eza, Pilloni:2016obd, Gong:2016jzb, Guo:2020oqk, Guo:2019twa}.\\[0.5ex]
  From Fig.~\ref{fig:spectrum}, one sees that other than the vector or charged states, most of the new structures have positive charge parity and mass larger than 3.9\,GeV.
  In $e^+e^-$ collisions, to have a relatively large production rate, such positive $C$-parity states need to be produced together with a negative $C$-parity light meson (such as $\rho$ and $\phi$). Thus, the highly excited states with masses larger than 4\,GeV are beyond the scope of the BESIII experiment, as mentioned above; they are also beyond the scope of the JLab 12\,GeV program \cite{Dudek:2012vr}, whose maximum reach is about 3.9\,GeV for the mass of $X$ in $ e^-p\to e^- p X$.\\[0.5ex]
  So far the only charmonium-like structure that has been observed in lepto-production is a negative $C$-parity structure with a mass of about 3.86\,GeV (COMPASS Collaboration \cite{Aghasyan:2017utv}), which nevertheless shows the feasibility of studying $XYZ$ states in lepto-/photo-production processes. Its quantum numbers are consistent with $J^{PC}=1^{+-}$ and it could be closely related to $h_c(2P)$, and positive $C$-parity states around that mass.
  One expects that all these hidden-charm states may be studied in detail at EicC through both hidden-charm and open-charm final states and the necessary simulations are being explored.\vspace*{0.5ex}
  %{\color{blue}FK: for some statements here, open-charm (and open-bottom later) mesons really need to be efficiently detected. Still don't have an answer whether this can be done at EicC.}
  % Therefore, in addition to the hidden-charm channels, the $XYZ$ charmonium-like states, including the highly excited ones beyond the capability of BESIII and JLab or those cannot be produced through the $B$ meson decays, can be studied through open-channel final states.

  \item \emph{Hidden-charm pentaquarks}.
  One sees a rich $XYZ$ spectrum above the lowest thresholds for a pair of charm and anti-charm mesons in Fig.\,\ref{fig:spectrum}, indicating the important role played by the open-charm meson-meson channels.
  This being true, no matter how the interaction happens, similar processes should also be able to produce a rich spectrum in the hidden-charm baryon sector above the $\Lambda_c\bar D$ threshold.
  Indeed, hidden-charm pentaquark states above 4\,GeV have been predicted to emerge from the interactions between a charmed meson plus baryon pair \cite{Wu:2010jy, Wang:2011rga, Yang:2011wz, Wu:2012md, Xiao:2013yca, Uchino:2015uha, Karliner:2015ina} and the first discovery of such states was made by the LHCb Collaboration in 2015 \cite{Aaij:2015tga}, after these theoretical predictions. The measurement was updated in 2019 \cite{Aaij:2019vzc}, reporting three narrow hidden-charm pentaquarks $P_c(4312)$, $P_c(4440)$ and $P_c(4457)$.
  There is also a hint for a narrow $P_c(4380)$ \cite{Du:2019pij}, and models predict \cite{Xiao:2013yca, Liu:2019tjn, Du:2019pij} the existence of more states (their heavy-quark spin partners).\\[0.5ex]
  Nevertheless, all these observations need to be confirmed in an independent experiment. There is a proposal to search for the $P_c$ states in Hall C at JLab \cite{Meziani:2016lhg}. The GlueX experiment in Hall D at JLab measured near-threshold $J/\psi$ photoproduction, but no signal was seen in an energy range covering the LHCb $P_c$ masses \cite{Ali:2019lzf}. The non-observation implies a small branching fraction for $P_c\to J/\psi p$, roughly in the range $(0.05-2)$\% \cite{Cao:2019kst}.
  Then the dominant decay modes of the $P_c$ should be open-charm channels, similar to the case of the $\chi_{c1}(3872)$ and $Z_c(3900)$, which include $\bar D^{(*)}\Lambda_c$ and $\bar D^{(*)}\Sigma_c$ \cite{Lin:2017mtz,Lin:2019qiv,Du:2019pij,Dong:2020nwk}.
  A search in the $\Lambda_b\to \Lambda_c \bar D^0K^-$ was performed at LHCb.  No signal has yet been found \cite{Piucci:2019vsk}.\\[0.5ex]
  At EicC, the $P_c$ can be sought in both exclusive processes, such as $\gamma p\to P_c\to J/\psi p$, $\bar D^{(*)}\Lambda_c$, $\bar D^{(*)}\Sigma_c$,  and semi-inclusive processes.
  Similar hidden-charm states with $s$ quarks will also be sought in order to establish SU(3) flavour multiplets.\vspace*{0.5ex}

 \item \emph{Bottom hadrons}.
  The c.m.\ energy range of EicC extends to 20\,GeV, Fig.\,\ref{EicCcomparison}; so, it covers the region for open and hidden-bottom hadrons as well.
  Since both the charm and bottom quark masses are much larger than $\Lambda_{\rm QCD}$, hadronic systems with a $c$-quark replaced by a $b$-quark should have similar properties -- a consequence of the so-called heavy-quark flavour symmetry. Although for systems with a pair of heavy ($c$, $b$) quarks, heavy quark flavour symmetry does not hold, the general pattern for the spectra should still be similar.
  From Fig.\,\ref{fig:spectrum}, one sees that the number of bottomonia and bottomonium-like states is much smaller than their charm analogues. Many  more \textit{should} exist.\\[0.5ex]
  Similarities and differences between the bottomonium(-like) spectrum and the charmonium(-like) spectrum should contain important information for understanding the role of EHM in excited hadrons.
  However, existing facilities have limited ability to search for the hidden-bottom states. The latest few discoveries of bottomonium(-like) states, which include the $Z_b(10610)$, $Z_b(10650)$ \cite{Belle:2011aa} and $\Upsilon(10753)$ \cite{Abdesselam:2019gth}, were all made in $e^+e^-$ collisions at the Belle experiment. Belle-II \cite{Kou:2018nap} does not have enough energy to produce positive $C$-parity states above the open-bottom thresholds as that would require them to be produced in association with a $\rho$, $\omega$ or $\phi$ meson. (The production rates of the radiative processes should be smaller by two orders of magnitude.)\\[0.5ex]
  Similar to the $P_c$ states, hidden-bottom pentaquark states are also predicted to exist \cite{Wu:2010rv, Xiao:2013jla, Wu:2017weo, Azizi:2017bgs, Shen:2017ayv, Yamaguchi:2017zmn, Lin:2018kcc, Anwar:2018bpu, Huang:2018wed, Yang:2018oqd, Gutsche:2019mkg, Peng:2019wys}.
  Such states are out of reach to \={P}ANDA \cite{Lutz:2009ff} and are difficult to detect in LHC experiments owing to large backgrounds since they need to be produced in prompt processes.\\[0.5ex]
  The EicC energy region is ideal for the study of bottom hadrons, and will play a unique role in the study of bottomonium-like structures, hidden-bottom pentaquarks and excited bottom hadrons, so long as the open-bottom hadrons can be efficiently detected.
  %{\color{blue}FK: again the open-bottom hadrons need to be efficiently detected.}
\end{itemize}

\begin{figure}[t]
  \centering
  \includegraphics[width=0.48\textwidth]{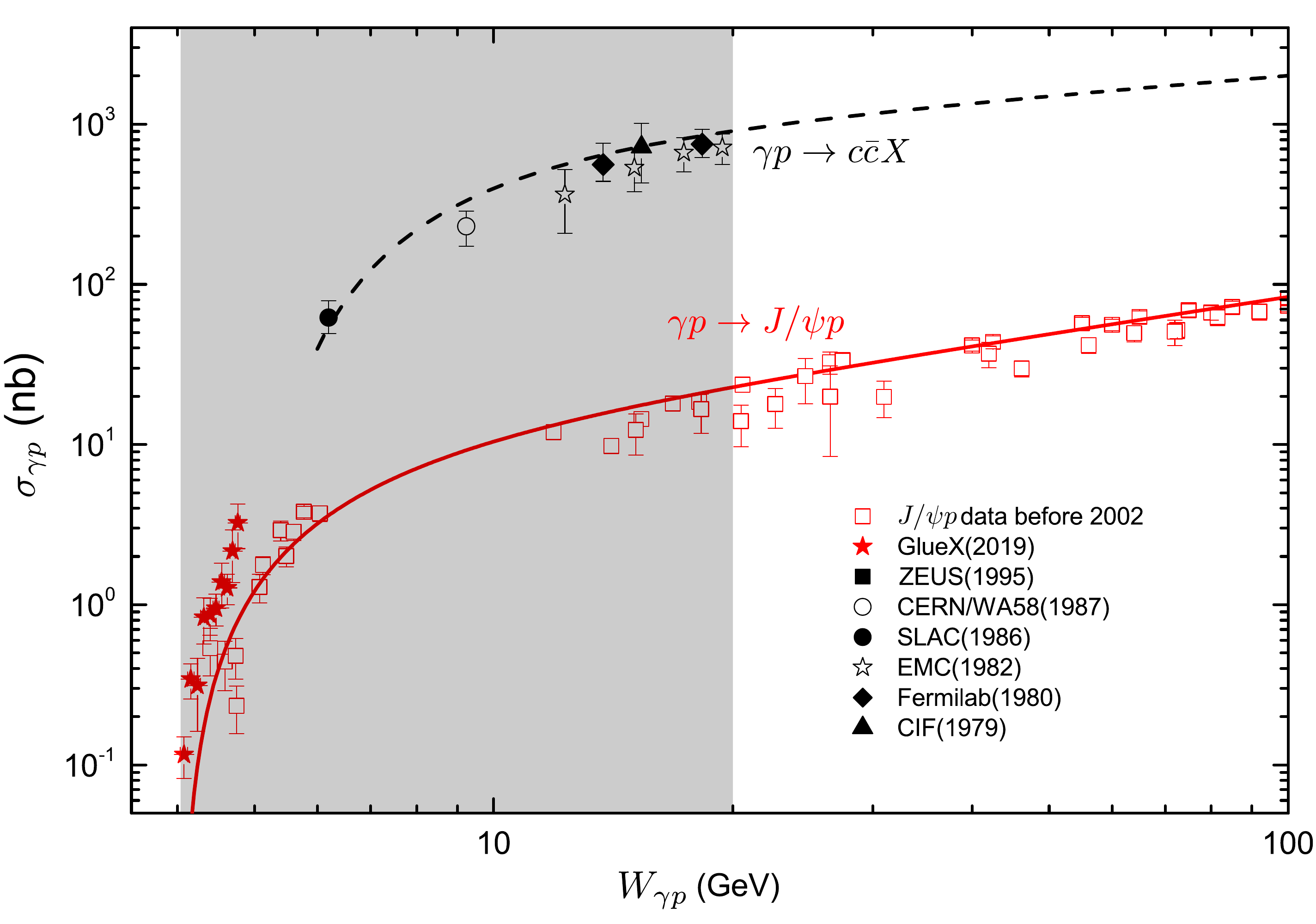}\hfill
  \includegraphics[width=0.48\textwidth]{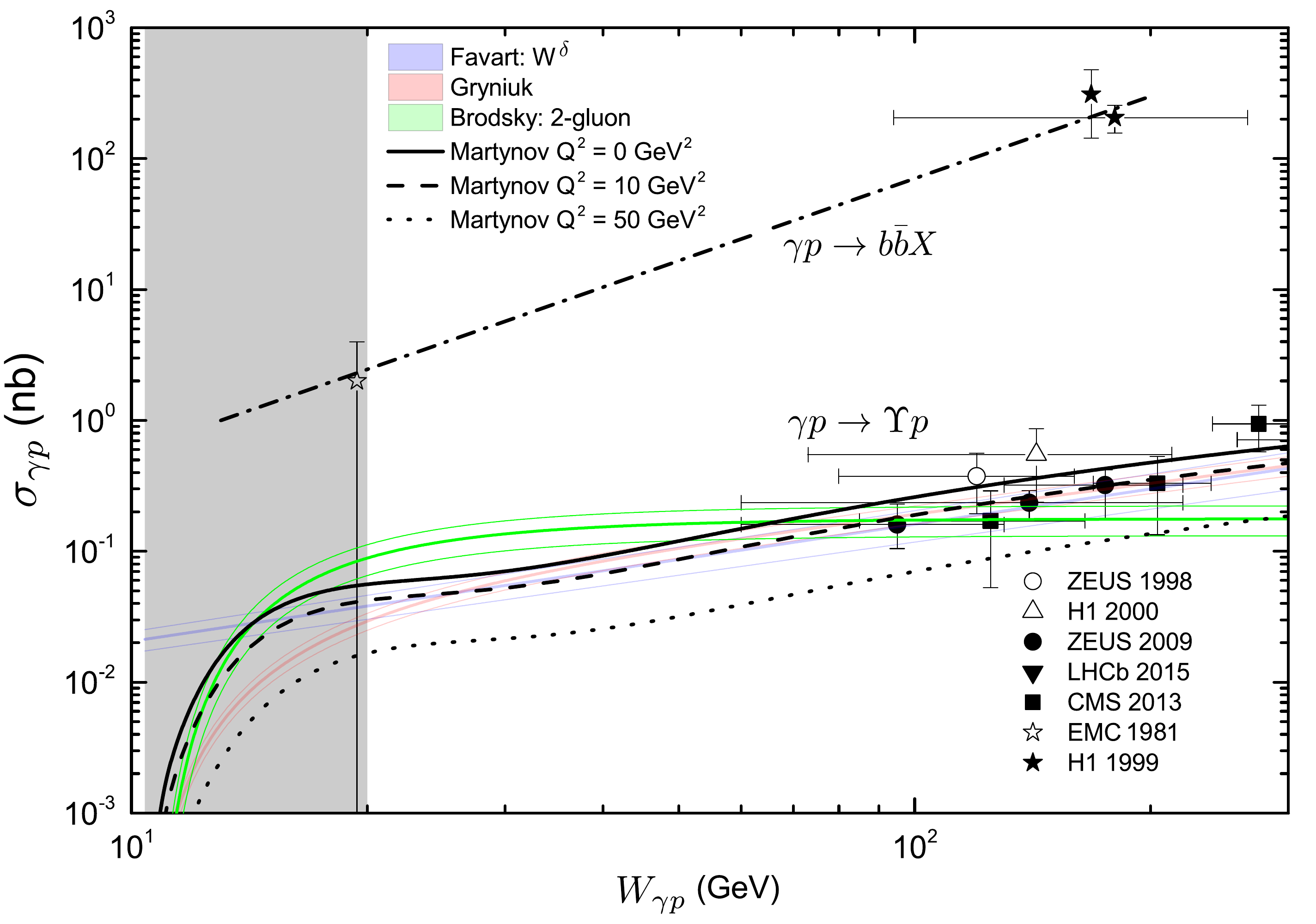}
  \caption{\label{fig:xection_QQbar}
  The c.m.\ energy dependence of the cross-sections for the inclusive and exclusive processes $\gamma p\to c\bar c X$ and $\gamma p\to J/\psi \, p$ (left) and $\gamma p\to b\bar b X$ and $\gamma p\to \Upsilon p$ (right). The shaded areas correspond to the EicC energy coverage. The $J/\psi \,p$ data before 2002 are from Refs.\,\cite{Gittelman:1975ix, Camerini:1975cy, Binkley:1981kv, Denby:1983az, Frabetti:1993ux, Adloff:2000vm, Chekanov:2002xi} and the GlueX data are from Ref.\,\cite{Ali:2019lzf}.
  The $c\bar c X$ data are from Fermilab \cite{Atiya:1979hx, Clark:1980ed}, EMC \cite{Aubert:1982tt}, SLAC \cite{Adamovich:1986gx}, and ZEUS \cite{Derrick:1995sc}.
  The $\Upsilon p$ data are from ZEUS \cite{Breitweg:1998ki, Chekanov:2009zz}, H1 \cite{Adloff:2000vm}, CMS \cite{CMS:2016nct} and LHCb \cite{Aaij:2015kea}. The $b\bar b X$ data are from EMC \cite{Aubert:1981gx} and H1 \cite{Adloff:1999nr}. The data in the charm sector were fitted with the model in Ref.\,\cite{Gryniuk:2016mpk}. The models used in fitting the bottom sector data include the deeply virtual meson production model by Favart {\it et al.} \cite{Favart:2015umi}, the 2-gluon exchange model by Brodsky {\it et al.} \cite{Brodsky:2000zc}, the parametrization by Gryniuk {\it et al.} \cite{Gryniuk:2016mpk}, and the dipole pomeron model by Martynov {\it et al.} ($Q^2 =$0, 10, 50\,GeV$^2$) \cite{Martynov:2001tn,Martynov:2002ez}.
  The figure is adapted from Ref.\,\cite{EicCWP}.}
\end{figure}

It should also be noted that for the commonly considered mechanisms of producing hidden-flavour exotic hadron candidates \cite{Liu:2008qx, Galata:2011bi, Lin:2013mka, Lin:2013ppa, Huang:2013mua, Adolph:2014hba, Wang:2015jsa, Wang:2015lwa, Kubarovsky:2015aaa, Karliner:2015voa, Huang:2016tcr, Blin:2016dlf, Aghasyan:2017utv, Meziani:2016lhg, Joosten:2018gyo, Paryev:2018fyv, Wang:2019krd, Goncalves:2019vvo, Wang:2019zaw, Cao:2019gqo, Wu:2019adv, Winney:2019edt, Xie:2020niw, Paryev:2020jkp, Yang:2020eye}, the kinematics of the eletro-/photoproduction processes at the EicC are free from the ambiguity in interpreting some hidden-charm resonance signals, which is introduced by triangle singularities that could occur in $b$-flavoured hadron decays or $e^+e^-$ collisions.
Moreover, the EicC's polarised beams enable determination of the quantum numbers of hadron resonances, such as spin and parity.

\subsection{Estimates of the production rates at EicC}

There have been various model estimates and suggestions for the photoproduction rates of hidden-charm and hidden-bottom mesons and baryons\,\cite{Liu:2008qx, Galata:2011bi, Lin:2013mka, Lin:2013ppa, Huang:2013mua, Adolph:2014hba, Wang:2015jsa, Wang:2015lwa, Kubarovsky:2015aaa, Karliner:2015voa, Huang:2016tcr, Blin:2016dlf, Aghasyan:2017utv, Meziani:2016lhg, Joosten:2018gyo, Paryev:2018fyv, Wang:2019krd, Goncalves:2019vvo, Wang:2019zaw, Cao:2019gqo, Wu:2019adv, Winney:2019edt, Xie:2020niw, Paryev:2020jkp, Yang:2020eye}.
For a given state, the estimate depends on both the assumed mechanism and on the branching fractions, which are rarely known to the desired precision. Some estimates of the production rates are sketched here.

% \begin{table}[b]
%   \centering
%   \caption{Estimates of the production rates for selected exotic states in exclusive processes at EicC \cite{EicCWP, EicCWPEL}.
%   An integrated luminosity of 50\,fb$^{-1}$ is assumed. Both the $J/\psi$ and $\Upsilon$ are reconstructed from their decays into lepton pairs, and the branching fractions have been accounted for in the event estimates. }
%   \label{tab:QQbar_prod_exclusive}
% \begin{tabular}{l | c  c   c}
% \hline \hline
%      Exotic states & Production, decay processes & Detection efficiency & Expected events
%     \\\hline

%   $P_c(4440)$   &   $ep \to e P_c(4440)$, $P_c(4440)\to p  J/\psi$         &   $\sim$30\% &   20$-$2200 \\

%     $P_b(\text{narrow})$   &   $ep \to e P_b(\text{narrow})$, $P_b{\rm (narrow)}\to p  \Upsilon$    &   $\sim$30\%  &    0$-$20  \\

%     $P_b(\text{wide})$   &   $ep \to e P_b(\text{wide})$, $P_b{\rm (wide)}\to p  \Upsilon$       &   $\sim$30\% &  0$-$200 \\

%     $\chi_{c1}(3872)$  &   $ep \to e\chi_{c1}(3872)p$, $\chi_{c1}(3872) \to \pi^+\pi^- J/\psi$  &  $\sim$50\% &   0$-$90  \\

%   $Z_c(3900)^+$   &   $ep \to eZ_c^+(3900)n$, $Z_c^+(3900) \to \pi^+  J/\psi$      &   $\sim$60\%  &   90$-$9300 \\\hline
%     \hline
% \end{tabular}
% \end{table}

Measurements exist of cross-sections for the exclusive production of the $J/\psi$\,\cite{Gittelman:1975ix, Camerini:1975cy, Binkley:1981kv, Denby:1983az, Frabetti:1993ux, Adloff:2000vm, Chekanov:2002xi, Ali:2019lzf} and $\Upsilon$\,\cite{Breitweg:1998ki, Adloff:2000vm, CMS:2016nct, Aaij:2015kea, Chekanov:2009zz}, as well as for the inclusive processes $\gamma p \to c\bar c X$\,\cite{Atiya:1979hx, Clark:1980ed, Aubert:1982tt, Adamovich:1986gx, Derrick:1995sc} and $\gamma p \to b\bar b X$\,\cite{Aubert:1981gx, Adloff:1999nr}, where $X$ represents anything else in the final state. The data together with model fits are shown in Fig.\,\ref{fig:xection_QQbar}.
This figure shows that the photoproduction cross section of the exclusive process $\gamma p\to J/\psi p$ is on the order of 10\,nb for $W_{\gamma p}$, the c.m.\ energy of the $\gamma p$, within 10 to 20\,GeV.
The eletroproduction process $e^-p\to e^- J/\psi p$ is about two orders of magnitude smaller. Thus, one expects
% $\mathcal{O}(5\times 10^6)$
a few millions of
$J/\psi$ events at EicC.
% assuming an integrated luminosity of 50\,fb$^{-1}$.
The cross-section for the inclusive process $\gamma p\to c\bar c X$ is about 50 times larger.
Since almost all excited charmed mesons (baryons) decay into $D$ ($\Lambda_c$) and their antiparticles, emitting light hadrons and/or photons, one can expect that there must be many more $D$ and $\Lambda_c$ events. This provides the opportunity to study hidden-charm mesons and baryons, not only in final states involving the $J/\psi$, but also in open-charm final states.
A rough estimate using a vector-meson dominance model for the exclusive process productions of hidden-charm states leads to the expectation of $\mathcal{O}(10-10^3)$ events for $P_c$, $\mathcal{O}(10^0-10^2)$ events for $\chi_{c1}(3872)$, and $\mathcal{O}(10^2-10^4)$ events for $Z_c(3900)^+$, considering the decay chains of the hidden-charm states into $J/\psi\to l^+l^-$ and light hadrons and assuming a reasonable detection efficiency~\cite{EicCWP, EicCWPEL}.

% Rough estimates for the discovery potential of a few selected hidden-charm states via exclusive processes \cite{EicCWP, EicCWPEL} are listed in Table\,\ref{tab:QQbar_prod_exclusive}. The estimates were made using a vector-meson dominance model.

\begin{figure}[tb]
  \centering
  \includegraphics[height=5cm]{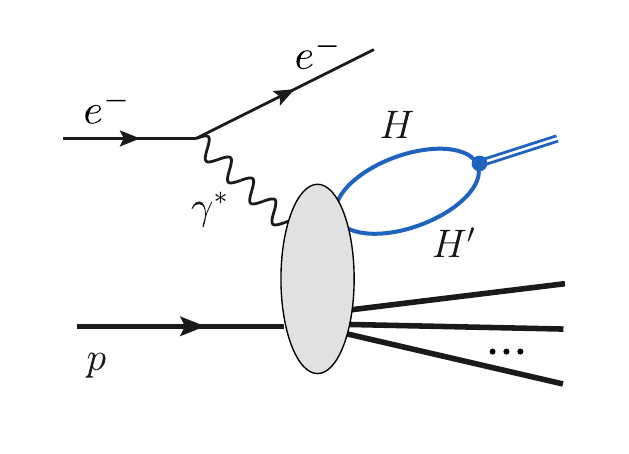}\hspace{1cm}
  \includegraphics[height=5cm]{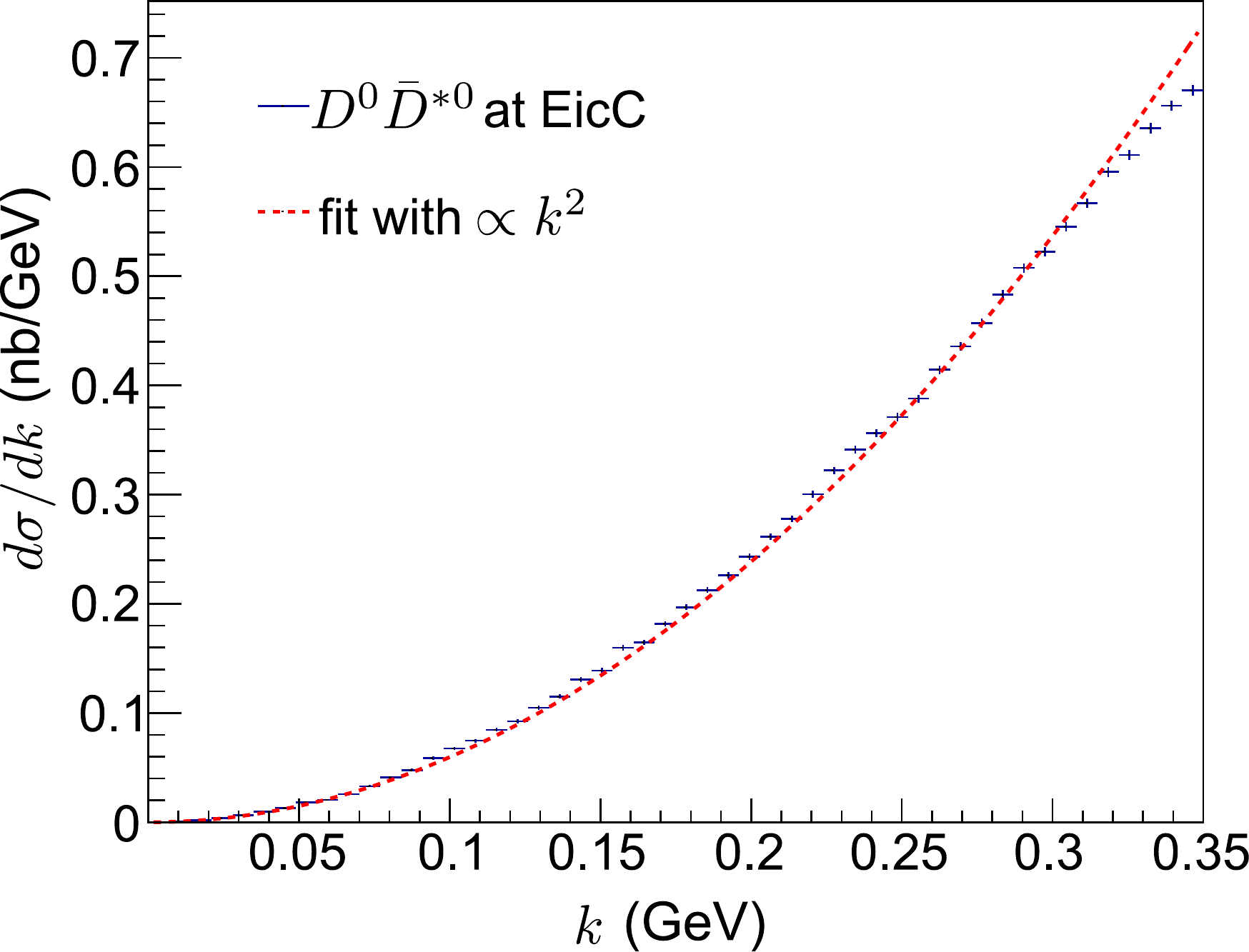}
  \caption{\emph{Left panel}.  Mechanism considered in Ref.\,\cite{Yang:2020} for producing charmonium-like states and hidden-charm pentaquarks that couple strongly to open-charm channels.
  \emph{Right panel}. Differential cross-section for the semi-inclusive production of $D^{0}\bar D^{*0}$ \cite{Yang:2020} generated using Pythia \cite{Sjostrand:2006za}.}
  \label{fig:semi-inclusive}
\end{figure}

The cross-sections for semi-inclusive production of hidden-charm states that couple strongly to a charm meson-baryon pair may be estimated by considering the mechanism shown in the left panel of Fig.\,\ref{fig:semi-inclusive} \cite{Yang:2020}: a pair of open-charm hadrons (denoted as $HH'$ in the figure) are produced, which then merge to form the hidden-charm state of interest.
As long as the hidden-charm state couples dominantly to the considered hadron pair, this mechanism should provide an important contribution to the production, owing to unitarity.
Such a mechanism has been considered before in estimating the production of charmonium-like states at hadron colliders \cite{Bignamini:2009sk, Artoisenet:2010uu, Guo:2013ufa, Guo:2014sca, Albaladejo:2017blx} and the resulting cross-sections are in agreement with experimental measurements when the momentum integration range in the $HH'$ Green function extends up to a few hundreds of MeV \cite{Artoisenet:2010uu, Guo:2014sca, Albaladejo:2017blx}.
The mechanism may be applied to the $\chi_{c1}(3872)$ and $Z_c(3900)$, which are known to couple strongly to the $D\bar D^*$ pair (and their antiparticles) \cite{Gokhroo:2006bt, Aubert:2007rva, Adachi:2008sua, Ablikim:2013xfr, Ablikim:2015swa, Zyla:2020}.
The differential cross section for the $HH'$ pair is simulated using the Pythia event generator~\cite{Sjostrand:2006za}, which nicely follows a distribution proportional to $k^2$ in the small $k$ region, where $k$ is the magnitude of the c.m.\ momentum of the hadron pair.

The right panel of Fig.\,\ref{fig:semi-inclusive} shows such a distribution for the $D^0\bar D^{*0}$ at EicC with the energies for the electron and proton beams taken as 3.5 and 20\,GeV, respectively.
The coupling of the hidden-charm state to the $HH'$ pair can be computed in a model assuming the state to be a $HH'$ hadronic molecule, where the $HH'$ Green function is modeled using a loop integral regulated by a  Gaussian form factor with a cutoff in the range between 0.5 and 1\,GeV.  In this way, the integrated cross-sections for the semi-inclusive production of the $\chi_{c1}(3872)$ and $Z_c(3900)$ are estimated to be $\mathcal{O}( 0.01\,{\rm nb})$ and $\mathcal{O}( 0.5\,{\rm nb})$, respectively; and those for each of the $P_c$ states, including those seen with LHCb and their spin partners, predicted in hadronic molecular models \cite{Xiao:2013yca, Liu:2019tjn, Du:2019pij}, are estimated to be $\mathcal{O}( 1\,{\rm pb})$ at EicC. Thus, EicC would have an excellent chance to observe them \cite{Yang:2020}.

For the bottom sector, one sees that the photoproduction cross section for $\Upsilon p$ in the range of $W_{\gamma p}\in [15,20]$\,GeV is of $\mathcal{O}(10^{-2}\,\text{nb})$. Correspondingly, the $e^- p \to e^- \Upsilon p$ cross section is of $\mathcal{O}(0.1\,\text{pb})$ (see also Ref.\,\cite{Xu:2020uaa}).
% Assuming an integrated luminosity of 50\,fb$^{-1}$, $\mathcal{O}(10^4)$ $\Upsilon$ events are expected through such an exclusive process.
Yet, the cross-section in the near-threshold region could be enhanced owing to the possible existence of hidden-bottom pentaquarks. The study in Ref.\,\cite{Cao:2019gqo} shows that there is good reason to search for $P_b$ at EicC.
The inclusive $b\bar b X$ cross section is two orders-of-magnitude higher, leading to the expectation of millions of  $B$-mesons  and $\Lambda_b$-baryons. It is desirable to search for the hidden bottom systems in $B^{(*)}\Lambda_b/\Sigma_b^{(*)}$ final states, if the weak decay particles can be efficiently detected.

% Double polarization observables \cite{Winney:2019edt}

% Identify signals from non-resonant background: \cite{Yang:2020eye}

% The cross section of the semi-inclusive production of $c\bar c p$ is about 50 times larger, see the left panel of Fig.~\ref{fig:sec2_4_xsect_ccbar}. The cross section for the electroproduction process is about two orders of magnitude smaller due to a factor of $\alpha$. One may estimate that the event number of $J/\psi$ produced from the exclusive process is about  $\mathcal{O}(5\times10^6)$.

%   Bottom hadrons: The shaded areas in Fig.~\ref{fig:sec2_4_xsect_ccbar} correspond to the EicC energy region. In the right panel, the experimental data are taken from LHCb (filled triangles~\cite{Aaij:2015kea}), ZEUS (empty circles~\cite{Breitweg:1998ki}, filled circles~\cite{Chekanov:2009zz}), H1 (empty triangles~\cite{Adloff:2000vm}), and CMS (filled squares~\cite{CMS:2016nct}). The models used to fit the data include, the empirical formula for the DVMP [to be defined] model (Favart)~\cite{Favart:2015umi}, the 2-gluon exchange model by Brodsky {\it et al.}~\cite{Brodsky:2000zc}, the parametrization by Gryniuk {\it et al.}~\cite{Gryniuk:2016mpk}, and the dipole pomeron model by Martynov {\it et al.} ($Q^2 =$0, 10, 50~GeV$^2$)~\cite{Martynov:2001tn,Martynov:2002ez}.

\section{Epilogue}
\label{epilogue}
Having uncovered the explicit source for $\lesssim 2$\% of the mass of visible matter, attention is now shifting to searches for the origin and the explanation of the remaining $\gtrsim 98$\%.  That emergent hadronic mass (EHM) is very probably to be found in the strong interaction sector of the SM, \emph{i.e}.\ QCD.  This is the purview of hadron physics, which is therefore on the cusp of an exciting period of discovery.

Existing facilities have revealed states and features of strongly interacting systems that were scarcely imagined in the previous millennium; and new facilities and apparatus promise to take science further into the heart of strongly interacting matter.

Hadron theory is challenged.  Nevertheless, diverse paths of progress are being followed.  In key directions, the tools are reaching a level of maturity that will enable some of the secrets locked in nonperturbative phenomena to be understood.

Looming largest amongst the challenges to hadron theory is gluon and quark confinement; but from a modern perspective, this problem does not stand alone.  As sketched herein, confinement and EHM are very likely intimately connected.  Their common origin may well be the QCD trace anomaly, whose strength is expressed in the emergence of GeV-size mass-scales for dressed-gluons and -quarks, even in the absence of a Higgs mechanism.

The discussion herein has drawn fundamental connections between the QCD trace anomaly and the spectra and structure of hadrons, which can be tested and strengthened by an array of experiments that are enabled by modern accelerator technology.  It has simultaneously argued that an EicC with the parameters under discussion would be ideally placed in the accelerator landscape to capitalise on modern theory advances and play a leading role in building a bridge across the last frontier within the SM, \emph{viz}.\ in locating the source of the bulk of visible mass.

\begin{acknowledgements}
We are grateful for constructive comments from
X.~Cao,
Z.-F.~Cui,
O.~Denisov,
F.~Gao,
S.~Goloskokov,
Y.-T.~Liang
and Z.~Yang.
This work is supported in part by:
the Chinese Academy of Sciences (CAS), under Grant Nos.\,XDB34030300, QYZDB-SSW-SYS013;
Jiangsu Province \emph{Hundred Talents Plan for Professionals};
the National Natural Science Foundation of China (NSFC), under Grant Nos.~11835015, 11947302, 11961141012 and 11621131001 (the Sino-German Collaborative Research Center CRC110 ``Symmetries and the Emergence of Structure in QCD");
and by the CAS Center for Excellence in Particle Physics (CCEPP).
\end{acknowledgements}

\newpage

\section*{Abbreviations}
%\addcontentsline{toc}{section}{\mbox{\hspace*{0.0\parindent}{Abbreviations}}}
\addcontentsline{toc}{section}{\mbox{\hspace*{1.5em}{Abbreviations}}}

\label{abbreviations}
The following abbreviations are used in this manuscript:\\[-2ex]
\noindent
\begin{longtable}{ll}
ATLAS & LHC detector  \\
\babar\ & detector at SLAC \\
Belle (Belle-II) & detector at Japan's high energy accelerator research complex in Tsukuba \\
BEPC & Beijing Electron Positron Collider \\
BESIII & detector at BEPC \\
BNL & Brookhaven National Laboratory \\
BRST & Bechi-Rouet-Stora-Tyutin (gauge transformation in quantised theory) \\
CDF & detector at the FNAL \\
CERN & European Laboratory for Particle Physics \\
%BS & Bethe-Salpeter \\
%BSE & Bethe-Salpeter equation\\
%CLS & coordinated lattice simulations \\
CLAS & detector in Hall-B at JLab\\
CLEO & detector at the Cornell University electron storage ring \\
COMPASS & detector at CERN \\
COMPASS++/AMBER & planned upgrade and expansion of COMPASS \\
c.m. & center of mass \\
CMS & detector at LHC \\
%CSM & continuum Schwinger-function method \\
%DA & distribution amplitude \\
%PDF & parton distribution function \\
D0 & detector at the Fermi National Accelerator Facility \\
DCSB & dynamical chiral symmetry breaking \\
DGLAP & Dokshitzer-Gribov-Lipatov-Altarelli-Parisi (evolution equations) \\
DIS & deep inelastic scattering \\
DF & distribution function \\
DSE & Dyson-Schwinger equation\\
DVCS & deeply virtual Compton Scattering \\
DVMP & deeply virtual meson production \\
DY & Drell-Yan (process) \\
EIC & electron ion collider in the USA \\
EicC & electron ion collider in China \\
%EFT  & effective field theory\\
EHM & emergent hadronic mass\\
EMC & particle physics collaboration at CERN \\
FNAL (Fermilab) & Fermi National Accelerator Facility \\
GlueX & experiment in Hall D at JLab \\
GPD & generalised parton distribution \\
H1 & detector at a facility in Hamburg \\
ISR & initial state radiation \\
JLab & Thomas Jefferson National Accelerator Facility\\
LHC & large hadron collider \\
LHCb & LHC beauty experiment \\
lQCD \hspace*{1em}  & lattice-regularised quantum chromodynamics\\
MARATHON & JLab experiment E12-010-103 \\
NG & Nambu-Goldstone (boson or mode) \\
%PFF & parton fragmentation function\\
\={P}ANDA & detector proposed for a new facility in Germany \\
PDA & parton distribution amplitude \\
PDG & Particle Data Group \\
PI & process independent \\
%QED & quantum electrodynamics\\
pQCD & perturbative quantum chromodynamics\\
QCD & quantum chromodynamics \\
RGI & renormalisation group invariant\\
RPP & Review of Particle Physics tabulation of empirical information\\
%SM & Standard Model (of Particle Physics)\\
%RBC & Riken-Brookhaven-Columbia lattice-QCD collaboration \\
SLAC & Stanford Linear Accelerator Center \\
SM & Standard Model of particle physics \\
%SPM & Schlessinger point method \\
%UKQCD & United Kingdom lattice-QCD collaboration
VMD & vector meson dominance \\
ZEUS & detector at a facility in Hamburg
\end{longtable}

%\bibliographystyle{../../zProc/z10/z10KITPC/h-physrev4}
%\bibliography{../../CollectedBiB}   % name your BibTeX data base

%\addcontentsline{toc}{toc}{\mbox{\hspace*{0.0\parindent}{References}}~\dotfill\hspace*{1.2em} }
\addcontentsline{toc}{section}{\mbox{\hspace*{1.5em}{References}}}

\bibliographystyle{cj} %%%-- Correct for FBS
\bibliography{CollectedBiBFBS}

\end{document}